\documentclass[preprints,review,accept,pdftex,moreauthors]{mdpi} 
\newcommand*\aap{Astron. Astrophys.}
\newcommand{\aaps}{Astron. Astrophys. Suppl. Ser.}
\newcommand*\aj{Astron. J.}
\newcommand*\apj{Astrophys. J.}
\newcommand*\apjl{Astrophys. J. Lett.}
\newcommand*\apjs{Astron. Astrophys. Suppl. Ser.}
\newcommand{\apss}{Astrophys. Space Sci. }
\newcommand{\araa}{Annu. Rev. Astron. Astrophys.}
\newcommand{\jcap}{J. Cosmol. Astropart. Phys.}
\newcommand*\mnras{Mon. Not. R. Astron. Soc.}
\newcommand{\na}{New Astron.}
\newcommand{\nat}{Nature}
\newcommand{\pasp}{Publ. Astron. Soc. Pac. }
\newcommand{\prc}{Phys. Rev. C}
\newcommand{\prd}{Phys. Rev. D}
\newcommand{\prl}{Phys. Rev. Lett.}
\newcommand*\physrep{Phys. Rep.}
\newcommand{\ssr}{Space Sci. Rev.}
\usepackage{longtable}
\firstpage{1} 
\pubvolume{1}
\issuenum{1}
\articlenumber{0}
\pubyear{2026}
\copyrightyear{2026}
\externaleditor{Firstname Lastname} 
\datereceived{22 February 2026} 
\daterevised{21 April 2026} 
\dateaccepted{24 April 2026} 
\datepublished{ } 

\Title{High-Synchrotron-Peaked BL Lacs as Multi-Messenger Sources: Connecting Ultra-High-Energy Cosmic Rays and Neutrinos}

\Author{Luiz Augusto Stuani Pereira $^{1,2,}$* and Rita C. Anjos $^{3,4,5,6,7,}$*}

\AuthorNames{Luiz Augusto Stuani Pereira and Rita C. Anjos}

\address{%
$^{1}$ \quad Unidade Acadêmica de Física, Universidade Federal de Campina Grande (UAF-UFCG), R. Aprígio Veloso, 882, Campina Grande 58429-900, PB, Brazil\\
$^{2}$ \quad Instituto de Física, Universidade de São Paulo (IFUSP), R. do Matão, 1371, São Paulo 05508-090, SP, Brazil\\
$^{3}$ \quad Departamento de Engenharias e Exatas, Universidade Federal do Paraná (UFPR), Rua Pioneiro,
  \linebreak  Palotina 85950-000, PR, Brazil\\
$^{4}$ \quad Programa de Pós-Graduação em Física, Departamento de Física, Universidade Estadual de Londrina (UEL), Rodovia Celso Garcia Cid Km 380, Londrina 86057-970, PR, Brazil\\
$^{5}$ \quad Programa de Pós-Graduação em Física e Astronomia, Universidade Tecnológica Federal do Paraná (UTFPR), Av. Sete de Setembro, 3165, Curitiba 80230-901, PR, Brazil\\
$^{6}$ \quad Programa de Pós-Graduação em Física Aplicada, Universidade Federal da Integração Latino-Americana, Av. Tarquínio Joslin dos Santos, 1000, Foz do Iguaçu 85867-670, PR, Brazil\\
$^{7}$ \quad Departamento de Física, Universidade Federal do Espírito Santo, Núcleo de Astrofísica e Cosmologia (Cosmo-Ufes), Av. Fernando Ferrari, 514, Vitória 29075-910, ES, Brazil}

\corres{Correspondence: luizstuani@uaf.ufcg.edu.br (L.A.S.P.); ritacassia@ufpr.br (R.C.A.)}

\abstract{High-synchrotron-peaked (HSP) BL Lac objects are extreme particle accelerators whose synchrotron emission peaks at high frequencies, typically in the UV-to-X-ray band \mbox{($\nu_{\rm peak} > 10^{15}$ Hz;} $\nu_{\rm peak} \geq 10^{17}$ for EHSPs), implying electron Lorentz factors of order $10^5$--$10^6$. Their relative proximity ($z \lesssim 0.5$), clean radiation environments, and favorable Hillas parameters make them prime candidates for ultra-high-energy cosmic ray (UHECR) acceleration beyond $10^{19}$ eV and for neutrino production above 100 TeV. The 2017 association of IceCube-170922A with the flaring blazar TXS 0506+056 provided compelling evidence for blazars as neutrino sources, while an archival neutrino flare from 2014–2015 with no clear electromagnetic counterpart (13 events) revealed additional complexity in the emission mechanism. This review examines HSP physical properties, identifies them through WISE-based infrared selection (the 2WHSP and 3HSP catalogs, $\sim$2000 sources), and contrasts leptonic synchrotron self-Compton models with hadronic alternatives. We assess the observational evidence linking HSPs to high-energy neutrinos and UHECRs, finding that extreme baryonic loading ($L_p/L_e \sim 10^3$--$10^5$) strains energetic budgets, Auger composition measurements favor heavy nuclei over proton-dominated scenarios, and the near-isotropy of UHECR arrival directions is difficult to reconcile with rare beamed sources. Potential resolutions involving magnetic reconnection, structured jets, and duty cycle effects are discussed. Next-generation facilities, including IceCube-Gen2, KM3NeT, CTAO, IXPE, and AugerPrime/TA $\times$ 4, will probe key observables to either establish HSP BL Lacs as sources of the highest-energy cosmic particles or redirect the search toward alternative \mbox{accelerator classes.}}


\begin{document}

\section{Introduction}
\label{sec:intro}

\textls[-15]{Active Galactic Nuclei (AGNs) rank among the most energetic persistent sources of electromagnetic radiation in the observable Universe, powered by the gravitational accretion of matter onto supermassive black holes (SMBHs), harboring masses in the range \mbox{$10^6$--$10^{10}\,M_{\odot}$ \cite{Lynden1969, Rees1984, Salpeter1964, Zeldovich1964}.} The release of gravitational potential energy in the accretion flow drives a rich phenomenology spanning over twenty decades of the electromagnetic spectrum. Within the diverse AGN zoo, approximately 10--15\% of objects are classified as ``radio-loud,''} distinguished by the presence of powerful, collimated, relativistic jets that channel energy, momentum, and magnetic flux from the immediate vicinity of the SMBH out to kiloparsec and even megaparsec \mbox{scales \cite{Blandford1979, Blandford1977, Begelman1984, Bridle1984}.} The physical mechanism responsible for jet launching is widely attributed to the extraction of rotational energy from either the spinning black hole \cite{Blandford1977} or the inner accretion disk \cite{Blandford1982}, mediated by large-scale organized magnetic fields threading the ergosphere or the disk surface.

When these relativistic jets happen to be oriented at a small viewing angle \linebreak  ($\theta_\mathrm{obs} \lesssim 10^\circ$--$15^\circ$) between the jet axis and the observer's line of sight, special relativistic effects, namely Doppler boosting and aberration, dramatically amplify the observed luminosity by factors of $\delta^{3+\alpha}$ (where $\delta = [\Gamma(1 - \beta\cos\theta_\mathrm{obs})]^{-1}$ is the Doppler factor and $\alpha$ the spectral index) and compress the observed variability timescales \cite{Blandford1979, Lind1985}. We note that this viewing angle is distinct from the jet opening angle $\theta_\mathrm{jet}$, which characterizes the lateral extent of the jet cone and has a median value of $\sim$0.1~rad ($\approx$5.7$^\circ$) from VLBI \mbox{observations \cite{Pushkarev2009, Jorstad2005}.} A source is observed as a blazar when the line of sight falls within the jet cone, i.e. $\theta_\mathrm{obs} \lesssim \theta_\mathrm{jet}$. Doppler boosting becomes significant when $\theta_\mathrm{obs} \lesssim 1/\Gamma$; for typical blazar bulk Lorentz factors $\Gamma \sim 10$--$20$, this corresponds to $\theta_\mathrm{obs} \lesssim 3^\circ$--$6^\circ$, so the most extreme boosting occurs for viewing angles well within the jet cone and significantly smaller than the $10^\circ$--$15^\circ$ upper limit commonly quoted for blazar classification \cite{Urry1995, Hovatta2009}. These oriented objects are collectively known as blazars \cite{Urry1995, Antonucci1993} and constitute the most extreme subclass of AGNs. Blazars are further subdivided into flat-spectrum radio quasars (FSRQs), which display prominent broad emission lines, and BL Lacertae (BL Lac) objects, characterized by featureless or nearly featureless optical spectra with equivalent widths $\text{EW} < 5$\,\AA\ \cite{Stickel1991, Stocke1991, Marcha1996}. This spectral dichotomy reflects fundamental differences in the accretion regime and the radiative environment of the jet \cite{Ghisellini2009, Ghisellini2011}.

The broadband spectral energy distribution (SED) of blazars is dominated by non-thermal emission and exhibits a characteristic double-humped structure extending from radio wavelengths to very-high-energy (VHE; $E > 100$\,GeV) $\gamma$-rays \cite{Fossati1998, Donato2001, Ghisellini2017}. The low-energy component, peaking at frequencies between the infrared and the X-ray band, is universally attributed to synchrotron radiation emitted by relativistic electrons (and possibly positrons) gyrating in the jet's magnetic field \cite{Urry1995, Rybicki1986}. The origin of the high-energy component, peaking in the MeV---TeV regime, remains the subject of the intense and ongoing theoretical debate that lies at the heart of multi-messenger astrophysics. Based on the rest-frame frequency of the synchrotron peak ($\nu_{\text{peak}}$), blazars are classified into low-synchrotron-peaked (LSP; $\nu_{\text{peak}} < 10^{14}$\,Hz), intermediate-synchrotron-peaked \linebreak  (ISP; $10^{14} < \nu_{\text{peak}} < 10^{15}$\,Hz), and high-synchrotron-peaked (HSP; $\nu_{\text{peak}} > 10^{15}$\,Hz) sources \cite{Padovani1995, Abdo2010, Giommi2012}. This classification scheme, rooted in the original ``blazar sequence'' phenomenology \cite{Fossati1998, Ghisellini1998}, has proven to be a powerful organizational framework, although its interpretation as a single physical sequence governed solely by jet power remains \mbox{debated \cite{Padovani2007, Giommi2012b, Finke2013}.}

HSP BL Lac objects occupy a unique position in this landscape. Their SEDs indicate that the jet environment operates as an extreme particle accelerator, capable of energizing electrons to Lorentz factors of $\gamma_e \sim 10^5$--$10^6$ \cite{Chang2017, Costamante2001a, Tavecchio2010b}. While leptonic models, specifically synchrotron self-Compton (SSC) scenarios, in which the same population of ultrarelativistic electrons up-scatters its own synchrotron photon field to $\gamma$-ray energies, have achieved considerable success in reproducing the broadband SEDs of many HSP sources \cite{Maraschi1992, Tavecchio1998, Ghisellini1996, Katarzynski2001, Finke2008}, they encounter significant difficulties when confronted with certain observational phenomena. A particularly striking challenge is posed by so-called ``orphan flares'': episodes in which a prominent VHE $\gamma$-ray outburst occurs without any accompanying enhancement in the X-ray synchrotron emission. The archetypal example is the 2002 orphan flare observed in the TeV blazar 1ES~1959 + 650 \cite{Krawczynski2004, Halzen2005, Daniel2005}, but similar events have since been reported in other sources, including Mrk~421 and PKS~2155$-$304 \cite{Blazejowski2005, Aharonian2009}. Such decoupled variability behavior is inherently difficult to reconcile with one-zone SSC models, in which the X-ray and $\gamma$-ray emissions are inextricably linked through the same electron population, and provides compelling, if circumstantial, evidence for the contribution of a distinct hadronic component to the high-energy emission \cite{Bottcher2005, Bottcher2013, Reimer2005}. Constraints on the minimum jet power of TeV BL~Lac objects in the $p\gamma$ model \cite{Xue2019a}, two-zone lepto-hadronic emission frameworks applied to TXS~0506+056 \cite{Xue2019b}, and time-dependent lepto-hadronic modeling of \mbox{Mrk~421 \cite{Xue2025}} have further developed these approaches, demonstrating that multi-zone and time-dependent treatments are essential for self-consistent modeling of blazar multi-messenger emission.

The hadronic interpretation posits that the high-energy hump in the SED originates, at least in part, from processes initiated by protons (and heavier nuclei) accelerated to ultra-relativistic energies within the jet \cite{Mannheim1993, Mucke2001, Mucke2003, Aharonian2000a}. These processes include proton-synchrotron radiation \cite{Aharonian2000a, Mucke2001}, photohadronic ($p\gamma$) interactions producing pions whose decay products (secondary electrons, positrons, photons, and neutrinos) cascade through the radiation field \cite{Mannheim1993, Kelner2008}, and in denser environments, proton--proton ($pp$) collisions \cite{Berezinsky1990, Romero2018}. This hadronic paradigm connects HSP blazars to two of the most enduring open questions in astroparticle physics: the origin of ultra-high-energy cosmic rays (UHECRs, $E \gtrsim 10^{18}$~eV) and the production of astrophysical high-energy neutrinos. Several extreme HSPs have been modeled in this framework, studying UHECR acceleration in HSP jets, $\gamma$-ray emission from line-of-sight interactions of escaping UHECRs with background photon fields, and the resulting cosmogenic neutrino flux \cite{Das2020}.

The question of UHECR origin has persisted for over six decades since their discovery by \cite{Linsley1963}. The celebrated ``Hillas Criterion'' \cite{Hillas1984} provides a necessary (though not sufficient) condition for particle acceleration, requiring that the magnetic rigidity of the source \mbox{($B \times R$,} where $B$ is the magnetic field strength and $R$ the characteristic size) be sufficient to confine particles up to the observed energies of $\sim$$10^{20}$\,eV. This criterion restricts the viable astrophysical accelerators to a select few source classes, among which the relativistic jets of AGNs (particularly blazars) feature prominently \cite{Hillas1984, Waxman1995, Norman1995}. Observations by the Pierre Auger Observatory and the Telescope Array have established that the UHECR spectrum features a suppression above $\sim$$4 \times 10^{19}$\,eV \cite{Abraham2008, AbuZayyad2013}, consistent with the Greisen--Zatsepin--Kuzmin (GZK) limit \cite{Greisen1966, Zatsepin1966}, and that the composition appears to become progressively heavier with increasing energy \cite{Aab2014, Bellido2018}. Crucially, unlike high-luminosity FSRQs that are bathed in intense external photon fields from the broad line region (BLR) and the dusty torus \cite{Sikora1994, Dermer2009}, HSP BL Lacs are characterized by radiatively inefficient accretion flows and a relative paucity of dense ambient photon fields \cite{Ghisellini2009}. This photon-poor environment theoretically permits UHECR-accelerating HSP jets to escape the source without suffering catastrophic photohadronic or photodisintegration energy losses, making HSPs a favorable candidate for UHECR accelerators \cite{Murase2012, Murase2014, Tavecchio2015b, Resconi2017}.

If protons are indeed accelerated to ultra-high energies in HSP jets, the very same $p\gamma$ interactions responsible for UHECR energy losses must also produce high-energy neutrinos. Charged pions ($\pi^\pm$) generated in photohadronic collisions decay via $\pi^+ \to \mu^+ + \nu_\mu \to e^+ + \nu_e + \bar{\nu}_\mu + \nu_\mu$ (and the charge-conjugate chain for $\pi^-$), yielding neutrinos with energies typically a low percentage of the parent proton energy \cite{Kelner2008, Atoyan2001, Stecker1991}. The detection of a diffuse sky flux of astrophysical neutrinos by the IceCube Neutrino Observatory in 2013 \cite{IceCube2013, IceCube2015}, with energies extending from $\sim$TeV to several PeV, opened an entirely new observational window onto the high-energy Universe and marked the dawn of neutrino astronomy. This landmark discovery was followed by the watershed multi-messenger event of \mbox{22 September 2017,} when the high-energy muon neutrino IceCube-170922A was found to be spatially and temporally coincident with the flaring $\gamma$-ray blazar \mbox{TXS~0506+056,} with a significance of $\sim$$3\upsigma$ \cite{IceCube2018, IceCube2018b}. A subsequent archival analysis of IceCube data revealed an independent $3.5\upsigma$ excess of neutrinos from the direction of TXS~0506+056 in 2014--2015: the best-fit Gaussian-shaped time profile had a width of \mbox{$\sim$110 days,} while the best-fit box-shaped window spanned 158~days and contained $13\pm5$ events above the background \cite{IceCube2018b}. Both of these estimates refer to the same statistically significant excess and are complementary parameterizations reported in that study. Although TXS~0506+056 is classified as an intermediate-synchrotron-peaked blazar \cite{Padovani2019}, this event provided the first suggestive direct evidence linking blazar jets to neutrino production and sparked intense theoretical efforts to model its multi-messenger emission \cite{Keivani2018, Murase2018, Cerruti2019, Gao2019}. More recently, a growing body of statistical evidence has implicated the broader blazar population in the diffuse neutrino flux \cite{Buson2022}, including analyzes highlighting correlations with specific \mbox{HSP subsamples.}

Identifying individual point sources within the diffuse neutrino flux requires robust statistical power, which, in turn, demands complete, deep, and well-characterized catalogs of potential electromagnetic counterparts. The construction of such catalogs for HSP BL Lacs has been a major community effort. The 1WHSP \cite{Arsioli2015}, 2WHSP \cite{Chang2017}, and 3HSP \cite{Chang2019} catalogs were developed precisely to address this need, providing the largest multi-frequency-selected samples of confirmed and candidate HSP blazars to date. These catalogs employ a multi-step selection procedure, combining radio, infrared (from \textit{WISE}), and X-ray survey data to identify sources exhibiting non-thermal SEDs consistent with the HSP classification. They serve as the foundational basis for population-level analyzes, including neutrino stacking searches that seek a cumulative signal from the aggregate HSP population \cite{Giommi2020a, Plavin2020, Plavin2021}, positional cross-correlations with UHECR arrival directions \cite{Resconi2017, Abbasi2023}, and broadband SED modeling campaigns that constrain the physical parameters of the jets \cite{Costamante2018, Acciari2020}.

The preference for HSP BL~Lacs over other blazar subclasses, namely FSRQs, LSP, and ISP BL~Lacs, as primary UHECR accelerators rests on several complementary physical arguments. While FSRQs possess more powerful jets ($L_\mathrm{jet} \sim 10^{46}$--$10^{48}$ erg~s$^{-1}$) and, in principle, a larger magnetic flux $BR$, their emission regions are typically located at parsec-scale distances from the black hole, where the field has already declined significantly, placing standard FSRQ parameters only marginally above the UHECR threshold. The compact, high-field regions of extreme HSPs ($B \sim 10$--$100$~G, $R \sim 10^{14}$--$10^{15}$~cm), by contrast, place them firmly above the acceleration threshold even for protons (see Section~\ref{sec:uhecr_connection}). Furthermore, the Hillas criterion provides only a necessary condition; the survival of UHECRs against energy losses during propagation through the source environment provides a second, equally decisive discriminant~\cite{Murase2012, Murase2014}.

The critical distinction between HSPs and denser-field sources concerns not only proton photo-meson losses, which are not always catastrophic for protons across a single blazar emission zone~\cite{Murase2014}, but also the photodisintegration of heavy nuclei on soft photon fields. In FSRQs and LSP BL~Lacs, the intense external radiation from the broad line region ($U_\mathrm{BLR} \sim 10^{-2}$~erg~cm$^{-3}$) and the dusty torus efficiently destroys intermediate and heavy nuclei ($A \gtrsim 4$) above the Giant Dipole Resonance threshold ($\epsilon_\mathrm{th} \sim 10$--$30$~MeV in the nucleus rest frame), with photodisintegration timescales of $t_\mathrm{dis} \lesssim 10^{3}$--$10^{4}$~s for iron at $E \sim 10^{20}$~eV, far shorter than the dynamical escape timescale \mbox{$t_\mathrm{dyn} \sim R/c \sim 10^{5}$--$10^{6}$~s \cite{Globus2015, Rodrigues2021}.} HSPs, lacking strong BLR and torus components ($U_\mathrm{rad} \sim 10^{-5}$--$10^{-3}$~erg~cm$^{-3}$), allow both protons and heavy nuclei to traverse the emission region without catastrophic losses, enabling UHECR escape~\cite{Murase2012, Fang2018}. This also reveals an inherent tension for FSRQs as UHECR sources: the same dense photon fields that make them efficient neutrino factories (high $f_{p\gamma}$) simultaneously suppress UHECR escape, so a source class cannot simultaneously maximize both neutrino production efficiency and UHECR release~\cite{Murase2014, Tavecchio2015b}.

Two further arguments reinforce the HSP preference. First, the Pierre Auger Observatory composition measurements, indicating a progressive transition toward heavier nuclei above $\sim$$10^{18.5}$~eV~\cite{Aab2014}, are more naturally accommodated by HSPs than by FSRQs: if FSRQs dominated UHECR production, their intense photon fields would selectively destroy heavy nuclei, yielding a proton-dominated spectrum at the highest energies, in tension with observations \cite{Rodrigues2021}. Second, HSPs exhibit negative or flat cosmological evolution \cite{Ajello2014}, making them locally abundant within the GZK horizon ($D \lesssim 100$--$200$~Mpc for protons above $6\times10^{19}$~eV), whereas FSRQs, though more luminous, follow strong positive evolution and are predominantly located at $z \gtrsim 1$, far beyond the GZK horizon \cite{Fang2018, Rodrigues2021}. Together, these arguments establish HSP BL~Lacs as occupying a physically motivated sweet spot: a sufficient Hillas parameter to accelerate particles to $\sim$$10^{20}$~eV, sparse enough radiation fields to permit UHECR escape even for heavy nuclei, enough local abundance within the GZK horizon to contribute meaningfully to the observed flux, and predicting a mixed-to-heavy composition broadly consistent with Auger data.

In this review, we provide a comprehensive and critical assessment of the role of HSP BL Lac objects as multi-messenger sources, situated at the crossroads of $\gamma$-ray astronomy, cosmic ray physics, and neutrino astrophysics. This review is organized as follows. In Section~\ref{sec:properties_selection}, we detail the physical properties and defining characteristics of HSP blazars, discuss the observational selection methods and biases inherent in modern catalogs such as 2WHSP and 3HSP, and examine the source demographics. Section~\ref{sec:emission_models} contrasts leptonic and hadronic emission models in a quantitative framework, focusing on the physical conditions required for efficient neutrino and UHECR production in HSP jets. Section~\ref{sec:neutrino_observations} reviews the current observational status of the blazar--neutrino connection, including the detailed case study of TXS\,0506 + 056, the constraints derived from non-detections of individual HSP sources, and the results of population-level stacking analyzes. Section~\ref{sec:uhecr_connection} discusses the connection to UHECRs, including spectral and composition constraints from cosmic ray observatories. Section~\ref{sec:future_outlook} outlines the transformative prospects offered by next-generation facilities, including IceCube-Gen2~\cite{IceCubeGen2_2021}, KM3NeT~\cite{KM3NeT2016}, the Cherenkov Telescope Array Observatory (CTAO)~\cite{CTA2019}, and the next-generation VHE neutrino detectors, as well as the role of time-domain multi-messenger programs in resolving the origin of UHECRs and the nature of cosmic neutrino sources. Finally, Section~\ref{sec:conclusions} summarizes the evidence  for and against HSP BL Lac objects as UHECR and neutrino sources and identifies the key measurements that will deliver a definitive  answer within the next decade.


\section{Physical Properties of HSP Blazars and Selection Methods in \linebreak  Modern Catalogs}
\label{sec:properties_selection}

The identification and characterization of HSP BL Lacs is fundamental to the search for sources of high-energy neutrinos. Unlike the more numerous low-synchrotron-peaked (LSP) objects, HSPs offer a direct view into the most energetic particle acceleration zones, often with minimal interference from external photon fields. This section reviews the physical characteristics that define this population and the multi-frequency selection techniques used to construct the catalogs currently employed in neutrino stacking analyzes.

\subsection{Physical Characteristics and Spectral Classification}

Identifying and characterizing the HSP BL Lac population is a prerequisite for any systematic search for their high-energy neutrino emission. Compared to the more abundant LSP blazars, HSPs provide a cleaner probe of extreme particle acceleration, as their jets are largely free of the dense external photon fields that complicate both the interpretation of broadband spectra and the survival of ultra-high-energy particles \cite{Ghisellini2009, Murase2014}. In this section, we first review the physical properties that define HSP. as a distinct class, and then describe the multi-frequency selection strategies used to build the  catalogs that currently underpin neutrino stacking analyzes and UHECR correlation studies, most notably 2WHSP \linebreak  and 3HSP.

The defining feature of blazars is their broadband SED, dominated by non-thermal emission from a relativistic jet closely aligned with the line of sight. This emission displays a characteristic double-humped structure in the $\nu F_{\nu}$ representation, spanning from radio wavelengths to VHE $\gamma$-rays \cite{Fossati1998, Ghisellini1998, Donato2001}. The low-energy component, extending from radio to X-ray wavelengths, is universally attributed to synchrotron radiation from relativistic electrons gyrating in the jet magnetic field \cite{Urry1995, Rybicki1986}. The spectral shape and peak frequency of this component depend on the maximum electron energy, the magnetic field strength, and the bulk Doppler factor of the emitting region \cite{Tavecchio1998, Ghisellini2002}. The high-energy component, peaking in the $\gamma$-ray regime (GeV to TeV), is generally attributed to inverse Compton (IC) scattering in leptonic models, either on the synchrotron photons themselves (SSC; \citet{Maraschi1992}, \linebreak  \citet{Bloom1996}) or on external photon fields from the disk, BLR, or dusty torus (EC; \citet{Sikora1994}, \citet{Dermer1993}). In hadronic scenarios, this component instead arises from proton-synchrotron radiation, proton-induced cascades via photomeson production, or Bethe--Heitler pair production \cite{Mannheim1993, Aharonian2000a, Mucke2001, Bottcher2013}. Lepto-hadronic models, where both channels contribute simultaneously, have also been widely \mbox{explored \cite{Dimitrakoudis2012, Petropoulou2015a, Cerruti2019}.}

The position of the synchrotron peak frequency, $\nu_{\text{peak}}^{S}$, serves as the primary metric for blazar classification. This scheme evolved from the original LBL/HBL dichotomy of \cite{Padovani1995} and the ``blazar sequence'' of \cite{Fossati1998, Ghisellini1998}, and was later formalized by the \textit{Fermi}-LAT collaboration \cite{Abdo2010} and extended by recent multi-frequency catalogs \cite{Giommi2012, Chang2017, Chang2019}. BL Lac objects are categorized based on the rest-frame position of this peak, as summarized in Table~\ref{tab:blazar_classes}. Although originally proposed as a single physical sequence driven by jet power, this interpretation remains debated \cite{Padovani2007, Giommi2012b, Meyer2011, Finke2013}.

\begin{table}[H]
\caption{ummary of BL Lac spectral classifications based on the rest-frame synchrotron peak frequency ($\nu_{\text{peak}}^{S}$). The extreme-HSP subclass was introduced by \cite{Costamante2001a} and further characterized \mbox{by \cite{Biteau2020, Foffano2019}.}}
\centering
\begin{tabularx}{\textwidth}{CCC}
\toprule
\textbf{Class} & \textbf{Frequency Range ($\nu_{\text{peak}}^{S}$)} & \textbf{Typical X-Ray Spectrum} \\
\midrule
LSP/LBL  & $<$$10^{14}$\,Hz ($<$0.4\,eV)       & Soft (steep, IC-dominated) \\
ISP/IBL  & $10^{14}$--$10^{15}$\,Hz             & Intermediate (transition) \\
HSP/HBL  & $>$$10^{15}$\,Hz ($>$4\,eV)          & Hard (flat, synchrotron tail) \\
EHSP/EHBL & $>$$10^{17}$\,Hz ($>$0.4\,keV)       & Very hard ($\Gamma_X < 2$) \\
\bottomrule
\end{tabularx}
\label{tab:blazar_classes}
\end{table}

\begin{figure}[H]

\includegraphics[width=0.7\linewidth]{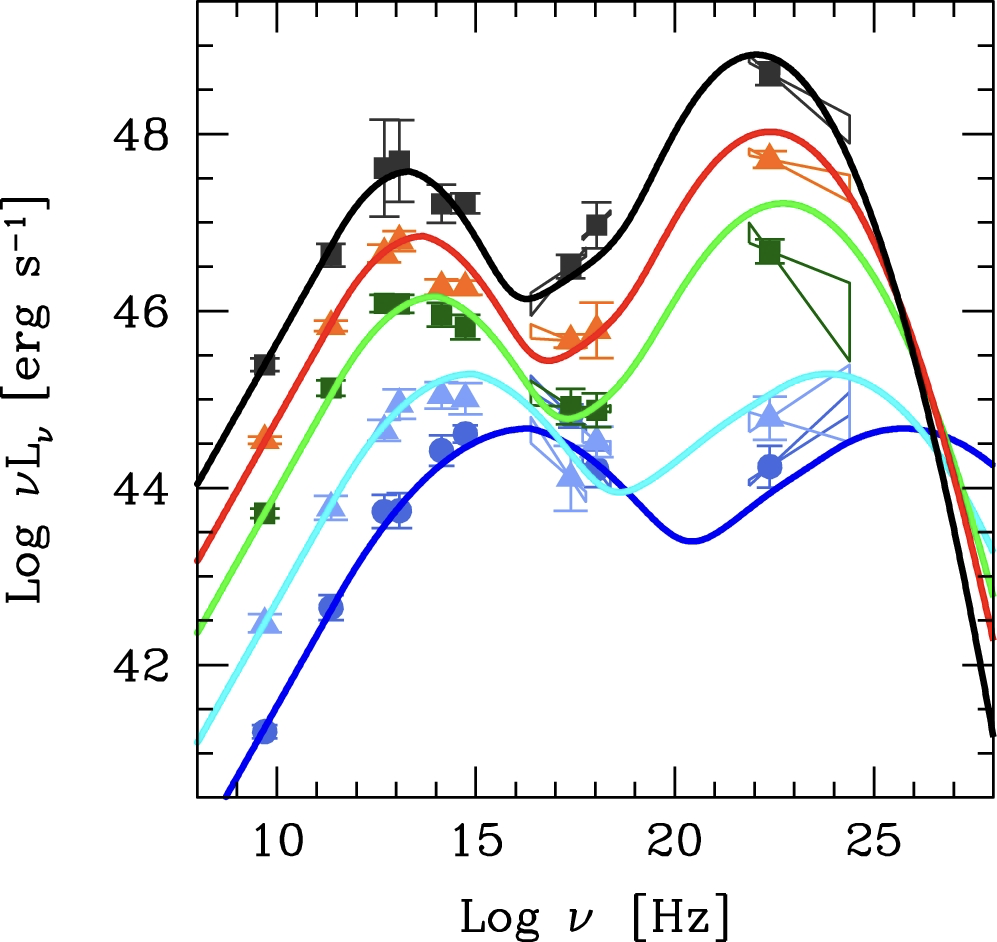}
\caption{The \textit{Fermi} blazar sequence for the combined sample of BL Lacs and FSRQs, constructed from the 3LAC flux-limited catalog (747 blazars with known redshift). Sources are binned by $\gamma$-ray luminosity in the 0.1--100\,GeV band: $\log(L_{\gamma}/\text{erg\,s}^{-1}) > 48$ (black), $47$--$48$ (red), $46$--$47$ (orange), \linebreak  $45$--$46$ (green), $44$--$45$ (cyan), and $< 44$ (blue). Solid curves are phenomenological fits consisting of two smoothly joined broken power laws plus a flat radio power law (see Table~\ref{tab:blazar_classes} in \mbox{\citet{Ghisellini2017}} for the fitting parameters). The progressive shift of both the synchrotron and high-energy peak frequencies toward higher values at lower luminosities, where HSP BL Lacs dominate, is clearly visible, along with the decreasing Compton dominance. Extracted from \citet{Ghisellini2017}.}
\label{fig:blazar_sequence}
\end{figure}

HSP BL Lacs ($\nu_{\text{peak}}^{S} > 10^{15}$\,Hz) represent the most efficient electron accelerators among AGNs, requiring Lorentz factors of $\gamma_e \sim 10^5$--$10^6$ to account for the observed peak \mbox{position \cite{Tavecchio1998, Costamante2001a, Tavecchio2010b}.} As illustrated in Figure \ref{fig:blazar_sequence}, these sources populate the low-luminosity end of the blazar sequence, where the synchrotron peak shifts to UV and X-ray frequencies, and the Compton dominance approaches unity \cite{Ghisellini2017}. A subset of this population, termed \textit{extreme} HSPs (EHSPs), exhibits peaks at even higher frequencies ($\nu_{\text{peak}}^{S} \geq 10^{17}$\,Hz), pushing the synchrotron emission well into the hard-X-ray domain \cite{Costamante2001a, Costamante2018}. Well-known EHSPs include 1ES\,0229 + 200 \cite{Aharonian2007_0229, Kaufmann2011}, 1ES\,0347 $-$ 121 \cite{Aharonian2007_0347}, and RGB\,J0710 + 591 \cite{Acciari2010_RGBJ0710}. These sources show remarkably hard VHE spectra, with photon indices $\Gamma_{\text{VHE}} < 2$ even after correction for EBL absorption \cite{Costamante2018, Biteau2020}. Fitting these spectra with standard one-zone SSC models is challenging, as this typically demands very low magnetic fields ($B \lesssim 10^{-2}$\,G) and/or large Doppler factors ($\delta \gtrsim 50$), values that are difficult to reconcile with independent constraints from radio VLBI observations and equipartition arguments \cite{Tavecchio2009b, Kaufmann2011, Aliu2014}. This tension has motivated alternatives ranging from multi-zone SSC setups and spine--sheath EC models \cite{Ghisellini2005, Tavecchio2014_spine} to hadronic scenarios in which the hard $\gamma$-ray emission arises from proton-initiated cascades \cite{Cerruti2015} or proton-synchrotron radiation \cite{Aharonian2000a, Mucke2003}.
\vspace{6pt}

Physically, HSPs are associated with radiatively inefficient accretion flows, commonly modeled as advection-dominated accretion flows (ADAFs), where the mass accretion rate falls well below the Eddington limit ($\dot{m} \lesssim 10^{-2}\,\dot{m}_{\text{Edd}}$; \citet{Narayan1994, Narayan1995}). In this regime, the gas cannot cool efficiently, and most of the gravitational energy is carried into the black hole rather than radiated, replacing the standard thin disk \cite{Shakura1973} with a geometrically thick, optically thin flow \cite{Ghisellini2009, Ghisellini2011}. Unlike high-luminosity quasars, HSPs therefore lack a luminous accretion disk, a significant BLR, and a prominent dusty \mbox{torus \cite{Plotkin2011, Giommi2012}.} Consequently, the blazar emission region is typically located outside the BLR radius, rendering inverse Compton scattering off external photon fields (EC) weak or negligible and making the SSC process the dominant high-energy emission \mbox{mechanism \cite{Ghisellini2009, Tavecchio1998}.} The absence of strong emission lines (equivalent width $<$5\,\AA; \citet{Stickel1991, Stocke1991}) and the lack of a ``Big Blue Bump'' in the optical/UV spectrum confirm that the non-thermal jet emission outshines the thermal components across the electromagnetic spectrum. The one exception is the host galaxy itself, whose starlight often dominates the optical flux during low-jet-activity states \cite{Nilsson2007, Sbarufatti2005}. Careful subtraction of this host galaxy contribution is important when building SEDs, since neglecting it can shift the inferred synchrotron peak frequency and bias the derived jet parameters \cite{Nilsson2007, Aleksic2015_Mrk421}.

However, an important caveat must be noted. A recently identified class of ``masquerading BL~Lacs'' \cite{Padovani2019, Padovani2022} comprises intrinsically flat-spectrum radio quasars (FSRQs) whose broad emission lines are overwhelmed by the boosted non-thermal jet continuum, causing them to mimic the featureless optical spectra of genuine BL~Lac objects. These sources harbor a hidden BLR whose photon field, though observationally obscured, can provide a dense target for $p\gamma$ interactions and significantly enhance neutrino production efficiency relative to true BL~Lacs \cite{Padovani2019, Padovani2022}. The prototype of this class is TXS~0506+056 itself, which was reclassified as a masquerading BL~Lac by \citet{Padovani2019} on the basis of its high luminosity and accretion rate inconsistent with a genuine BL~Lac object. This reclassification has important implications for multi-messenger astrophysics: if the most significant blazar--neutrino associations preferentially involve masquerading BL~Lacs rather than true HSPs, this would suggest that a hidden BLR photon field is a necessary condition for efficient neutrino production in blazar jets, and that the energetic difficulties encountered when modeling neutrino production in genuine HSPs (Section~\ref{sec:emission_models}) may reflect the absence of this dense external photon target rather than a fundamental problem with the hadronic scenario \cite{Padovani2022, Murase2014}.

In the context of spectral fitting, the emission of well-known HSPs such as Mrk\,421 ($z = 0.031$) and Mrk\,501 ($z = 0.034$) (the first extragalactic sources detected at TeV \mbox{energies \cite{Punch1992, Quinn1996})} is typically modeled using one-zone SSC frameworks during quiescent \mbox{states \cite{Tavecchio1998, Abdo2011_Mrk421, Abdo2011_Mrk501, Aleksic2015_Mrk421}.} In these models, a single spherical emission region of radius $R$, filled with a tangled magnetic field $B$ and moving with a bulk Lorentz factor $\Gamma$, contains a relativistic electron population (described by a broken power-law or log-parabola distribution) that produces synchrotron radiation. These synchrotron photons then serve as the seed field for IC scattering, making the two SED humps tightly coupled \cite{Tavecchio1998, Ghisellini2002}. The typical SSC parameters for HSPs are $B \sim 0.01$--$0.1$\,G, $R \sim 10^{15}$--$10^{16}$\,cm, and $\delta \sim 10$--$30$ \cite{Tavecchio1998, Tavecchio2010b, Finke2008}, yielding radiation energy densities of $U_{\text{rad}} \sim 10^{-5}$--$10^{-3}$\,erg\,cm$^{-3}$---orders of magnitude below the $U_{\text{BLR}} \sim 10^{-2}$\,erg\,cm$^{-3}$ found in FSRQs \cite{Ghisellini2010}.

\subsection{Multi-Frequency Selection and the ``Blazar Strip''}
\label{subsec:selection_methods}

Historically, single-band surveys have introduced selection biases that have dichotomized the blazar population. Radio-selected samples (e.g., the 1 Jy catalog) were dominated by LBLs and FSRQs, while the first X-ray surveys (e.g., \textit{Einstein} and \textit{ROSAT}) preferentially detected high-peaked sources \cite{Padovani1995, Perlman1998}. To overcome these biases and identify the complete HSP population—which is often radio-faint yet X-ray bright—modern selection methods utilize a multi-frequency approach. The most effective strategy involves cross-matching deep radio catalogs (e.g., NVSS, FIRST, SUMSS) with X-ray catalogs (e.g., \textit{ROSAT}, \textit{Swift}-XRT, \textit{XMM-Newton}) and imposing specific criteria on the broadband spectral slope \cite{Giommi1999, Giommi2005, Arsioli2015}.

A critical parameter for this classification is the radio-to-X-ray spectral slope, $\alpha_{\rm rx}$, defined as \cite{Stocke1991, Giommi2005}
\begin{equation}
    \alpha_{\rm rx} = \frac{\log(F_{\text{radio}} / F_{\text{X}})}{\log(\nu_{\text{X}} / \nu_{\text{radio}})}\,,
    \label{eq:alpha_rx}
\end{equation}
where typical reference frequencies are 1.4 GHz for radio and 1 keV for X-rays. HSPs are characterized by a flat spectral slope, typically requiring $\alpha_{\rm rx} \lesssim 0.75$ (or equivalently $\alpha_{\rm rx} \lesssim 0.56$ when using different flux density normalizations) to ensure the synchrotron peak lies above $10^{15}$ Hz \cite{Giommi2005, Chang2017}. This threshold ensures that the spectral slope between radio and X-ray is flatter than would be expected if the synchrotron peak were located in the infrared--optical regime \cite{Giommi2005, Padovani2007}.

Furthermore, the infrared band provides a powerful discriminator. Using data from the \textit{Wide-field Infrared Survey Explorer} (WISE), blazars occupy a distinct region in the color--color space defined by the W1 (3.4 $\upmu$m), W2 (4.6 $\upmu$m), and W3 (12 $\upmu$m) magnitude bands, known as the ``WISE Blazar Strip'' \cite{DAbrusco2012, Massaro2011, Massaro2015}.

\begin{figure}[H]

\includegraphics[width=0.85\textwidth]{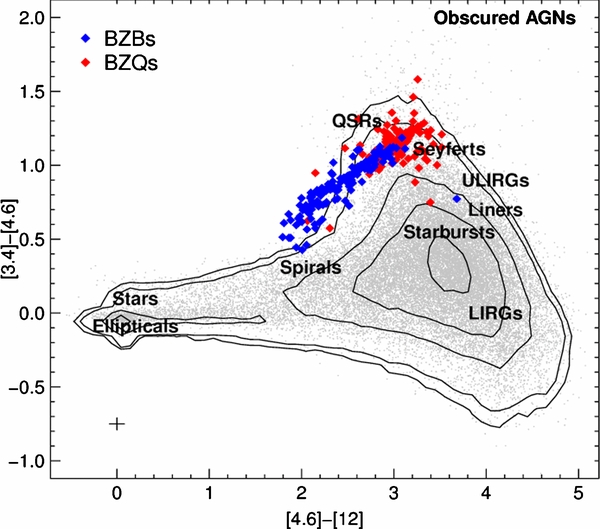}
\caption{WISE infrared color--color diagram showing the $[W1-W2]$ versus $[W2-W3]$ color space (in Vega magnitudes). BL Lac objects (BZBs; blue points) and flat-spectrum radio quasars (BZQs; red points) occupy a distinct diagonal region known as the ``WISE Blazar Strip,'' clearly separated from other extragalactic source populations. Background gray dots represent 453,420 generic WISE sources detected at high Galactic latitude. Isodensity contours (corresponding to 50, 100, 500, and 2000 sources per unit area) are overlaid to show the concentration of thermal-emission-dominated sources. The locations of quasars (QSRs), ultraluminous and luminous infrared galaxies (ULIRGs and LIRGs), starbursts, spirals, stars, and elliptical galaxies are labeled. BL Lacs show larger scatter than FSRQs due to contamination from host galaxy starlight in the W1 and W2 bands. Extracted from \citet{DAbrusco2012}.}
\label{fig:wise_colors}
\end{figure}

For HSPs, infrared selection is complicated by contamination from the host galaxy. Unlike LBLs and FSRQs, where the non-thermal jet emission dominates across all wavelengths, the jet emission in HSPs is often radiatively inefficient in the optical/IR band. Consequently, the thermal emission from the giant elliptical host galaxy can contribute significantly or even dominate the W1 and W2 bands, causing some HSPs to drift away from the standard blazar strip and toward the region occupied by passive elliptical galaxies in the WISE color--color plane \cite{Nilsson2007, Foffano2019}. This effect is visible in Figure \ref{fig:wise_colors} as the larger dispersion of BL Lac objects (which include the HSP population) compared to the tighter distribution \mbox{of FSRQs.}

\begin{figure}[H]

\includegraphics[width=0.85\textwidth]{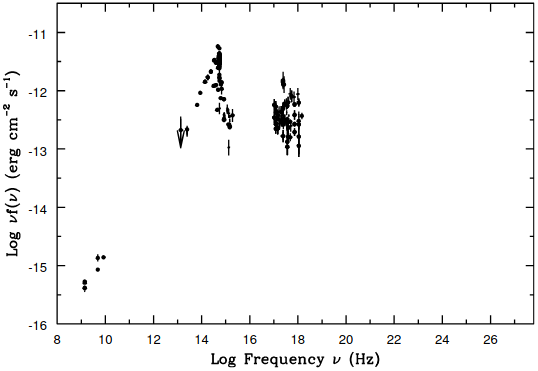}
\caption{Spectral energy distribution of 5BZG J0903 + 4055. This source was not selected for the 2WHSP catalog because it does not satisfy the radio--IR slope criterion due to host galaxy contamination in the IR band, despite being a bona fide HSP blazar with a highly variable X-ray spectrum. The SED clearly shows strong non-thermal emission extending from radio to $\gamma$-rays, with the synchrotron component peaking above $10^{15}$ Hz. This example illustrates why multi-frequency catalogs like 3HSP that relax IR slope criteria are necessary to achieve completeness in HSP samples, particularly for nearby sources where the host galaxy contribution is significant. Extracted from \citet{Chang2019}.}
\label{fig:hsp_host_contamination}
\end{figure}

A clear example of this host galaxy contamination issue is illustrated in Figure \ref{fig:hsp_host_contamination}, which shows the SED of 5BZG~J0903 + 4055. This source was not included in the 2WHSP catalog despite being a genuine HSP because it does not satisfy the radio--IR slope criterion \linebreak  ($0.05 < \alpha_{1.4~\text{GHz}-3.4~\upmu \text{m}} < 0.45$) due to host galaxy contamination in the IR band. Several other well-studied extreme HSPs are similarly affected. Mrk~421 ($z = 0.031$) and Mrk~501 ($z = 0.034$), the two closest and best-monitored TeV blazars, both show significant host galaxy contributions in the optical and near-IR bands during low-jet-activity states, requiring careful elliptical galaxy template subtraction before reliable SED fitting can be performed \cite{Nilsson2007, Aleksic2015_Mrk421}. The nearby extreme HSP 1ES~0229 + 200 ($z = 0.14$) similarly exhibits host galaxy contamination at optical wavelengths, complicating the determination of its synchrotron peak frequency and intrinsic jet parameters \cite{Kaufmann2011}. PKS~2155$-$304 ($z = 0.116$), one of the brightest and most studied HSPs in the Southern sky, also shows measurable host galaxy starlight in quiescent states \cite{Nilsson2007, Foffano2019}. In general, host galaxy contamination is most severe for nearby sources ($z \lesssim 0.2$) where the host galaxy is spatially resolved and its optical flux is not diluted by the angular diameter distance, as well as for sources caught in low-jet-activity states where the non-thermal continuum is suppressed \cite{Sbarufatti2005, Foffano2019}. Nevertheless, its highly variable X-ray spectrum, with luminosity ranging between $2.4$ and \mbox{$7.3 \times 10^{44}$ erg s$^{-1}$} in the 0.3--10.0 keV band \cite{Giommi2020a}, along with its clear non-thermal emission, confirms its HSP nature. This demonstrates why relaxing the IR slope criteria, as conducted in the 3HSP catalog, is essential for achieving completeness in HSP samples.

To account for this effect, the contribution of the host galaxy can be statistically subtracted using \cite{Nilsson2007}
\begin{equation}
    F_{\text{jet}}(\lambda) = F_{\text{obs}}(\lambda) - F_{\text{host}}(\lambda)\,,
    \label{eq:host_subtraction}
\end{equation}
where $F_{\text{host}}$ is modeled as an elliptical galaxy template (typically an old stellar population spectrum) scaled to the source redshift \cite{Nilsson2007, Foffano2019}. Recent efforts have employed machine learning algorithms (e.g., Random Forests, Neural Networks) to disentangle these components, using WISE colors in conjunction with radio and X-ray fluxes to probabilistically classify candidates, even when host galaxy contamination is significant \cite{DAbrusco2012, Massaro2012, DAbrusco2019}.

\subsection{The 2WHSP and 3HSP Catalogs}
\label{sec:catalogs}

In neutrino astronomy, the completeness of the source catalog is as important as the accuracy of individual source parameters. Stacking analyzes, which sum the neutrino signal from many sub-threshold sources, rely on the assumption that the catalog accurately represents the underlying population density and spatial distribution.

\begin{figure}[H]

\includegraphics[width=0.9\textwidth]{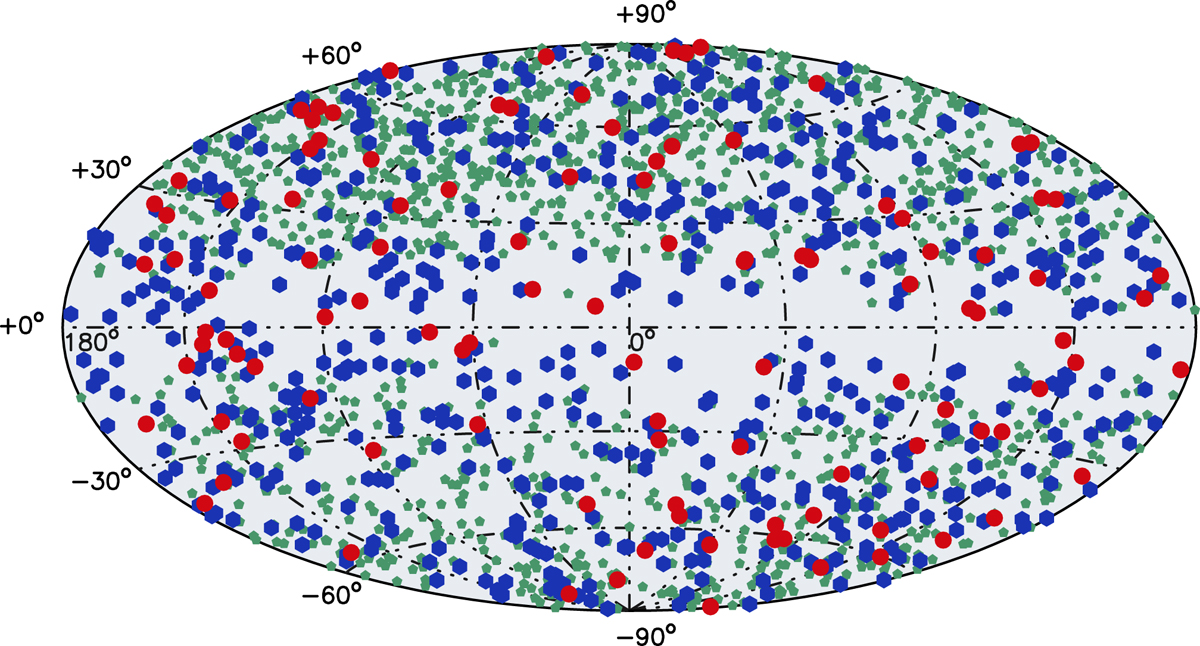}
\caption{All-sky distribution of the 2013 sources in the 3HSP catalog in Galactic coordinates (Hammer--Aitoff projection). Sources are color-coded by their synchrotron peak frequency: extreme HSPs ($\nu_{\text{peak}} > 10^{17}$ Hz) are shown in red, standard HSPs ($10^{15}$ Hz $< \nu_{\text{peak}} < 10^{17}$ Hz) in blue, and lower-confidence candidates in green. The avoidance of the Galactic plane due to X-ray absorption and source confusion is apparent, with source density dropping sharply at $|b| < 20^\circ$. The concentration at high Galactic latitudes demonstrates the extragalactic nature of the HSP population. The spatial distribution shows no significant clustering on large angular scales, consistent with an isotropic extragalactic source population, though some concentrations correspond to well-studied sky regions with deeper X-ray coverage. Extracted from \citet{Chang2019}.}
\label{fig:3hsp_skymap}
\end{figure}

The 2WHSP (2nd World High Synchrotron Peaked) catalog \cite{Chang2017} represents a major step forward, assembling 1691 HSP sources through systematic cross-matching of multi-frequency databases and calculating the SED slope to estimate $\nu_{\text{peak}}^{S}$. This was succeeded by the 3HSP (3rd High Synchrotron Peaked) catalog \cite{Chang2019}, which is currently the benchmark for HSP studies and serves as the primary input for IceCube neutrino stacking analyzes. Figure~\ref{fig:3hsp_skymap} shows the all-sky distribution of the 2013 sources in 3HSP, revealing their concentration at high Galactic latitudes and the characteristic avoidance of the Galactic plane due to X-ray absorption.

The 3HSP catalog improves upon 2WHSP through several key enhancements:
\begin{enumerate}
    \item Increasing the sample size to 2013 sources, representing a $\sim$$20\%$ growth.
    \item Imposing a strict cut of $\nu_{\text{peak}}^{S} > 10^{15}$ Hz to ensure sample purity, with improved SED fitting procedures.
    \item Introducing a ``Figure of Merit'' (FOM) for the detectability of each source in the TeV band. The FOM is defined as the ratio of the expected synchrotron peak flux to the sensitivity limit of current Imaging Atmospheric Cherenkov Telescopes (IACTs) such as H.E.S.S., MAGIC, VERITAS, and the future CTAO, serving as a proxy for the potential high-energy output and TeV detectability.
    \item Providing updated redshift estimates for $\sim$$70\%$ of the sample, though many sources still lack spectroscopic confirmation.
\end{enumerate}

A comparison of the primary catalogs used in the field is provided in Table \ref{tab:catalogues}.

\begin{table}[H]
\caption{Comparison of major catalogs used for HSP selection and neutrino stacking analyzes.}
	\begin{adjustwidth}{-\extralength}{0cm}
	\begin{tabularx}{\fulllength}{>{\raggedright\arraybackslash}m{2.8cm}>{\raggedright\arraybackslash}m{1.5cm}>{\raggedright\arraybackslash}m{4.5cm}>{\raggedright\arraybackslash}m{7.5cm}}
\toprule
\textbf{Catalog} & \textbf{N$_{\rm src}$} & \textbf{Primary Selection} & \textbf{Relevance to Neutrinos} \\
\midrule
FGL-DR3 \cite{Abdollahi2020} & 5064 & $\gamma$-ray detection (\textit{Fermi}-LAT) & Gold standard for confirmed GeV $\gamma$-ray emitters; flux-limited bias in GeV band; includes all blazar types. \\
2WHSP \cite{Chang2017} & 1691 & Radio + X-ray + $\alpha_{\rm rx}$ & First large dedicated HSP sample; basis for initial IceCube stacking analyzes. \\
3HSP \cite{Chang2019} & 2013 & Radio + X-ray + SED fitting & Largest current HSP sample; includes ``dark'' TeV accelerators not detected by \textit{Fermi}. \\
\bottomrule
\end{tabularx}	\end{adjustwidth}
\label{tab:catalogues}
\end{table}

The distinction between $\gamma$-ray-selected samples (like the \textit{Fermi} 4FGL) and multi-frequency-selected samples (like 3HSP) is critical. While 4FGL sources are guaranteed GeV emitters, the sample is flux-limited in the 0.1--300 GeV range where \textit{Fermi}-LAT operates. The 3HSP catalog, by selecting based on the synchrotron peak position rather than $\gamma$-ray flux, includes a large population of ``dark'' or ``hidden'' accelerators—sources that are bright in X-rays (indicating the presence of high-energy electrons with $\gamma_{\rm e} \sim 10^5$--$10^6$) but have not yet been detected in $\gamma$-rays by \textit{Fermi}-LAT \cite{Giommi2020a}.

This non-detection can arise from several physical scenarios:
\begin{itemize}
    \item The inverse Compton peak is shifted to TeV energies ($E > 100$ GeV), where \textit{Fermi}-LAT sensitivity drops, and the observed flux is strongly attenuated by $e^+e^-$ pair production on the EBL \cite{Franceschini2017}.
    \item The source is in a low state during \textit{Fermi} observations, as HSPs are known to exhibit significant flux variability on timescales of days to years \cite{Aleksic2015}.
    \item The magnetic field is relatively high ($B \sim 1$ G), leading to efficient synchrotron cooling and suppression of the IC component relative to the synchrotron peak \cite{Tavecchio1998}.
\end{itemize}

Cross-matching the 3HSP catalog with the \textit{Fermi}-LAT 4FGL-DR3 catalog reveals that only $\sim$$30$--$40\%$ of the 2013 HSP sources have a confirmed GeV $\gamma$-ray counterpart \cite{Chang2019, Abdollahi2020}, implying that $\sim$$60$--$70\%$ of the HSP population, corresponding to $\sim$$1200$--$1400$ sources, remains undetected by \textit{Fermi}-LAT. Among these GeV-undetected HSPs, the Figure of Merit (FOM) defined in the 3HSP catalog suggests that $\sim$$10$--$15\%$ have predicted synchrotron peak fluxes sufficient for detection by current IACTs such as H.E.S.S., MAGIC, and VERITAS, meaning that the majority are genuinely ``dark'' accelerators with no current high-energy counterpart at any wavelength \cite{Chang2019, Costamante2018}. This non-detection fraction is physically significant: \citet{Giommi2020a} found that IceCube high-energy neutrino error circles contain a statistically significant excess of 3HSP sources lacking 4FGL counterparts, suggesting that these GeV-dark HSPs may contribute disproportionately to the diffuse neutrino flux relative to their GeV-bright counterparts. Critically, if the neutrino flux scales with the UHECR luminosity (and, by extension, the total jet power rather than just the $\gamma$-ray luminosity), these ``hidden'' TeV sources in 3HSP could contribute significantly to the diffuse astrophysical neutrino flux observed by IceCube, even if they remain below the GeV detection \mbox{threshold \cite{Giommi2020a, Padovani2016}.} This motivates the use of multi-frequency-selected catalogs over $\gamma$-ray-selected samples for neutrino point-source searches and stacking analyzes.

It is worth noting that contamination in the 3HSP sample remains a concern. Photometric redshift uncertainties, host galaxy misclassification, and foreground Galactic stars can introduce spurious sources. Conservative estimates suggest a contamination fraction of $\sim$$5$--$10\%$ \cite{Chang2019}, which must be accounted for when interpreting the non-detection of neutrino point sources or weak stacking signals.

Having established the observational properties, selection criteria, and catalog construction for the HSP population, we now turn to the physical models that predict their multi-messenger signatures, contrasting leptonic and hadronic emission scenarios.


\section{Emission Models and Neutrino Production Mechanisms}
\label{sec:emission_models}

The interpretation of the multi-messenger emission from HSP blazars relies heavily on the underlying particle acceleration and radiation mechanisms. While the electromagnetic spectrum can often be adequately described by leptonic models, the production of high-energy neutrinos necessitates the presence of relativistic hadrons. This section contrasts these two frameworks, detailing the mathematical foundations of the particle distributions and the specific conditions required within HSP jets to sustain efficient neutrino production.

\subsection{The Leptonic Standard: Synchrotron Self-Compton}

The standard paradigm for the high-energy emission of BL Lac objects, particularly HSPs, is the one-zone synchrotron self-Compton (SSC) model \cite{Maraschi1992, Tavecchio1998, Ghisellini2009}. This model assumes that non-thermal radiation originates from a single, homogeneous, spherical region (the ``blob'') of radius $R$, permeated by a tangled magnetic field $B$. The blob moves relativistically down the jet with a bulk Lorentz factor $\Gamma$ at a small angle $\theta$ to the observer's line of sight, resulting in a Doppler boosting factor $\delta = [\Gamma(1 - \beta \cos\theta)]^{-1}$.

In this scenario, a population of relativistic electrons ($e^-$) is accelerated and injected into the emitting region. The acceleration mechanism is not uniquely constrained by the SSC framework itself; viable candidates include Fermi-type diffusive shock acceleration (DSA;~\citet{Blandford1978, Drury1983}), which naturally produces power-law particle distributions, and magnetic reconnection in the jet~\cite{Sironi2015, Petropoulou2019}, which can operate in magnetically dominated ($\upsigma > 1$) plasmas and generates harder injection spectra ($s \sim 1$--$1.5$) more consistent with extreme HSP observations (see Section~\ref{sec:reconnection} for a detailed discussion). The injection function, $Q_e(\gamma_e)$ [particles cm$^{-3}$ s$^{-1}$], describes the rate at which fresh electrons are supplied and is typically parameterized as a broken power-law or a power-law with an exponential cutoff \cite{Massaro2004}, as follows:
\begin{equation}
    Q_e(\gamma_{\rm e}) = K_{\rm e} \gamma_{\rm e}^{-s_{\rm e}} \exp\left(-\frac{\gamma_{\rm e}}{\gamma_{\rm e,\text{max}}}\right), \quad \text{for } \gamma_{\rm e} > \gamma_{\rm e,\text{min}}\,,
    \label{eq:electron_injection}
\end{equation}
where $K_{\rm e}$ is the normalization constant, $s_{\rm e}$ is the spectral index, and $\gamma_{\rm e}$ is the electron Lorentz factor. The exponential cutoff represents the maximum energy attainable through the acceleration mechanism, typically limited by radiative cooling or escape from the acceleration region \cite{Massaro2004}. The steady-state electron distribution, $N_{\rm e}(\gamma_{\rm e})$, is then determined by the balance between this injection and radiative cooling losses.

The emission spectrum is characterized by two broad non-thermal components. The low-energy hump (radio to X-rays) arises from synchrotron radiation emitted by electrons spiraling in the magnetic field. The peak frequency of this emission in the observer's frame is given by \cite{Rybicki1986, Ghisellini2009b}
\begin{equation}
    \nu_{\text{syn}}^{\text{obs}} \approx 3.7 \times 10^6 B \gamma_{\text{peak}}^2 \delta \quad \text{[Hz]}\,.
    \label{eq:synchrotron_freq}
\end{equation}
For HSPs, the defining characteristic is a synchrotron peak frequency \mbox{$\nu_{\text{syn}}^{\text{obs}} > 10^{15}$ Hz \cite{Padovani1995, Abdo2010}.} This requires either extremely efficient particle acceleration (yielding $\gamma_{\text{peak}} \sim 10^5$--$10^6$) or significant Doppler boosting in environments with relatively low magnetic fields \mbox{($B \sim 0.1$~G).}

For a typical HSP with $\nu_{\text{syn}}^{\text{obs}} = 10^{17}$ Hz, $B = 0.1$~G, and $\delta = 20$, the required electron Lorentz factor is
\begin{equation}
    \gamma_{\text{peak}} \approx 1.3 \times 10^5 \left(\frac{\nu_{\text{syn}}}{10^{17}\text{ Hz}}\right)^{1/2} 
    \left(\frac{B}{0.1\text{ G}}\right)^{-1/2} \left(\frac{\delta}{20}\right)^{-1/2}\,.
    \label{eq:gamma_peak_calculation}
\end{equation}
This places extreme demands on particle acceleration mechanisms, requiring efficient energy gain rates that exceed radiative losses \cite{Ghisellini2009, Tavecchio2009b}.

The high-energy hump ($\gamma$-rays) is produced via the inverse Compton (IC) process, where the same population of relativistic electrons up-scatters the synchrotron photons they generate (hence ``Self-Compton'') \cite{Jones1968, Ghisellini1985}. In the Thomson regime, the IC peak frequency scales as $\nu_{\text{IC}} \propto \gamma_{\text{peak}}^2 \nu_{\text{syn}}$ \cite{Blumenthal1970, Rybicki1986}. However, for the ultra-relativistic electrons present in HSPs, the scattering often occurs in the Klein--Nishina regime, where the cross-section is suppressed, causing the IC spectrum to steepen and the peak energy to shift to lower values relative to the Thomson prediction \cite{Blumenthal1970, Tavecchio1998, Moderski2005}.

While SSC models successfully reproduce the tight temporal correlation often observed between X-ray and TeV $\gamma$-ray fluxes, exemplified by the variability of sources like \mbox{Mrk 421 \cite{Fossati2007},} they face significant challenges. The most notable is the ``orphan'' TeV flare phenomenon, such as the 2002 event in 1ES 1959 + 650, where the $\gamma$-ray flux increased dramatically without a corresponding X-ray counterpart \cite{Krawczynski2004, Bottcher2005, Bottcher2007, Aleksic2015}. Such events are difficult to reconcile with a simple one-zone leptonic model in which both spectral components are derived from the same electron population. Furthermore, in pure leptonic models, the $\gamma$-ray emission is entirely attributed to leptonic processes (synchrotron and inverse Compton), with any hadronic contribution to the electromagnetic output neglected. This does not strictly preclude some level of neutrino production if a sub-dominant hadronic component is present, but the predicted neutrino flux in such scenarios is typically far below the sensitivity of current and near-future detectors \cite{Petropoulou2015a, Rodriguez2024}. Detectable neutrino fluxes therefore require the explicit consideration of hadronic or lepto-hadronic scenarios in which the proton luminosity is sufficiently large to produce a measurable signal, motivating the investigation of such models for multi-messenger sources.

\subsection{Hadronic Interactions and the ``Proton Blazar''}

In the ``Proton Blazar'' scenario \cite{Mannheim1993, Mucke2003, Aharonian2002}, protons are accelerated alongside electrons, likely via the same shock acceleration mechanisms. The proton injection function, $Q_{\rm p}(\gamma_{\rm p})$ [particles cm$^{-3}$ s$^{-1}$], is modeled as a power-law with an exponential cutoff as follows:
\begin{equation}
    Q_p(\gamma_p) = K_p \gamma_p^{-s_p} \exp\left(-\frac{\gamma_p}
{\gamma_{p,\rm max}}\right), \quad \gamma_p > \gamma_{p,\rm min}\,.
    \label{eq:proton_injection}
\end{equation}
\textls[-25]{To satisfy the energetic requirements for neutrino production in HSPs, hard spectral indices are often assumed ($s_p \approx 1.5$) along with high minimum Lorentz factors\mbox{ ($\gamma_{p,\rm min} \approx 100$) \cite{Cerruti2015, Mucke2003}.}} The maximum proton energy is typically constrained by the Larmor radius (Hillas criterion) or by radiative cooling losses \cite{Aharonian2000a}.

It is worth noting that many blazar--neutrino models assume a sharp cutoff at $\gamma_{p,\rm max}$, which is a mathematical convenience rather than a physically motivated choice. An exponential-cutoff power-law spectrum, as written in Equation~(\ref{eq:proton_injection}), is physically more appropriate since it reflects the gradual steepening of the acceleration spectrum near the maximum energy set by the balance between acceleration and loss rates \cite{Drury1983, Blasi2013}. Moreover, a sharp cutoff implicitly assumes that protons above $\gamma_{p,\rm max}$ are lost within the source, whereas in a more self-consistent picture, the highest-energy protons, those approaching the Hillas limit, can escape the source before losing their energy and be injected into the extragalactic medium as UHECRs \cite{Murase2012, Rodrigues2021}. In a rigidity-dependent diffusion model, the escape probability scales as $P_\mathrm{esc} \propto (E/Z)^\delta$, where $\delta \sim 1/3$--$1$ depending on the diffusion regime \cite{Kotera2011, AlvesBatista2019}, so that the highest-rigidity particles escape most efficiently while lower-energy protons remain confined and contribute to neutrino production via $p\gamma$ interactions. This picture naturally connects the neutrino production model to the UHECR phenomenology discussed in Section \ref{sec:uhecr_connection}: the same proton population that produces PeV neutrinos through interactions with the synchrotron photon field also seeds the extragalactic cosmic ray flux at the highest energies, with the relative contributions to neutrinos and UHECRs determined by the competition between interaction and escape timescales \cite{Murase2012, Fang2018}. A fully self-consistent multi-messenger model should therefore treat neutrino production and UHECR escape as two aspects of the same hadronic process rather than modeling them independently with artificially imposed spectral cutoffs.

The production of high-energy neutrinos can proceed via two primary hadronic channels: interactions with radiation fields ($p\gamma$) or interactions with ambient matter ($pp$).

\subsubsection{Photomeson Production ($p\gamma$)}

In the low-density environment of blazar jets, photomeson production is generally considered the dominant cooling mechanism \cite{Stecker1968, Atoyan2001}. The interaction threshold depends on the target photon energy $\epsilon'$ in the comoving frame of the jet, as follows:
\begin{equation}
    E'_p \epsilon' \gtrsim \frac{m_\pi m_p c^4}{2} \approx 
    0.2~{\rm GeV}^2\,,
\end{equation}
where $m_\pi$ is the pion mass, $m_p$ is the proton mass, and primed quantities denote comoving frame values. The corresponding observed proton energy is $E_{p,\rm obs} = \Gamma E'_p$, and the observed target photon energy is $\epsilon_{\rm obs} = \delta \epsilon'$, where $\Gamma$ is the bulk Lorentz factor and $\delta$ is the Doppler factor of the jet. The process proceeds primarily via the $\Delta^+$ resonance \cite{Muecke1999, Hummer2010}, as follows:
\begin{equation}
    p + \gamma \rightarrow \Delta^+ \rightarrow \begin{cases} 
    p + \pi^0 \rightarrow p + 2\gamma & (2/3) \\
    n + \pi^+ \rightarrow n + \upmu^+ + \nu_\upmu \rightarrow n + e^+ + \nu_e + \bar{\nu}_\upmu + \nu_\upmu & (1/3)
    \end{cases}
    \,.
    \label{eq:delta_decay}
\end{equation}

The photomeson production rate peaks when the product $E'_p \epsilon'$ matches the $\Delta^+$ resonance at $\sim$$0.3~\mathrm{GeV}^2$ in the center-of-mass frame \cite{Muecke1999, Hummer2010}. Note that the product $E'_p \epsilon'$ has dimensions of energy squared, as both $E'_p$ and $\epsilon'$ are energies. All quantities are defined in the comoving frame of the jet. The resonance condition gives
\begin{equation}
    E'^{\rm res}_p \approx \frac{0.3~{\rm GeV}^2}
    {\epsilon'}\,,
\end{equation}
which, for comoving-frame X-ray synchrotron photons at $\epsilon' \sim 1~{\rm keV} = 10^{-6}~{\rm GeV}$, yields
\begin{equation}
    E'^{\rm res}_p \approx 
    \frac{0.3~{\rm GeV}^2}{10^{-6}~{\rm GeV}} 
    = 3\times10^{5}~{\rm GeV} = 3\times10^{14}~{\rm eV}
    \approx 3\times10^{14}
    \left(\frac{\epsilon'}{1~{\rm keV}}\right)^{-1}
    ~{\rm eV}\,,
\end{equation}
corresponding to PeV-range protons interacting with keV X-ray photons in the comoving frame \cite{Atoyan2003, Mannheim1992}. The corresponding observed resonance proton energy is $E^{\rm res}_{p,\rm obs} = \Gamma E'^{\rm res}_p$, and the observed target photon energy relates to the comoving value via $\epsilon_{\rm obs} = \delta\epsilon'$, so that
\begin{equation}
    E^{\rm res}_{p,\rm obs} \approx 
    \frac{0.3~{\rm GeV}^2\,\Gamma\delta}
    {\epsilon_{\rm obs}}\,.
\end{equation}
For typical HSP parameters ($\Gamma \sim \delta \sim 15$) and observed X-ray photons at $\epsilon_{\rm obs} \sim 1$~keV,
\begin{equation}
    E^{\rm res}_{p,\rm obs} \approx 
    \frac{0.3~{\rm GeV}^2 \times 225}
    {10^{-6}~{\rm GeV}} \approx 6.75\times10^{7}~
    {\rm GeV} \approx 6.75\times10^{16}~{\rm eV}\,.
\end{equation}
This supports that X-ray photons from HSPs are efficient targets for sub-PeV-to-PeV protons in the comoving frame, while UV/optical photons ($\epsilon' \sim$ few eV) are more relevant for multi-PeV and EeV protons via the same resonance condition \cite{Aharonian2000a, Hummer2010}.

The cooling timescale for protons via this channel is given by \cite{Kelner2008}
\begin{equation}
    t_{{\rm p} \upgamma}^{-1}(\gamma_{\rm p}) = \frac{c}{2\gamma_{\rm p}^2} \int_{\epsilon_{\text{th}}}^{\infty} d\epsilon \frac{n_\gamma(\epsilon)}{\epsilon^2} \int_{0}^{2\gamma_{\rm p} \epsilon} d\epsilon' \upsigma_{{\rm p} \upgamma}(\epsilon') \kappa_{{\rm p} \upgamma}(\epsilon') \epsilon'\,,
    \label{eq:pgamma_cooling}
\end{equation}
where $\epsilon_{\text{th}} = m_\pi m_{\rm p} c^4 / (2 E_{\rm p})$ is the threshold photon energy, $\upsigma_{{\rm p} \upgamma}$ is the cross-section (peaking at $\sim$$500~\upmu$b), and $\kappa_{{\rm p} \upgamma} \approx 0.2$ is the inelasticity.

\subsubsection{Proton--Proton Interactions ($pp$)}

Although typically sub-dominant in the rarefied jet regions of BL Lacs, proton--proton 
interactions can become significant if the jet interacts with denser clouds of gas or 
in specific lepto-hadronic models with high baryon loading \cite{Barkov2012, Araudo2013}. 
The threshold energy for pion production in $pp$ collisions is $E_{\text{th}} \approx 1.22$ GeV \cite{Kafexhiu2014}. The interaction produces a spray \mbox{of pions, as follows:}
\begin{equation}
    p + p \rightarrow N + N + \xi_{\pi^0}\pi^0 + \xi_{\pi^\pm}(\pi^+ + \pi^-)\,,
    \label{eq:pp_interaction}
\end{equation}
where $N$ represents a nucleon (proton or neutron). Unlike the $p\gamma$ case, the charge 
ratio is roughly symmetric, with $\pi^0 : \pi^+ : \pi^- \approx 1:1:1$ at high energies. 
The resulting neutrino flavor ratio at the source is $(\nu_e : \nu_\mu : \nu_\tau) \approx (1:2:0)$, similar to the $p\gamma$ channel after \mbox{full decay.}

The cooling timescale for $pp$ interactions is determined by the density of the target gas, 
$n_{\text{gas}}$ \cite{Kelner2006}:
\begin{equation}
    t_{\rm pp}^{-1} \approx c n_{\text{gas}} \upsigma_{\rm pp} \kappa_{\rm pp}\,,
    \label{eq:pp_timescale}
\end{equation}
where $\upsigma_{\rm pp} \approx 40$ mb is the inelastic cross-section (slowly increasing with energy above the pion production threshold at $E_{\rm p} \gtrsim 1$ GeV), and $\kappa_{\rm pp} \approx 0.5$ is the inelasticity (the fraction of proton energy transferred to secondary particles). For typical HSP jet densities \mbox{($n_{\text{gas}} \lesssim 10^{3}$ cm$^{-3}$),} $t_{\rm pp}$ is much longer than the dynamical timescale \mbox{($t_{\rm dyn} \sim R/c \sim 10^{3}$--$10^{4}$ s),} making $p\gamma$ the preferred route for neutrino production unless specific high-density target scenarios are invoked, such as jet interactions with stellar winds or red giant envelopes \cite{Barkov2012, Araudo2013}, cloud--jet collisions within the broad line region \cite{Araudo2013}, or enhanced gas densities in the inner parsec-scale jet environment \cite{Romero2018}.

\subsection{Lepto-Hadronic Solutions and the Efficiency Problem}
\label{sec:lepto_hadronic}

In lepto-hadronic scenarios, the electromagnetic SED is primarily reproduced by leptonic processes (synchrotron and SSC), while a sub-dominant hadronic component is introduced to account for VHE $\gamma$-rays and, crucially, the neutrino flux. However, applying these models to HSP BL Lacs reveals a severe tension between the predicted neutrino luminosity and the total energy budget of the jet.

\subsubsection{The Energetic Crisis in One-Zone Models}

The efficiency of photomeson production, $f_{p\gamma}$, represents the fraction of proton energy converted into secondary particles per dynamical crossing time $t_{\rm dyn} \sim R/c$. It is related to the cooling rate of Equation~(\ref{eq:pgamma_cooling}) via $f_{p\gamma}(\gamma_p) = t_{\rm dyn}/t_{p\gamma}(\gamma_p)$ \cite{Waxman1997, Halzen2002}, giving the following explicit $\gamma_p$-dependent expression:
\begin{equation}
    f_{p\gamma}(\gamma_p) = \frac{R}{2\gamma_p^2}
    \int_{\epsilon'_{\rm th}}^{\infty} d\epsilon'\,
    \frac{n_\gamma(\epsilon')}{\epsilon'^2}
    \int_{0}^{2\gamma_p\epsilon'} d\bar{\epsilon}\,
    \upsigma_{p\gamma}(\bar{\epsilon})\,
    \kappa_{p\gamma}(\bar{\epsilon})\,\bar{\epsilon}\,,
    \label{eq:fpgamma}
\end{equation}
where $\epsilon'_{\rm th} = m_\pi m_p c^4/(2\gamma_p m_p c^2)$ is the threshold photon energy in the comoving frame for a proton of Lorentz factor $\gamma_p$, $n_\gamma(\epsilon')$ is the comoving-frame target photon number density, $\upsigma_{p\gamma}(\bar{\epsilon})$ is the interaction cross-section (peaking at $\sim 500~\upmu$b at the $\Delta^+$ resonance), and \mbox{$\kappa_{p\gamma} \approx 0.2$} is the inelasticity. In the simplified limit where the cross-section is approximated by a step function at the $\Delta^+$ resonance peak, Equation~(\ref{eq:fpgamma}) reduces to the following \mbox{energy-averaged expression:}
\begin{equation}
    f_{p\gamma} \approx R\,\hat{n}_\gamma\,
    \upsigma_{p\gamma}\,\kappa_{p\gamma}\,,
    \label{eq:fpgamma_simple}
\end{equation}
where $\hat{n}_\gamma$ is the photon number density evaluated at the resonance threshold energy $\epsilon'_{\rm th}(\gamma_p)$. The full expression in Equation~(\ref{eq:fpgamma}) makes explicit that $f_{p\gamma}$ is a decreasing function of $\gamma_p$ for soft photon spectra (since higher-energy protons interact with lower-energy, less abundant photons above the threshold), with important consequences for the shape of the resulting neutrino spectrum. The neutrino luminosity $L_\nu$ is then related to the proton luminosity $L_p$ by 
\cite{Kelner2008, Halzen2002}
\begin{equation}
    L_\upnu \approx \frac{3}{8} f_{{\rm p} \upgamma} L_{\rm p}\,.
    \label{eq:neutrino_luminosity}
\end{equation}

In HSPs, the target photon field is dominated by internal synchrotron radiation. While the peak frequency is high ($\nu > 10^{15}$ Hz), the photon number density $n_\gamma$ is relatively low compared to LSPs, which benefit from external fields (BLR/torus). Typical one-zone models for HSPs yield very low efficiencies, often $f_{{\rm p} \upgamma} \lesssim 10^{-4}$ \cite{Petropoulou2015a}. Consequently, to produce a neutrino flux detectable by IceCube (e.g., $L_\nu \sim 10^{45}$ erg s$^{-1}$ during a flare), the required proton luminosity must exceed $L_{\rm p} \sim 10^{49}$ erg s$^{-1}$ (implied by Equation~(\ref{eq:neutrino_luminosity})). This value is often orders of magnitude higher than the Eddington luminosity of the central black hole, posing a significant challenge to the energy budget of the source \cite{Cerruti2019, Gao2019}.

\subsubsection{The Cascade Constraint}

A secondary limitation arises from the electromagnetic cascades initiated by the $p\gamma$ process. The pairs ($e^\pm$) produced via the Bethe--Heitler process ($p + \gamma \to p + e^+ + e^-$) and pion decay chains radiate via synchrotron and inverse Compton mechanisms. In sources with high proton luminosities, this cascade emission can fill the ``dip'' between the synchrotron and high-energy humps in the SED (typically in the hard-X-ray-to-soft-$\gamma$-ray band) \cite{Murase2014}. Since HSPs are characterized by a clean separation between these humps (``hard'' X-ray spectra), the non-observation of this cascade emission places stringent upper limits on the allowable proton luminosity, often rendering simple one-zone models incapable of explaining neutrino events like TXS 0506+056 without violating X-ray constraints \cite{Keivani2018, Reimer2019}.

This raises the question of whether X-ray and cascade constraints can serve as decisive discriminators between leptonic and hadronic emission models in HSPs. In principle, the answer is yes: the clean separation between the synchrotron and IC humps in HSP SEDs, the so-called ``valley'' in the hard-X-ray-to-soft-$\gamma$-ray band ($\sim$$10$~keV--$100$~MeV), provides a sensitive window for detecting or constraining cascade emission from hadronic processes \cite{Zech2017, Cerruti2015}. Any hadronic component producing neutrinos via $p\gamma$ interactions inevitably generates secondary pairs through Bethe--Heitler production and pion decay, whose synchrotron emission fills this valley. The non-observation of excess emission in this band by instruments such as \textit{Swift}-BAT, \textit{NuSTAR}, and \textit{INTEGRAL} therefore places stringent upper limits on proton luminosity, often constraining $L_p/L_e \lesssim 10^3$--$10^4$ in one-zone models \cite{Reimer2019, Petropoulou2015a}. In this sense, the clean SED separation that makes HSPs observationally distinctive also makes them particularly constraining for hadronic models, and the very feature that suggests a photon-poor environment favorable for UHECR escape simultaneously limits the cascade emission that would betray hadronic activity.

However, this constraining power cuts both ways. The clean SED separation is also what makes HSPs favorable as UHECR accelerators: the low photon density in the valley implies low $p\gamma$ optical depth, meaning that protons accelerated to ultra-high energies can escape the source without catastrophic energy losses \cite{Murase2012, Fang2018}. There is therefore an inherent tension between the conditions required for efficient neutrino production (high photon density, high $p\gamma$ optical depth) and those required for efficient UHECR escape (low photon density, low optical depth). HSPs sit at the low-efficiency end of this trade-off, making them better UHECR factories than neutrino factories, consistent with the population constraints discussed in Section~\ref{sec:neutrino_observations} \cite{Murase2014, Tavecchio2015b}.

Regarding neutrino associations in sources with clean SED separation, the observational record is sparse but instructive. The most significant association, TXS~0506+056, is not a true HSP but a masquerading BL~Lac with a hidden BLR \cite{Padovani2019}, suggesting that its efficient neutrino production arises precisely from the additional photon target provided by the hidden BLR rather than from the SSC photon field alone. Among genuine HSPs with clean SED separation, no statistically significant neutrino associations have been established for sources such as Mrk~421, Mrk~501, and 1ES~0229 + 200 despite their proximity and extensive monitoring \cite{Aartsen2017, Acciari2020}. This non-detection is consistent with the cascade constraint argument: in genuine HSPs, the photon valley is real and deep, limiting both cascade emission and neutrino production efficiency. The most promising candidates for neutrino-producing HSPs may therefore be extreme HSPs during rare flaring states when the photon density is temporarily enhanced, or sources embedded in environments with additional photon targets such as galaxy clusters or interacting gas clouds \cite{Fang2018, Barkov2012}, rather than HSPs in their typical quiescent states. Next-generation facilities including IceCube-Gen2 and simultaneous hard X-ray monitoring with NuSTAR will be decisive in testing whether cascade emission fills the SED valley during neutrino-active states, providing the clearest possible discriminant between leptonic and hadronic scenarios in HSP jets.

\subsubsection{Multi-Zone Lepto-Hadronic Models}

\begin{figure}[H]

\includegraphics[width=0.90\textwidth]{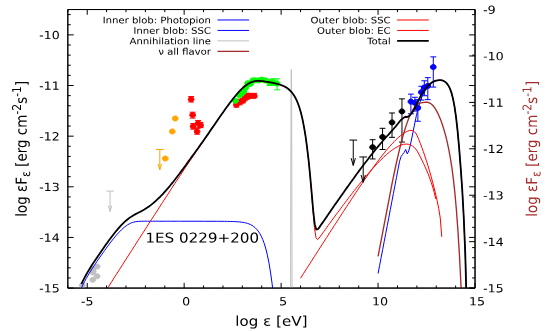}
\caption{Two-zone lepto-hadronic model for the extreme HSP 1ES 0229 + 200, showing the broadband spectral energy distribution from radio to TeV energies. The model includes contributions from an inner blob (blue solid line) producing VHE gamma-rays via photopion interactions with 511~keV annihilation-line photons from a pair plasma, and outer blobs generating X-ray synchrotron emission  and GeV-TeV gamma-rays via SSC (red dot-dashed line) and external Compton with seed photons from the dusty torus (DT, red dotted line). The total model (black solid line) successfully reproduces the multi-wavelength data without requiring extreme parameters. Inner blob parameters: $\Gamma_i = 3$; $R_i = 2.3 \times 10^{14}$ cm; $B_i = 50$ G. Outer blob parameters: $\Gamma_o = 5$; $R_o = 1.3 \times 10^{16}$ cm; $B_o = 0.18$ G. The brown line represents the predicted neutrino flux extending to PeV energies. Data points compiled from \citet{Costamante2018}. Figure extracted from \citet{Aguilar2022}.}
\label{fig:1es0229_twozone}
\end{figure}

While simple one-zone hadronic models face severe challenges in explaining the hard TeV spectra of extreme HSPs \cite{Cerruti2015}, multi-zone lepto-hadronic scenarios offer a viable alternative \cite{Zech2017, Petropoulou2020b, Tavecchio2014_spine}. Figure~\ref{fig:1es0229_twozone} shows a representative two-zone model applied to the prototypical extreme HSP 1ES 0229 + 200 \cite{Aguilar2022}. In this framework, an inner blob close to the supermassive black hole produces VHE $\gamma$-rays via photohadronic interactions with annihilation-line photons from a sub-relativistic pair plasma, while an outer blob generates X-rays through synchrotron emission and sub-TeV $\gamma$-rays via synchrotron self-Compton (SSC) and external Compton (EC) processes. This approach relaxes the extreme parameter values (such as very low magnetic field strengths and high minimum electron Lorentz factors) demanded by one-zone SSC models while naturally producing the hard spectral indices ($\Gamma \lesssim 2$) observed at TeV energies.

\subsubsection{Structured Jets: The Spine--Sheath Solution}

To resolve the efficiency and spectral issues that plague one-zone hadronic models, structured jet geometries are invoked, most notably the ``spine--sheath'' framework \cite{Ghisellini2005, Tavecchio2008}. In this scenario, the jet is composed of two coaxial regions:
\begin{itemize}
    \item A fast inner core (spine) with $\Gamma_s \sim 10$--20.
    \item A slower outer layer (sheath) with $\Gamma_{\text{sh}} \sim 2$--3.
\end{itemize}

For protons accelerated within the spine, the photon field produced by the sheath appears relativistically boosted. The relative Lorentz factor $\Gamma_{\rm rel}$ between the spine and the sheath is given by \cite{Tavecchio2008, 
Ghisellini2005}
\begin{equation}
    \Gamma_{\rm rel} = \Gamma_s\Gamma_{\rm sh}
    (1 - \beta_s\beta_{\rm sh})\,.
\end{equation}
In the ultra-relativistic limit $\beta_s, \beta_{\rm sh} \approx 1$, and for $\Gamma_s \gg \Gamma_{\rm sh}$, this expression reduces to \cite{Tavecchio2008}
\begin{equation}
    \Gamma_{\rm rel} \approx \frac{\Gamma_s}{2\Gamma_{\rm sh}}\,.
\end{equation}
\textls[-25]{For a spine with $\Gamma_s = 15$ and a sheath with $\Gamma_{\rm sh} = 3$, using $\beta_s = \sqrt{1 - 1/\Gamma_s^2} \approx 0.998$ and $\beta_{\rm sh} = \sqrt{1 - 1/\Gamma_{\rm sh}^2} \approx 0.943$, the exact expression gives \mbox{$\Gamma_{\rm rel} = 15 \times 3 \times (1 - 0.998 \times 0.943) \approx 2.7$,}} rather than the naive estimate $\Gamma_s/\Gamma_{\rm sh} = 5$, which overestimates the relative boost, or the approximate value $\Gamma_s/(2\Gamma_{\rm sh}) = 2.5$, which slightly underestimates it.

In the spine frame, sheath photons at $\epsilon_{\rm sh} = 1$~eV (observer frame) are blueshifted to $\epsilon'_{\rm sh} \approx \Gamma_{\rm rel}\epsilon_{\rm sh} \approx 2.7$~eV, placing them near the optimal energy for $p\gamma$ interactions with high-energy protons. The enhancement in photon energy density scales as $u' \propto \Gamma^2_{\rm rel}u$, where the corrected factor $\Gamma^2_{\rm rel} \approx 7.3$ provides a meaningful boost to the photohadronic interaction rate $f_{p\gamma}$ \cite{Tavecchio2008, Tavecchio2014}, though more modest than the factor of $\sim$$25$ implied by the original \mbox{naive estimate.}

Crucially, the sheath photons, which are typically in the IR/optical range in the observer frame, are boosted into the UV range in the spine frame, sitting closer to the threshold for the $\Delta^+$-resonance with high-energy protons. This mechanism provides a significant enhancement of $f_{p\gamma}$ by a factor of $\sim$$\Gamma^2_{\rm rel} \sim 6$ relative to interactions with internal synchrotron photons alone, allowing for meaningful neutrino production with sub-Eddington proton powers while maintaining the ``clean'' electromagnetic signature characteristic of HSPs \cite{Righi2017}. We note that while this enhancement is substantial, claiming an increase in $f_{p\gamma}$ by orders of magnitude requires additional conditions beyond the geometric boost alone, such as a particularly dense sheath photon field or favorable spectral overlap with the $\Delta^+$ resonance and should not be assumed generically.

\begin{figure}[H]

\includegraphics[width=0.80\textwidth]{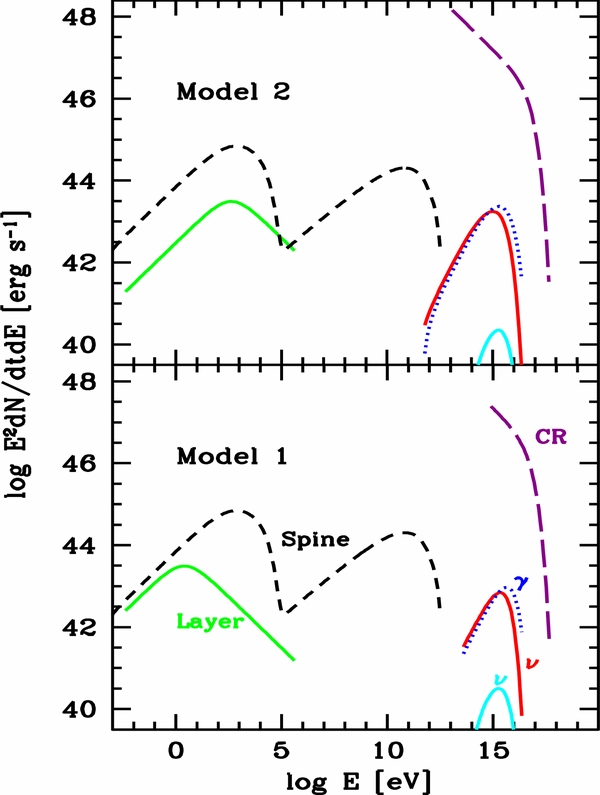}
\caption{Spectral energy distributions for two spine--sheath models of HBL blazars capable of producing the observed IceCube neutrino flux. \textbf{Top panel} (Model 2): Broad neutrino spectrum extending to $\sim$$10$ PeV. \textbf{Bottom panel} (Model 1): Narrow spectrum peaking below 1 PeV. In both models, the spine generates the dominant electromagnetic emission (black dashed line), resembling typical HBL SEDs, while the slower sheath produces a low-energy component (green solid line). The neutrino emission (red solid line) and cosmic ray spectrum (violet long-dashed line) result from photohadronic interactions of spine protons with the relativistically boosted sheath photon field. The cyan line shows neutrino emission from internal synchrotron photons alone—this is orders of magnitude lower, demonstrating the crucial enhancement provided by the spine--sheath geometry. The dotted blue line indicates photons from $\pi^0$ decay. Parameters: $\Gamma_s = 15$; $\Gamma_{\text{sh}} = 2$; $\Gamma_{\text{rel}} = 4$; $R = 10^{15}$ cm. Model 1: sheath photon peak $\epsilon_o = 2.5$ eV; Model 2: $\epsilon_o = 410$ eV. Extracted from \citet{Tavecchio2014}.}
\label{fig:spine_sheath_seds}
\end{figure}

Figure~\ref{fig:spine_sheath_seds} demonstrates how this geometry naturally produces both the observed electromagnetic spectrum and a substantial neutrino flux. Two representative models for HBL-type blazars are shown \cite{Tavecchio2014}: Model 1 produces a narrow neutrino spectrum peaking below 1 PeV (designed to lie below the IceCube upper limit at that energy), while Model 2 generates a broader spectrum extending to $\sim$$10$ PeV that can explain the entire observed IceCube flux. In both cases, the spine produces the dominant electromagnetic emission (black dashed line), resembling a typical HBL spectrum, while the layer provides a modest low-energy component (green solid line). The neutrino emission (red solid line) and the associated cosmic ray flux (violet long-dashed line) arise from photohadronic interactions between spine protons and the boosted sheath photon field. The cyan line shows the neutrino flux that would result from interactions with only the internal synchrotron photons, which is dramatically lower than the spine--sheath scenario, illustrating the critical role of the structured geometry in enabling efficient neutrino production.

\subsection{Particle Acceleration Mechanisms}\label{sec3.4}
\label{sec:reconnection}

Two primary mechanisms are invoked for particle acceleration in HSP blazar jets: diffusive shock acceleration (DSA) and magnetic reconnection. In the DSA \mbox{scenario \cite{Blandford1978, Drury1983},} particles gain energy by repeatedly crossing a shock front, naturally producing power-law energy distributions with spectral indices $s \sim 2$--$2.5$ consistent with standard blazar emission models. The maximum energy achievable via DSA is set by the balance between the energy gain rate and radiative or escape losses, and is given by the Hillas criterion discussed in Section~\ref{sec:uhecr_connection}. While DSA is well established as an efficient accelerator in mildly relativistic and non-relativistic shocks, it faces challenges in the highly magnetized ($\upsigma \gg 1$) environments inferred for HSP jets, where the magnetic pressure dominates over the particle pressure, and standard shock formation may be suppressed \cite{Sironi2015}.

Magnetic reconnection provides a complementary and increasingly favored mechanism. Recent particle-in-cell (PIC) simulations have demonstrated that reconnection in relativistic jets can efficiently accelerate particles to non-thermal distributions with hard spectral indices \cite{Sironi2015, Petropoulou2019}. Unlike DSA, reconnection-driven acceleration can operate in magnetically dominated ($\upsigma > 1$) plasmas and naturally produces the hard injection spectra ($s_p \sim 1$--$1.5$) required to explain extreme HSP emission \cite{Hakobyan2023} without requiring the fine-tuning of shock parameters that DSA demands in this regime.

In this scenario, the jet is threaded by anti-parallel magnetic field lines that reconnect in thin current sheets, converting magnetic energy into particle kinetic energy \cite{Giannios2009, Sironi2014}. The maximum particle energy achievable via reconnection is set by the condition that the particle Larmor radius does not exceed the size of the reconnection layer $R'$ \cite{Sironi2014, Werner2016}, as follows:
\begin{equation}
    \gamma'_{\rm max} \sim \frac{eB'R'}{mc^2}\,,
    \label{eq:gamma_max_reconnection}
\end{equation}
where $B'$ and $R'$ are the comoving magnetic field strength and reconnection layer size, respectively, and $m$ is the particle mass. For electrons, PIC simulations show that the maximum Lorentz factor scales with the electron magnetization parameter $\upsigma_e = B'^2/(4\pi n_e m_e c^2)$ as $\gamma'_{{\rm max},e} \sim \upsigma_e$ \cite{Sironi2014, Werner2016}, which, for blazar jets with $\upsigma_e \sim 10^4$--$10^6$, yields $\gamma'_{{\rm max},e} \sim 10^4$--$10^6$, consistent with the electron Lorentz factors required to explain HSP X-ray and TeV emission. For protons, the maximum energy in the observer frame is
\begin{equation}
    E_{{\rm max},p} = \Gamma\, eB'R' \propto 
    Z\upsigma^{1/2}\Gamma m_p c^2\,,
    \label{eq:emax_reconnection}
\end{equation}
where $\upsigma = B'^2/(4\pi \rho c^2)$ is the total plasma magnetization parameter, $\rho$ is the mass density, and the proportionality reflects the dependence of $B'$ and $R'$ on $\upsigma$ and $\Gamma$ for a given plasma configuration \cite{Sironi2014, Werner2018}. The two expressions in Equation~(\ref{eq:emax_reconnection}) are not generally numerically equal; the proportionality constant depends on the specific values of $B'$ and $R'$ adopted. For a numerical estimate, we adopt typical extreme HSP comoving-frame parameters inferred from proton-synchrotron modeling ($B' \sim 10$~G, $R' \sim 10^{14}$~cm; \citet{Cerruti2015}, \linebreak  \citet{Aguilar2022}) with $\Gamma \sim 15$, as follows:
\begin{equation}
    E_{{\rm max},p} = \Gamma\, ZeB'R' \approx 
    15 \times Z \times (4.8\times10^{-10}~{\rm esu}) 
    \times 10~{\rm G} \times 10^{14}~{\rm cm} 
    \approx Z \times 10^{20}~{\rm eV}\,,
    \label{eq:emax_numerical}
\end{equation}
This yields $E_{{\rm max},p} \sim 10^{20}$~eV for protons and $\sim 26\times10^{20}$~eV for iron nuclei ($Z = 26$), consistent with the highest-energy cosmic rays observed \cite{Fromm2017}. The $\upsigma$-parameterized form in Equation~(\ref{eq:emax_reconnection}) then implies $\upsigma^{1/2} \sim eB'R'/(m_p c^2) \sim (4.8\times10^{-10} \times 10 \times 10^{14})/(1.5\times10^{-3}) \sim 3\times10^6$ for these parameter values, corresponding to $\upsigma \sim 10^{13}$, an extremely high magnetization consistent with the inner jet regions of extreme HSPs where reconnection is expected to operate most efficiently \cite{Sironi2014, Werner2018}.

The acceleration timescale in reconnection is typically much shorter than that in shock acceleration, potentially explaining the rapid variability observed in some HSP flares. Furthermore, the naturally hard spectral indices produced by reconnection ($s \sim 1$--$1.5$) are consistent with the values required in hadronic models to reproduce extreme TeV spectra without invoking excessive fine-tuning of parameters.

Beyond DSA and magnetic reconnection, several additional acceleration mechanisms have been proposed as relevant in the context of relativistic blazar jets. \textit{Curvature and gradient drift acceleration} occurs when particles drift along curved or inhomogeneous magnetic field lines, gaining energy from the electric field induced by the drift motion \cite{Hirotani2016}. In the highly curved magnetic field geometry near the jet launching region or in the vicinity of the central black hole magnetosphere, curvature radiation losses and curvature drift acceleration can compete, with the maximum particle energy set by the balance between the two \cite{Hirotani2016, Levinson2018}. This mechanism is particularly relevant for particle acceleration in black hole magnetosphere gap regions of depleted plasma, where the electric field induced by the rotating spacetime can accelerate charges to ultra-high energies \cite{Blandford1977, Levinson2018}.

\textit{Stochastic (second-order Fermi) acceleration}, also known as magnetic mirror acceleration or turbulent acceleration, involves particles gaining energy through repeated scatterings off magnetic irregularities or plasma waves moving in random directions \cite{Fermi1949, Blandford1987}. Unlike first-order DSA, where particles systematically gain energy at a shock front, stochastic acceleration is a diffusive process in momentum space with an energy gain rate $\dot{E} \propto E$ and a characteristic acceleration timescale $t_\mathrm{acc} \sim \lambda_\mathrm{mfp}/(v_A^2/c)$, where $\lambda_\mathrm{mfp}$ is the particle mean free path and $v_A$ is the Alfv\'en speed \cite{Petrosian2012, Asano2014}. In blazar jets, stochastic acceleration by magnetohydrodynamic (MHD) turbulence has been invoked to explain the hard particle spectra and log-parabolic SEDs observed in HSPs \cite{Tramacere2011, Massaro2004}, as well as the rapid variability observed during flaring states \cite{Asano2014, Asano2014}. The mechanism naturally produces curved log-parabolic rather than pure power-law energy distributions, consistent with the spectral shapes commonly inferred from broadband SED fitting of HSP blazars \cite{Massaro2004, Tramacere2011}.

\textit{Converter acceleration} \cite{Derishev2003} is a mechanism specific to relativistic pair-plasma jets in which a neutron produced in a hadronic interaction can cross a shock front freely (being uncharged), decay back into a proton on the other side, and thus gain energy without the scattering limitations that constrain standard DSA. This mechanism can, in principle, accelerate protons to energies beyond the standard DSA limit and has been proposed as a route to UHECR energies in GRB and blazar jets \cite{Derishev2003}.

\textit{Shear acceleration} at the boundary between the fast jet spine and the slower jet \mbox{sheath \cite{Rieger2004, Rieger2019}} operates through the first-order Fermi mechanism, acting on the velocity gradient rather than on a shock, producing hard power-law spectra with indices \mbox{$s \sim 1$--$2$ \cite{Rieger2019, Webb2018}.} This mechanism is particularly attractive for HSPs since it operates naturally within the spine–-sheath geometry already invoked for neutrino production \mbox{(Section~\ref{sec:reconnection}),} and can accelerate both electrons and protons to high energies at the jet boundary without requiring a strong shock. \textit{Wakefield acceleration} by plasma waves \cite{Chen2002} has also been proposed, though it remains less studied in the blazar context.

The relative importance of these mechanisms in HSP blazar jets remains an open question. In practice, multiple mechanisms may operate simultaneously in different regions of the jet, with DSA dominating at strong internal shocks, reconnection operating in magnetically dominated regions, stochastic acceleration contributing throughout the turbulent jet volume, and shear acceleration active at the jet boundary \cite{Blandford2019}. Disentangling their contributions requires time-resolved multi-wavelength observations, polarimetric measurements with IXPE \cite{Weisskopf2022}, and detailed particle-in-cell simulations that can capture the microphysics of each process \cite{Sironi2015, Petropoulou2019}.


\section{Observational Status of the Blazar--Neutrino Connection}
\label{sec:neutrino_observations}

The search for the sources of high-energy astrophysical neutrinos has been one of the primary objectives of multi-messenger astronomy. While Active Galactic Nuclei (AGNs), and specifically blazars, have long been proposed as the most likely candidates for accelerating cosmic rays to PeV and EeV energies \cite{Mannheim1993, Halzen1997}, the observational evidence has historically been elusive. This section reviews the current status of the field, detailing the breakthrough association with TXS 0506+056, the puzzling ``orphan'' neutrino flares, and the stringent constraints derived from the non-detection of the broader HSP population.

\subsection{The 2017 Multi-Messenger Flare of TXS 0506+056}
\label{sec:txs}

The landscape of high-energy astrophysics was fundamentally altered on \linebreak  \mbox{22 September 2017} with the detection of the high-energy neutrino event IceCube-170922A. This event was identified by the IceCube Real-time Alert System as an extremely high-energy (EHE) track-like event originating from a neutrino interaction vertex within the ice. The reconstructed muon energy was estimated at $290$ TeV, with a 90\% confidence level lower limit of $183$ TeV, implying a primary neutrino energy likely in the PeV \mbox{range \cite{IceCube2018}.} Crucially, the event was reconstructed with a high angular resolution, yielding a best-fit direction with a 50\% error radius of $15'$ and a 90\% error radius of roughly $1^\circ$, which triggered an automated Gamma-ray Coordinate Network (GCN) notice to the \mbox{astronomical community.}

\subsubsection{Electromagnetic Counterparts and Identification}
\label{sec:txs1}

Within the error circle of IceCube-170922A lies the blazar TXS~0506+056 (also known as 3FGL~J0509.4+0541), a source previously cataloged by \textit{Fermi-LAT}. Spectroscopic observations obtained shortly after the alert determined the redshift to be $z=0.3365$, placing the source at a luminosity distance of approximately $1.8\,$Gpc \cite{Paiano2018}.

The association was solidified by the state of the source in the $\gamma$-ray band. The \textit{Fermi}-LAT Large-Area Telescope revealed that TXS~0506+056 was in a state of heightened activity, having brightened by a factor of $\sim$$6$ relative to its quiescent level in the weeks leading up to the neutrino's arrival. The source was located only $0.1^{\circ}$ from the best-fit neutrino direction, well within the positional uncertainty.

Following the alert, an extensive multi-wavelength campaign was initiated. While the VERITAS observatory observed the source immediately, it yielded only upper limits due to the source's soft spectrum at the time and absorption effects \cite{Abeysekara2018}. The MAGIC Cherenkov telescopes detected very-high-energy (VHE, $E > 100\,$GeV) $\gamma$-rays from the source during an observation window of $12\,$days after the neutrino alert (28 September to 4 October). This detection revealed a significant hardening of the spectrum during the flare, with photons detected up to $\sim$$400$~GeV \cite{Ansoldi2018}. This marked the first time VHE $\gamma$-rays were detected from this object, establishing a direct link between the neutrino event and the acceleration of particles to energies sufficient to produce the VHE emission.

\begin{figure}[H]

    \includegraphics[width=0.78\textwidth]{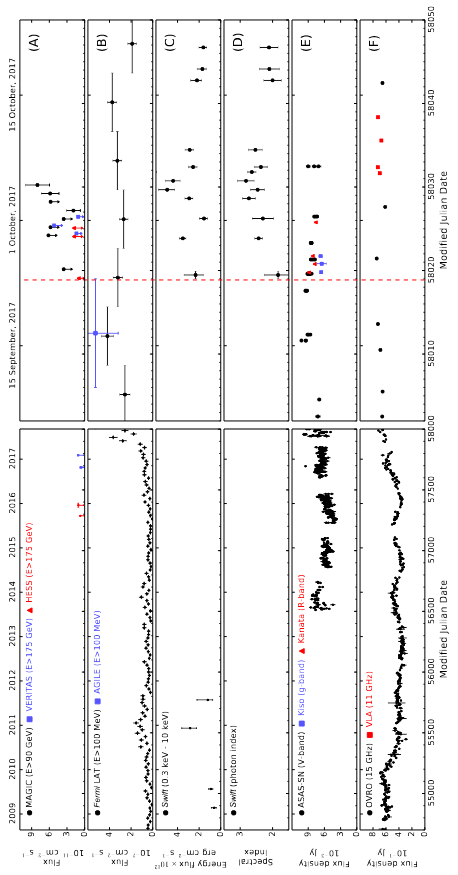}
    \caption{Multi-wavelength light curves of TXS~0506+056 surrounding the detection of IceCube-170922A (vertical dashed line at MJD~58018.6). (\textbf{Panel A}): High-energy (HE, 0.1--300 GeV) $\gamma$-rays from \textit{Fermi}-LAT (black points) and AGILE (blue points). (\textbf{Panel B}): Very-high-energy (VHE, \mbox{$>$90 GeV)} $\gamma$-rays from MAGIC, with red squares indicating detections during the follow-up campaign and gray arrows showing upper limits. (\textbf{Panel C}): X-ray flux (0.3--10 keV) from \textit{Swift}-XRT (black) and \textit{NuSTAR} (red, within the highlighted flare window). \textbf{Panel D}): Optical R-band magnitude. (\textbf{Panel E}): Radio flux density at 15 GHz from OVRO. (\textbf{Panel F}): Radio flux density at 37 GHz from Metsähovi. The $\gamma$-ray enhancement coinciding with the neutrino arrival is evident, while the X-ray, optical, and radio bands show comparatively modest variability. Extracted from \citet{IceCube2018}.}
    \label{fig:txs_lightcurves}
\end{figure}

Figure~\ref{fig:txs_lightcurves} displays the comprehensive multi-wavelength monitoring of TXS~0506+056 across six decades in frequency, from radio to VHE $\gamma$-rays, encompassing the critical period surrounding the neutrino detection. The temporal coincidence between the IceCube-170922A event (marked by the vertical dashed line) and the elevated $\gamma$-ray flux measured by Fermi-LAT and AGILE is striking. The subsequent MAGIC detection window (highlighted in panel B) reveals the presence of TeV photons, demonstrating that the source was actively accelerating particles to ultrarelativistic energies during this epoch. Notably, the X-ray flux (panel C) remains relatively stable, a feature that constrains hadronic cascade models, as discussed below.

\subsubsection{Statistical Significance}

The significance of this association was evaluated by combining the spatial coincidence with the temporal correlation of the neutrino arrival during a high-flux $\gamma$-ray state. Using a likelihood ratio test and accounting for the trial factor from scanning the entire \textit{Fermi}-LAT catalog, the combined spatial and temporal significance was determined to be \mbox{$3.0\upsigma$ \cite{IceCube2018}.} This association provided the first compelling evidence linking a specific blazar to high-energy neutrino production and, more generally, to hadronic particle acceleration.

\subsubsection{Theoretical Modeling and Constraints}

The simultaneous observation of photons from radio to VHE $\gamma$-rays and a high-energy neutrino enabled comprehensive SED modeling; however, simple one-zone lepto-hadronic models face severe challenges \cite{Keivani2018, Gao2019}.

In standard hadronic scenarios, neutrinos arise from charged pion decay following photomeson production. The production reaction proceeds via
\begin{equation}
    p + \gamma \rightarrow n + \pi^+,
\end{equation}
after which the charged pion decays as
\begin{equation}
    \pi^+ \rightarrow \mu^+ + \nu_\mu \rightarrow 
    e^+ + \nu_e + \bar{\nu}_\mu + \nu_\mu,
\end{equation}
while neutral pions produced in the complementary channel $p + \gamma \rightarrow p + \pi^0$ decay as $\pi^0 \rightarrow 2\gamma$, producing $\gamma$-ray cascades. Critical constraints emerge from secondary pair production via Bethe--Heitler processes ($p + \gamma \rightarrow p + e^+ + e^-$) and $\gamma\gamma$ absorption, whose synchrotron emission appears predominantly in X-rays.

Modeling TXS~0506+056 reveals the following fundamental tensions:
\begin{enumerate}
    \item Energetic Requirements: Producing the detected neutrino flux demands proton luminosities $L_p \sim 10^{47}$ erg s$^{-1}$, significantly exceeding Eddington limits for the estimated black hole mass \cite{Cerruti2019}.
    \item X-ray Cascade Limits: The enhanced proton luminosity required for neutrinos inevitably boosts secondary pair synchrotron emission, overshooting the X-ray flux measured by \textit{Swift}-XRT and \textit{NuSTAR} during the flare (Figure~\ref{fig:txs_lightcurves}, panel C) \cite{Keivani2018, Gao2019}.
\end{enumerate}

Reconciling these constraints requires either radiatively sub-dominant hadronic components (neutrinos from hadrons, $\gamma$-rays from leptonic SSC) or structured jet geometries (spine--sheath) where neutrinos and photons originate from spatially separated regions with different Doppler factors \cite{Petropoulou2020a, Ansoldi2018, Zheng2016}. Beyond the immediate multi-messenger modeling, TXS~0506+056 has also been studied as a source of UHECRs: \mbox{Das, Gupta, and Razzaque \cite{Das2022}} analyzed the implications of its multi-wavelength spectrum for cosmic-ray acceleration, while Saveliev and Alves Batista \cite{Saveliev2020} studied the intrinsic $\gamma$-ray spectrum accounting for intergalactic propagation effects and derived constraints on the extragalactic magnetic field strength along the line of sight.

\subsection{The 2014--2015 ``Orphan'' Neutrino Flare}

Following the identification of the 2017 event, the IceCube collaboration undertook a comprehensive time-dependent analysis of archival data collected between April 2008 and July 2017, centered on the coordinates of TXS~0506+056 \cite{IceCube2018b}. This retrospective search uncovered a significant neutrino excess preceding the 2017 alert by nearly three years.

\subsubsection{Characteristics of the Neutrino Excess}

\begin{figure}[H]
   
    \includegraphics[width=.98\textwidth]{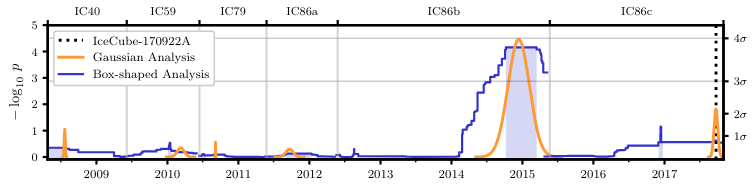}
    \caption{Time-dependent analysis revealing the 2014--2015 orphan neutrino flare from TXS~0506+056. The orange curve shows the analysis using a Gaussian-shaped time profile, with the central time $T_0$ and width $T_{\rm W}$ plotted for the most significant excess in each period. The blue curve uses a box-shaped time profile, tracing the outer edge of best-fitting time windows with significance indicated by height. The prominent blue band centered near 2015 represents the best-fitting 158-day window with \mbox{$13 \pm 5$ events} above background, corresponding to $3.5\upsigma$ significance and a spectral index of $\gamma_\upnu = 2.1 \pm 0.2$. The vertical dotted line in IC86c marks the IceCube-170922A event (September 2017). Data periods correspond to different detector configurations: IC40 (2008--2009), IC59 (2009--2010), IC79 (2010--2011), IC86a (2011--2012), IC86b (2012--2015), and IC86c (2015--2017). Extracted from \mbox{\citet{IceCube2018b}.}}
    \label{fig:neutrino_excess}
\end{figure}

The analysis identified $13 \pm 5$ neutrino events above background during a 158-day window from September 2014 to March 2015 (MJD 56937--57096), as shown in Figure~\ref{fig:neutrino_excess} \cite{IceCube2018b}. The excess reached $3.5\upsigma$ significance (independent of the 2017 observation), with a hard spectral index $\gamma_\upnu \sim 2.1 \pm 0.2$ and time-averaged neutrino luminosity \mbox{$L_{\upnu,\text{iso}} \approx (1$--$3) \times 10^{46}$ erg s$^{-1}$,} comparable to the contemporaneous $\gamma$-ray luminosity \cite{IceCube2018b}.

\subsubsection{The ``Dark'' Nature and Multi-Messenger Tension}

Unlike the 2017 event, the 2014--2015 neutrino flare occurred without electromagnetic enhancement. \textit{Fermi}-LAT data show TXS~0506+056 was in a quiescent state, with no corresponding radio or X-ray activity \cite{Garrappa2019}. This ``orphan'' neutrino flare challenges standard one-zone lepto-hadronic models: $p\gamma$ interactions produce both neutrinos ($\pi^\pm \to \mu^\pm \to \nu$) and $\gamma$-rays ($\pi^0 \to 2\gamma$) with probabilities governed by isospin symmetry. High neutrino flux requires comparable $\gamma$-ray emission; yet, none were observed.

\subsubsection{Theoretical Interpretations}

Several scenarios address this tension. In all cases, the central challenge is the same: $p\gamma$ or $pp$ interactions that produce charged pions ($\pi^\pm$), and hence neutrinos, inevitably also produce neutral pions ($\pi^0 \rightarrow 2\gamma$) whose decay $\gamma$-rays must be accounted for. Each model must therefore explain not only why neutrinos were detected but also where the associated $\pi^0$ decay $\gamma$-rays have gone:
\begin{enumerate}
\item High Internal Opacity: Neutrinos escape while $\gamma$-rays, including those from $\pi^0$ decay, are absorbed via $\gamma\gamma$ pair production ($\tau_{\gamma\gamma} \gg 1$) and cascade to lower energies. However, \citet{Reimer2019} showed that the resulting cascaded X-ray emission from the reprocessed $\pi^0$ decay photons would violate \textit{Swift}-XRT and MAXI limits, ruling out simple one-zone hidden models. The $\pi^0$ decay $\gamma$-rays therefore cannot simply be hidden by absorption without producing detectable secondary X-ray emission.
\item Structured Jets: Spine--sheath geometries decouple neutrino production (dense inner spine) from quiescent $\gamma$-rays (outer layer). In this scenario, $\pi^0$ decay $\gamma$-rays produced in the compact inner spine are subject to strong $\gamma\gamma$ absorption on the dense photon field of the spine itself ($\tau_{\gamma\gamma} \gg 1$ locally), and the cascaded emission is reprocessed into X-rays and soft $\gamma$-rays at flux levels below the detection threshold of \textit{Fermi}-LAT and contemporaneous X-ray instruments \cite{Petropoulou2020a, Tavecchio2014}. The spatial separation between the neutrino-producing spine and the leptonic outer layer further dilutes any observable electromagnetic signature of the hadronic activity.
\item Jet--Environment Interactions: In $pp$ collision scenarios driven by transient density enhancements from jet--star or jet--cloud collisions, the $\pi^0$ decay $\gamma$-rays are produced at the interaction site and are subject to strong internal absorption if the target cloud or stellar envelope is optically thick to $\gamma\gamma$ pair production \cite{Liu2019, Wang2023}. The cascaded emission is again reprocessed to energies below the \textit{Fermi}-LAT threshold, while the $pp$ channel avoids the strict isospin symmetry coupling between neutrino and $\gamma$-ray fluxes that makes the $p\gamma$ channel so constraining, since in $pp$ interactions, the pion charge ratio depends on the kinematics rather than being fixed by isospin symmetry alone. However, the energetics of producing sufficient target material transiently remain challenging \cite{Liu2019, Wang2023}.
\end{enumerate}

The 2014--2015 flare suggests that blazars produce neutrinos efficiently only during transient phases decoupled from electromagnetic activity \cite{Abdollahi2020}.

\subsection{Other Candidate Associations and Population Constraints}

While TXS 0506+056 remains the most statistically significant association to date, the search for correlations between high-energy neutrinos and blazars has yielded other tentative candidates. These associations, often based on temporal coincidences with electromagnetic flares, provide crucial test cases for lepto-hadronic models, even if their statistical significance does not yet reach the $5\upsigma$ threshold.

\begin{itemize}
    \item PKS 1424-418 and the ``Big Bird'' Event: One of the earliest potential associations involved the PeV-energy neutrino event IC35 (nicknamed ``Big Bird''), detected by IceCube on 4 December 2012. The event was a cascade-type interaction (shower) with a reconstructed energy of $2^{+0.4}_{-0.3}$ PeV, making it one of the highest-energy neutrinos ever observed \cite{Kadler2016}. 

    A temporal and spatial coincidence was identified with a major outburst of the flat-spectrum radio quasar (FSRQ) PKS 1424-418. During the epoch of the neutrino arrival, the source was undergoing a dramatic brightening in $\gamma$-rays (observed by \textit{Fermi}-LAT) and radio wavelengths. Kadler et al. (2016) \cite{Kadler2016} showed that the integrated electromagnetic energy output of the outburst was sufficient to explain the neutrino event energetically. However, the association is limited by the large angular uncertainty characteristic of cascade events ($\sim$$15^\circ$), resulting in a post-trial chance probability of approximately 5\%.

    \item GB6 J1040+0617: A more recent search found a spatial coincidence between the track-like neutrino event IceCube-141209A and the low-synchrotron-peaked object GB6 J1040+0617. Unlike the TXS 0506+056 case, this source was in a relatively modest state of activity. A comprehensive multi-wavelength analysis by Garrappa et al. (2019) \cite{Garrappa2019} demonstrated the difficulty of reconciling this association with standard emission models. To reproduce the observed neutrino flux, a single-zone lepto-hadronic model would require a proton luminosity exceeding the Eddington limit by several orders of magnitude ($L_p \gtrsim 10^{49}$ erg s$^{-1}$), suggesting that if the association is real, the neutrino production must occur in a highly efficient, likely multi-zone environment.
\end{itemize}

\subsubsection*{The Silence of the Brightest HSPs}

Perhaps as significant as these tentative detections is the non-detection of the brightest and closest high-synchrotron-peaked blazars. Sources such as Markarian 421 and \mbox{Markarian 501} are regularly monitored at TeV energies and exhibit extreme flaring states. Despite their high $\gamma$-ray flux and favorable viewing angles, stacking analyzes of these sources have yielded null results.

Several works placed early constraints on the blazar contribution to the IceCube-detected neutrino background prior to the IceCube collaboration's own analysis. \linebreak  \mbox{\citet{Wang2014}} used \textit{Fermi}-LAT observations to show that the blazar contribution to the IceCube neutrino flux is limited, while \citet{Wang2016} demonstrated that FSRQs alone cannot account for the observed diffuse neutrino emission, and \citet{Zhang2017} derived constraints on cosmic ray and PeV neutrino production in blazars from multi-wavelength considerations. These early constraints were subsequently corroborated and quantified more precisely by the IceCube collaboration's stacking analysis of the Fermi-2LAC catalog, which placed stringent upper limits on the contribution of the general blazar population to the diffuse astrophysical neutrino flux, constraining it to $\lesssim$$27\%$ of the total observed \mbox{flux \cite{Aartsen2017}.} Together, these works imply that efficient neutrino production is not a ubiquitous property of all blazars but may be restricted to specific subclasses (e.g., masked/extreme HBLs) or rare transient phases where the target photon density for $p\gamma$ interactions is temporarily enhanced \cite{Murase2014, Righi2017}.

\subsection{Population Constraints and the Diffuse Flux}
\label{sec:population_constraints}

While the identification of TXS~0506+056 provided  a proof-of-principle that blazars can produce high-energy neutrinos, the lack of additional significant associations has created tension between individual source models and the properties of the population as a \mbox{whole \cite{Palladino2019}.} To quantify this, the IceCube collaboration and independent groups performed ``stacking analyzes'', where the sub-threshold neutrino signals from the directions of known electromagnetic sources were summed to test for a collective \mbox{excess \cite{Aartsen2017, Padovani2016}.}

\subsubsection{Limits from Stacking Analyzes}

Extensive stacking searches using catalogs of blazars detected by \textit{Fermi}-LAT (e.g., the 2LAC, 3FHL, and 4FGL catalogs) have consistently yielded null results for the general population. Aartsen et al. (2017) \cite{Aartsen2017} analyzed the contribution of 862 2LAC blazars to the diffuse neutrino flux and found no significant excess. They derived an upper limit, stating that resolvable \textit{Fermi}-LAT blazars can account for at most $\sim$$27\%$ of the total diffuse astrophysical neutrino flux detected by IceCube in the TeV--PeV range. 

\begin{figure}[H]

    \includegraphics[width=0.9\textwidth]{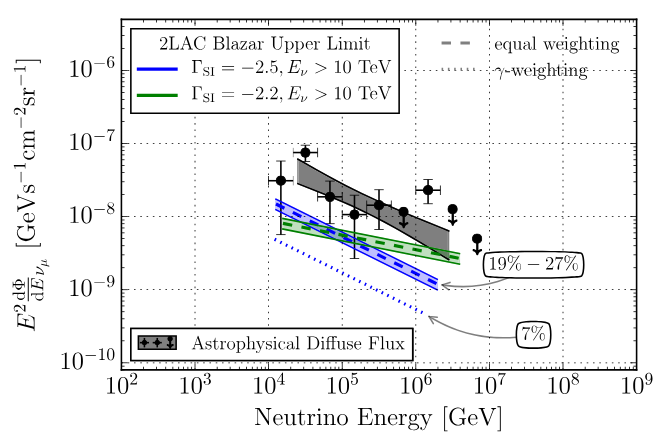}
    \caption{IceCube stacking analysis results for the \textit{Fermi}-2LAC blazar catalog. The neutrino flux $E^2 \Phi$ is plotted as a function of neutrino energy for different blazar subclasses. The horizontal gray band indicates the measured astrophysical diffuse neutrino flux. Colored arrows show 90\% confidence level upper limits for specific populations: flat-spectrum radio quasars (FSRQs, solid red), BL Lac objects (solid blue), low-synchrotron-peaked blazars (LSP, solid green), and high-synchrotron-peaked blazars (HSP, solid purple). Two spectral hypotheses are shown for each class, namely $\Gamma_{\mathrm{SI}} = -2.5$ (solid lines) and $\Gamma_{\mathrm{SI}} = -2.2$ (dashed lines), where harder spectra ($\Gamma = -2.2$) produce higher fluxes at multi-PeV energies. Dotted lines show the flux expected if neutrino production were proportional to $\gamma$-ray luminosity (``$\gamma$-weighting''). The null detection across all subclasses constrains the contribution of resolvable \textit{Fermi}-LAT blazars to $\lesssim$$27\%$ of the total diffuse flux, implying that the bulk of the neutrino background originates either from sources that are dark in $\gamma$-rays or from other source classes entirely. Extracted from \citet{Aartsen2017}.}
    \label{fig:stacking_limits}
\end{figure}

Figure~\ref{fig:stacking_limits} presents these constraints quantitatively, showing the 90\% confidence level upper limits on the neutrino flux contributions from different blazar subclasses. The horizontal gray band represents the measured diffuse astrophysical neutrino flux at \linebreak  $E^2 \Phi \approx 10^{-8}\,\mathrm{GeV\,cm^{-2}\,s^{-1}\,sr^{-1}}$, while the arrows indicate that even the most optimistic scenarios for individual subclasses (FSRQs, BL Lacs, LSPs, HSPs) fall well below this level. The best-fit values (solid points) are consistent with zero, demonstrating that the population of $\gamma$-ray-bright blazars cannot account for the majority of the diffuse neutrino background.

This limit has profound implications for the ``Blazar Sequence'' and hadronic emission models. If the majority of the diffuse neutrino flux does not come from the blazars we see in $\gamma$-rays, it must originate from the following:
\begin{enumerate}
    \item Other Source Classes: Such as starburst galaxies, radio galaxies (e.g., NGC~1068), tidal disruption events, or galaxy clusters \cite{Murase2016, Fang2018}.
    \item Unresolved/Faint Blazars: A population of sources that are too faint to be resolved individually by \textit{Fermi}-LAT but numerous enough to dominate the neutrino background. However, population studies of the $\gamma$-ray luminosity function suggest that the unresolved blazar contribution is likely sub-dominant \cite{Ajello2015}.
    \item Dark Neutrino Sources: Objects where the $\gamma$-ray emission is completely suppressed by internal absorption (``hidden accelerators'') leave only neutrinos and cascaded X-rays to escape \cite{Murase2016}.
\end{enumerate}

\subsubsection{Tension with the Diffuse Gamma-Ray Background}

There is also a multi-messenger tension related to the isotropic diffuse gamma-ray background (IGRB). Hadronic interactions ($p\gamma$ and $pp$) that produce neutrinos inevitably produce neutral pions ($\pi^0$), which decay into high-energy $\gamma$-rays. These $\gamma$-rays cascade down to lower energies (GeV-TeV) via interactions with the EBL and the CMB.

If the diffuse neutrino flux were entirely produced by hadronically emitting blazars that are transparent to $\gamma$-rays, the associated cascade emission would saturate or exceed the observed IGRB measured by \textit{Fermi}-LAT \cite{Murase2016}. This suggests that the sources responsible for the bulk of the neutrino flux must be opaque to $\gamma$-rays (``gamma-dark'') or have spectral indices significantly higher than the average blazar population.


\section{Observational Status of the Blazar--UHECR Connection}
\label{sec:uhecr_connection}

While the association between blazars and high-energy neutrinos has seen a breakthrough with TXS 0506+056, the direct link between HBLs and UHECRs ($E > 10^{18}$ eV) remains one of the most challenging frontiers in astroparticle physics. Unlike neutrinos and photons, cosmic rays are charged particles (protons and heavier nuclei) and are deflected by Galactic and extragalactic magnetic fields (GMF and EGMF), scrambling their arrival directions and delaying their arrival times relative to electromagnetic signals \cite{Kotera2011, AlvesBatista2019}.

Despite these challenges, HBLs are considered prime candidates for UHECR acceleration due to their low-luminosity and high-peaked SEDs, which suggest an environment where radiative losses are minimized, potentially allowing protons to reach energies up to $10^{20}$ eV via the Hillas criterion \cite{Hillas1984, Murase2012}.

The fundamental requirement for accelerating particles to ultra-high energies is encapsulated in the Hillas criterion, which states that the gyroradius of the particle must remain smaller than the size of the acceleration region, as follows:
\begin{equation}
    E_{\max} \lesssim \frac{ZeBR\beta_{\rm shock}}{\Gamma}\,,
    \label{eq:hillas}
\end{equation}
where $Z$ is the charge number, $B$ is the magnetic field strength, $R$ is the
characteristic size of the acceleration region in the observer frame, $\beta_{\rm shock}$ is the shock velocity in units of $c$, and $\Gamma$ is the bulk Lorentz factor of the jet. The factor $\Gamma^{-1}$ arises because the confinement condition must be satisfied in the comoving frame: the comoving frame size of the emission region is $R^{\prime}=R/\Gamma$, so the maximum energy in the comoving frame is $E^{\prime}_{\max} = ZeB^{\prime}R^{\prime}\beta_{\rm shock} = ZeBR\beta_{\rm shock}/\Gamma$. This comoving frame bound is the physically relevant one for particle confinement, and Equation~(\ref{eq:hillas}) expresses it directly in terms of observer frame quantities \cite{Hillas1984,Kotera2011}. We stress that this expression applies to the DSA/confinement picture. In the magnetic reconnection scenario discussed in Sections~\ref{sec3.4} and~\ref{sec5.3}, the maximum energy is instead set by the electric potential drop across a reconnection layer,
\begin{equation}
    E_{\max,\rm obs} = \Gamma\,ZeB^{\prime}R^{\prime},
    \label{eq:Emax_reconnection}
\end{equation}
which carries $\Gamma$ in the \emph{numerator} because both the comoving field $B^{\prime}$ and layer size $R^{\prime}$ are specified in the comoving frame, and the resulting comoving frame energy is then boosted once to the observer frame~\cite{Sironi2014,Werner2016,Werner2018}. The two scalings are physically distinct and should not be conflated.

\begin{figure}[H]

    \includegraphics[width=0.8\textwidth]{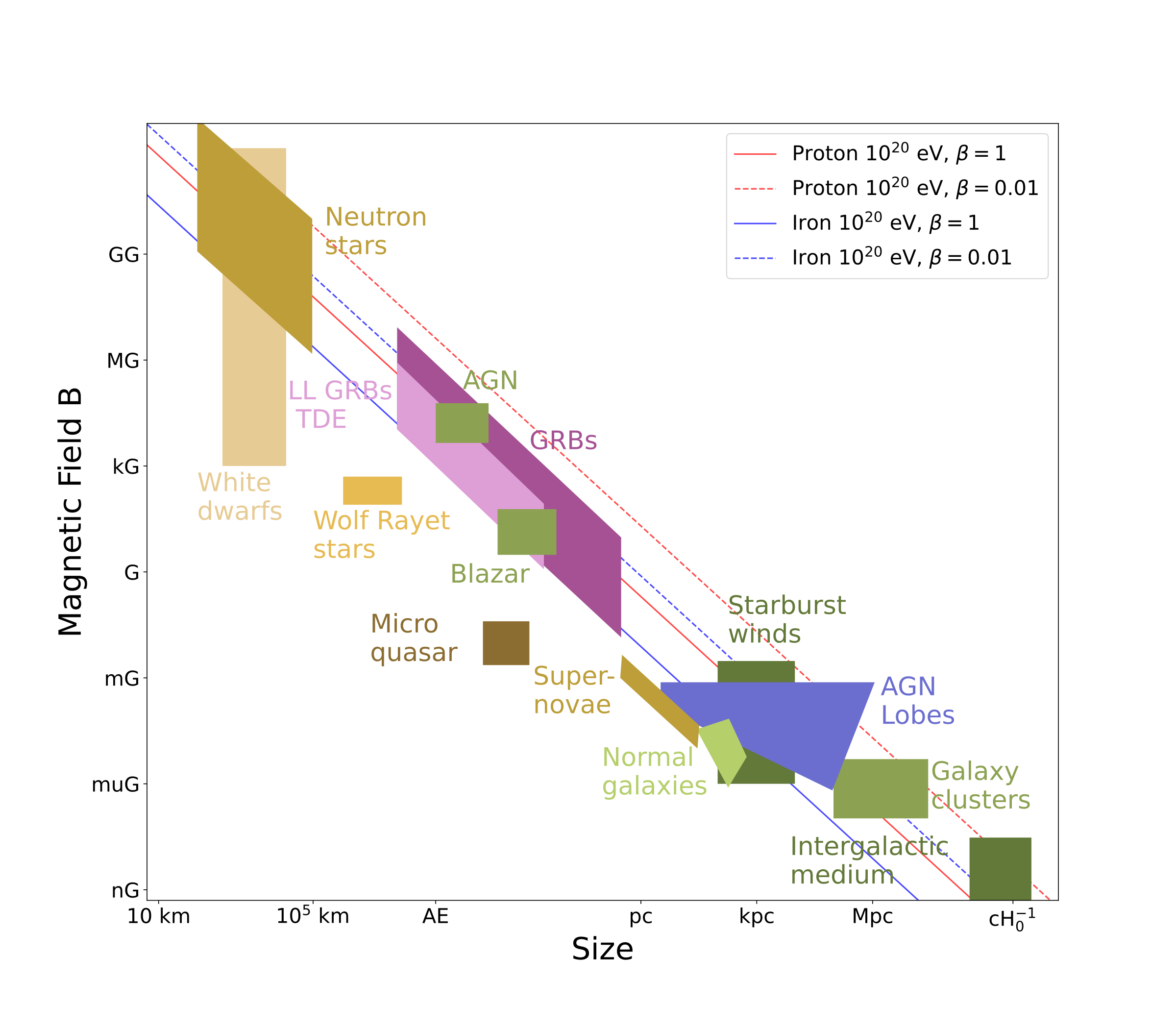}
    \caption{Hillas diagram showing magnetic field strength versus size for astrophysical objects and UHECR acceleration to $10^{20}$ eV. Diagonal lines indicate constant maximum energy: red (protons) and blue (iron nuclei) for relativistic ($\beta = 1$, solid) and non-relativistic ($\beta = 0.01$, dashed) shocks. Galactic sources (tan/orange/brown: white dwarfs, Wolf-Rayet stars, neutron stars, micro-quasars, supernovae) fall below the proton threshold. The olive green ``Blazar'' region ($B \sim 0.1$--$10$ G, $R \sim 10^{14}$--$10^{16}$ cm) places standard HSPs marginally at the proton threshold but above the iron threshold, while extreme HSPs ($B \sim 10$--$100$ G, $R \sim 10^{14}$--$10^{15}$ cm, left portion) firmly exceed both. The AGN region (olive green, overlapping blazars) includes radio galaxies. Purple GRBs and light pink LL GRBs/TDEs occupy high-field compact zones. Blue AGN lobes (lower right) satisfy Hillas via Mpc scales. Green starburst winds, light green normal galaxies, olive green galaxy clusters, and the dark green intergalactic medium occupy progressively larger, weaker-field environments. HSPs, particularly EHSPs, approach or exceed UHECR acceleration requirements, motivating anisotropy studies, composition measurements, and multi-messenger correlations. Extracted from \citet{HillasPlotGithub}.}
    \label{fig:hillas_plot}
\end{figure}

Figure~\ref{fig:hillas_plot} presents a Hillas diagram highlighting the parameter space relevant for UHECR acceleration to $10^{20}$ eV. The diagonal lines represent the constant maximum energy for both protons (red lines) and iron nuclei (blue lines) at $10^{20}$ eV, with solid lines showing $\beta = 1$ (relativistic shocks) and dashed lines showing $\beta = 0.01$ (non-relativistic shocks). Traditional Galactic sources, namely white dwarfs (light tan, upper left), Wolf Rayet stars (orange), neutron stars (tan/brown, upper left), microquasars (brown), and supernovae (yellow-tan, lower middle), fall below the $10^{20}$ eV proton threshold line (solid red) for typical \mbox{shock parameters.}

Extragalactic objects occupy the parameter space above the UHECR acceleration threshold. The ``Blazar'' region (olive green, center) represents the typical HSP parameter space, with magnetic fields $B \sim 0.1$--$10$~G and characteristic sizes $R \sim 10^{14}$--$10^{16}$~cm (spanning from $\sim$$7$~AU to $\sim$$700$~AU, or equivalently $\sim 3\times10^{-5}$ to $\sim 3\times10^{-3}$~pc), derived from standard synchrotron self-Compton modeling of X-ray and TeV SEDs. This region intersects the solid red proton line (indicating $\beta \sim 1$ relativistic shocks can just barely reach $10^{20}$ eV) and sits comfortably above the solid blue iron line (indicating iron nuclei can easily reach $10^{20}$ eV total energy).

For canonical HSP models with $B \sim 0.1$--$1$ G and $R \sim 10^{15}$--$10^{16}$ cm (the right side of the blazar box, inferred from day-scale variability and equipartition assumptions), the Hillas criterion places these sources marginally at or slightly below the threshold for $10^{20}$~eV proton acceleration when accounting for bulk Lorentz factors $\Gamma \sim 10$--$20$. However, a subset of extreme HSPs (EHSPs), whose hard TeV spectra require more extreme conditions, specifically higher magnetic fields ($B \sim 10$--$100$ G) in more compact regions ($R \sim 10^{14}$--$10^{15}$ cm, the left side of the blazar box), would comfortably exceed the UHECR threshold even for protons. These parameters are derived from proton-synchrotron and lepto-hadronic models applied to sources like 1ES 0229+200 \cite{Cerruti2015, Biteau2020}.

The broader ``AGN'' region (also olive green, center-upper) encompasses the full range of jet-powered active galaxies, including both blazars and their misaligned counterparts. ``AGN Lobes'' (blue, lower right) represent the extended radio lobes of powerful radio galaxies, with scales of 10 kpc to Mpc and weaker magnetic fields ($\sim$$\upmu$G), easily satisfying the Hillas criterion for iron and even protons due to their enormous sizes. ``Gamma-ray bursts'' (``GRBs'', purple, center) and ``low-luminosity GRBs with tidal disruption events'' (``LL GRBs TDE'', light pink) occupy high-field, compact parameter spaces. ``Starburst winds'' (green, right-center) extend to parsec-kpc scales with milli-Gauss fields. ``Normal galaxies'' (light green, lower right), ``galaxy clusters'' (olive green, far right), and the ``intergalactic medium'' (dark green, far lower right) represent progressively larger, weaker-field environments.

The following diagram demonstrates that while standard HSPs are marginal UHECR accelerators for protons, they become increasingly viable for heavier nuclei and are particularly promising if extreme conditions (higher $B$, smaller $R$) or alternative acceleration mechanisms like magnetic reconnection apply. The fact that the blazar region sits precisely at the threshold for $10^{20}$ eV proton acceleration, combined with the observation that EHSPs push toward higher fields and smaller sizes, makes this source class a compelling but challenging candidate requiring the multi-messenger observational strategies and composition studies discussed throughout this section.
\vspace{-24pt}

\subsection{Anisotropy Searches and Correlation Studies}

The primary observational strategy involves searching for anisotropies in the arrival directions of UHECRs and cross-correlating them with the positions of known HBLs.

\subsubsection{The Pierre Auger Observatory Results}

The Pierre Auger Observatory (PAO) has provided the most statistically significant evidence for UHECR anisotropy. PAO reported a large-scale dipole anisotropy above $8 \times 10^{18}$ eV at $>5\upsigma$ significance,  pointing away from the Galactic Center, confirming the extragalactic origin \cite{PierreAuger2017}.

\begin{figure}[H]
 
    \includegraphics[width=0.85\textwidth]{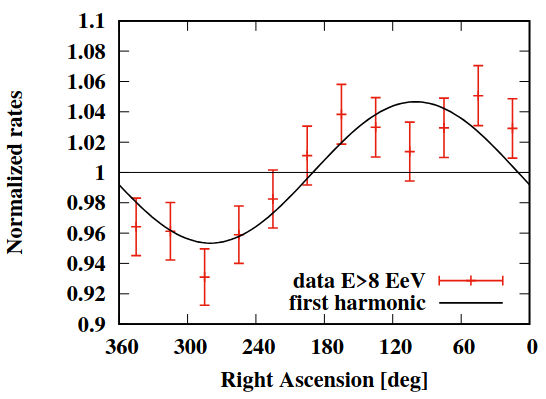}
    \caption{Right ascension distribution of cosmic rays with $E > 8 \times 10^{18}$ eV from Pierre Auger Observatory. Red points show normalized event rates; the black curve shows best-fit dipole modulation with amplitude $6.5^{+1.3}_{-0.9}\%$ and phase 
    $\alpha \approx 100^\circ \pm 10^\circ$ (Centaurus direction). This $>$$5.2\upsigma$ anisotropy points away from the Galactic Center, confirming extragalactic origin. The modest amplitude suggests multiple contributing sources or magnetic deflections smoothing point-source signatures. Extracted from \citet{PierreAuger2017}.}
    \label{fig:auger_anisotropy}
\end{figure}

Figure~\ref{fig:auger_anisotropy} shows the right ascension distribution above $8 \times 10^{18}$ eV. The dipole has an amplitude of $\sim$$6.5\%$ pointing toward $\alpha \approx 100^\circ$ (Centaurus), detected at $>$$5.2\upsigma$. This direction points \textit{away} from the Galactic Center ($\alpha \approx 266^\circ$), confirming its extragalactic origin. The moderate amplitude and broad angular scale suggest either multiple sources within the GZK horizon or significant magnetic deflections smoothing point-source signatures.

On intermediate scales, PAO found a correlation with starburst galaxies ($\sim$$4\upsigma$) and $\gamma$-ray AGNs ($\sim$$2.7\upsigma$) for $E > 39$~EeV \cite{PierreAuger2018}. While starburst galaxies currently fit better, HBLs remain viable if magnetic deflections are larger than the models predict. The best-fit angular scale of $\sim$$13^\circ$ is consistent with intermediate-mass nuclei at $\sim$$10^{19}$~eV in IGMF of $B \sim 1$~nG with a coherence length of $\lambda \sim 1$~Mpc \cite{Sigl2004, Kotera2011}.

\begin{figure}[H]

    \includegraphics[width=0.85\textwidth]{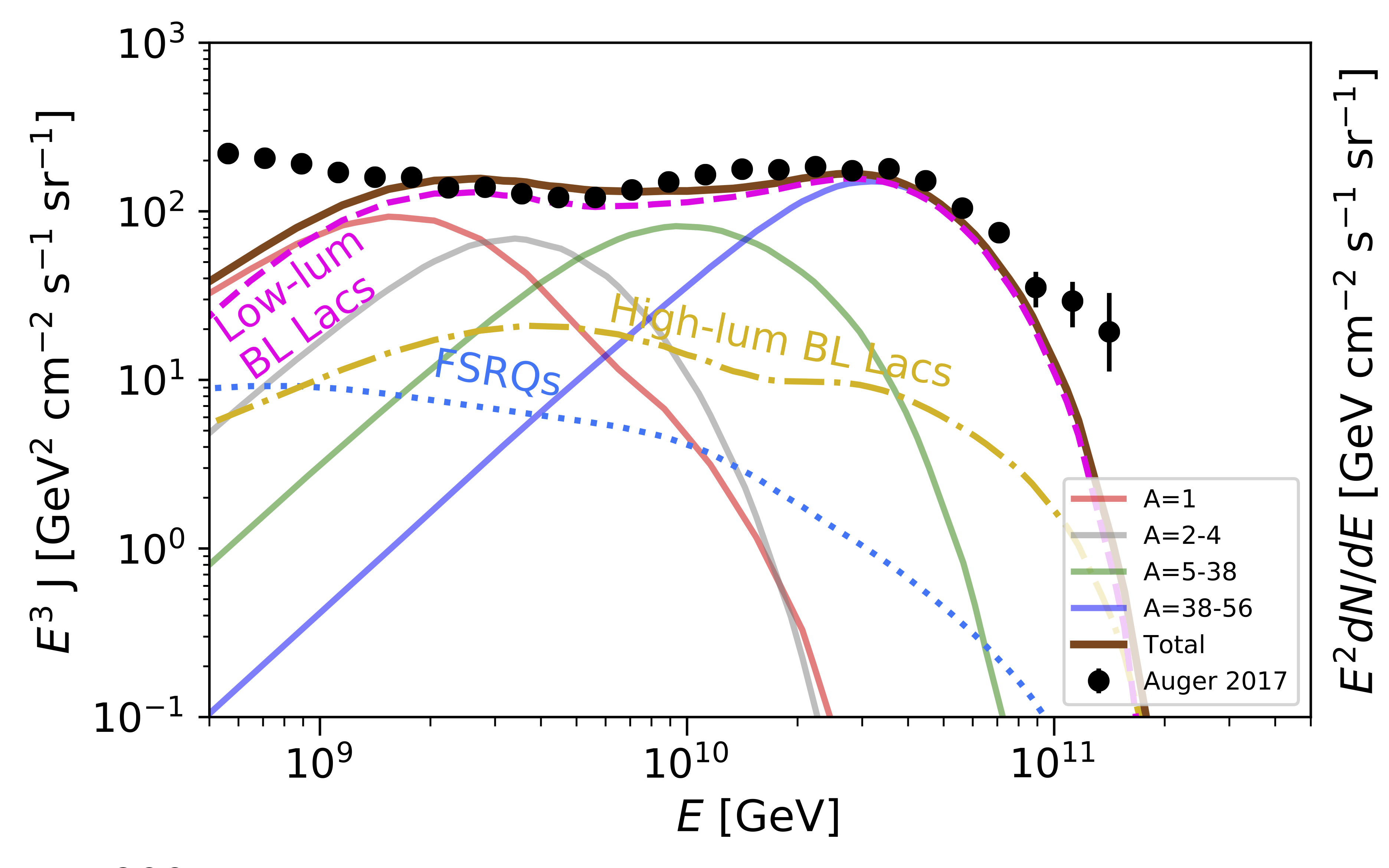}
    \caption{Simulated UHECR spectrum from the AGN population model of \citet{Rodrigues2021} compared to Pierre Auger Observatory data (black points). The total flux (pink) is dominated by low-luminosity BL~Lacs (blue curve), with subdominant contributions from high-luminosity BL~Lacs (yellow) and FSRQs (dotted blue). Individual  mass groups are shown for $A = 1$ (protons); $A = 2$--4; $A = 5$--38; and $A = 38$--56. The model reproduces the spectral shape at and above the ankle with baryonic loading $\xi_{\rm CR} \approx 380$ for low-luminosity BL~Lacs and a cosmic-ray acceleration efficiency of $\eta_{\rm acc} = 0.1$. Extracted from~\citet{Rodrigues2021}.}
    \label{fig:rodrigues_uhecr}
\end{figure}

A population-level explanation for the Auger spectrum consistent with a BL~Lac origin was advanced by \citet{Rodrigues2021}, who performed a self-consistent multi-messenger fit, including source interactions, extragalactic propagation, and blazar population evolution. As illustrated in Figure~\ref{fig:rodrigues_uhecr}, the dominant contribution to the UHECR flux at and above the ankle arises from \textit{low-luminosity} BL~Lacs, whose negative cosmological evolution ($z \lesssim 0.5$) and sparse radiation fields allow accelerated nuclei to escape the jet essentially unimpeded, naturally reproducing the rigidity-dependent spectral shape inferred by Auger. The composition above the ankle is predicted to be mixed to heavy (dominated by $A = 5$--56 groups), broadly consistent with the Auger $X_{\rm max}$ measurements, though the model yields a composition slightly heavier than that observed at $E \gtrsim 10^{10}$~GeV unless a subdominant proton-rich contribution from FSRQs or high-luminosity BL~Lacs is included. This framework places HSP~BL~Lacs at the center of UHECR phenomenology while simultaneously predicting a low source-neutrino flux from BL~Lacs themselves, since their photon-starved jets suppress photohadronic 
interactions.

\subsubsection{Telescope Array and the ``Hotspot''}

Telescope Array (TA) reported an event concentration with $E > 57$ EeV near \linebreak  $\alpha \approx 146.7^\circ$, $\delta \approx 43.2^\circ$ (Ursa Major) at $\sim$$3.4\upsigma$ significance \cite{Abbasi2014}, currently $\sim$$3.3\upsigma$ after 10 years \cite{Abbasi2018}.

This region contains prominent HBLs, including Mrk~421 ($\alpha = 166.1^\circ$, $\delta = +38.2^\circ$, $d \approx 140$ Mpc) and Mrk~180 ($\alpha = 177.4^\circ$, $\delta = +70.2^\circ$, $d \approx 200$ Mpc), bright TeV blazars demonstrating ultrarelativistic particle acceleration. However, direct correlation is complicated by composition: if dominated by intermediate/heavy nuclei (C, N, O, Fe) rather than protons, magnetic deflections could exceed $10^\circ$--$20^\circ$, washing out point-source \mbox{correlations \cite{Abbasi2018}.}

For a cosmic ray with energy $E$ and charge $Z$ in IGMF strength $B$ with a
coherence length $\lambda$, the RMS deflection angle is determined by the particle rigidity $\mathcal{R} = E/Z$ \cite{Kotera2008} as follows:
\begin{equation}
    \langle \theta_{\text{rms}} \rangle \approx 1.7^\circ
    \left(\frac{Z}{1}\right)
    \left(\frac{10^{19}\,\text{eV}}{E}\right)
    \left(\frac{B}{1\,\text{nG}}\right)
    \sqrt{\frac{D}{100\,\text{Mpc}}}
    \sqrt{\frac{\lambda}{1\,\text{Mpc}}}\,.
    \label{eq:magnetic_deflection}
\end{equation}
For iron ($Z=26$) at $E = 6\times10^{19}$~eV traversing 100--150~Mpc, this yields $\langle\theta_{\rm rms}\rangle \approx 1.7^\circ \times 26 \times (10^{19}/6\times10^{19})
\approx 7.4^\circ$ for $B = 1$~nG and $D = 100$~Mpc, rising to $\sim$$37^\circ$ for $B = 5$~nG. Deflections large enough to fully obscure HBL point-source correlations ($\theta \gtrsim 10^\circ$--$20^\circ$) therefore require either fields at the upper end of current EGMF estimates ($B \gtrsim 1$--$3$~nG; \citet{Sigl2004, Dolag2011}) or a composition dominated by intermediate-to-heavy nuclei at lower rigidity. The composition--magnetic field degeneracy therefore remains a fundamental challenge.

An equally important and often underappreciated consequence of magnetic deflection is the associated time delay of UHECRs relative to simultaneously emitted photons.
A cosmic ray deflected by angle $\theta$ while traversing a distance $d$ arrives delayed by~\cite{Waxman1996,Miralda1996}
\begin{equation}
    \Delta t \approx \frac{\theta^{2}\,d}{2c}\,,
    \label{eq:time_delay}
\end{equation}
Substituting the deflection angle from Equation~(\ref{eq:magnetic_deflection}) and typical parameters, this yields
\begin{equation}
    \Delta t \approx
    \frac{1}{2c}
    \left(\frac{Z}{1}\right)^{2}
    \left(\frac{10^{19}\,\text{eV}}{E}\right)^{2}
    \left(\frac{B}{1\,\text{nG}}\right)^{2}
    \left(\frac{D}{100\,\text{Mpc}}\right)^{3}
    \left(\frac{\lambda}{1\,\text{Mpc}}\right)
    \times 10^{3}\,\text{yr}\,,
    \label{eq:time_delay_explicit}
\end{equation}
where the dependence is on rigidity $\mathcal{R}=E/Z$, i.e., as $(Z/E)^{2}$, rather than on energy per nucleon $E/A$. For protons ($Z=1$) at $E=10^{19}$~eV traversing 100~Mpc in a field $B\sim1$~nG with a coherence length $\lambda\sim1$~Mpc, this gives $\Delta t \sim 10^{3}$ years. For iron ($Z=26$) at the same total energy, the rigidity is a factor of 26 lower, giving $\Delta t \sim 26^{2}\times10^{3}\approx 7\times10^{5}$~years. These time delays are vastly longer than any blazar monitoring timescale or human observational baseline, with profound implications for multi-messenger studies: even if a nearby HSP such as Mrk~421 or Mrk~501 is currently undergoing an extreme flaring state, the UHECRs produced during this flare will not arrive on Earth for thousands to millions of years. Conversely, the UHECRs arriving today were produced during a past activity state of the source that occurred long before any electromagnetic monitoring was possible~\cite{Kotera2011,Waxman1996}. This makes it essentially impossible to establish direct temporal correlations between UHECR arrival directions and the high-activity states of blazars observed in GeV--TeV $\gamma$-rays~\cite{Murase2012,AlvesBatista2019}. Any UHECR--blazar association must therefore rely on directional rather than temporal coincidence and must account for the smearing of arrival directions by magnetic deflections, as quantified in Equation~(\ref{eq:magnetic_deflection}). The time delay also implies that the UHECR flux from a given source reflects its time-averaged luminosity over the delay timescale rather than its instantaneous state, further diluting any correlation with current electromagnetic observations~\cite{Kotera2011,
AlvesBatista2019}.

The TA$\times$4 upgrade will determine whether the hotspot persists and constrains the composition of contributing particles \cite{Kawata2017}. If confirmed with light (proton/helium) composition, this would strengthen the case for nearby HBLs like Mrk~421 as UHECR accelerators.

\subsection{Constraints from Multi-Messenger Limits}

The non-detection of strong UHECR-HBL correlations constrains source properties and the intervening medium. In the HSP blazar cluster scenario \cite{Fang2018}, blazars embedded in galaxy clusters simultaneously produce UHECRs escaping above $\sim$$10^{18}$ eV, PeV neutrinos from hadronic interactions in the intracluster medium, and sub-TeV $\gamma$-rays from electromagnetic cascades. The cosmic ray spectrum in this picture decomposes into light (H, He) and medium-heavy (CNO, Si, Fe) components, with a compositional transition around $10^{19}$ eV driven by energy-dependent escape and cluster confinement, a feature consistent with Auger and TA compositional measurements. Neutrino production occurs on two distinct scales: hadronic interactions within clusters yield a PeV component in agreement with IceCube observations, while cosmogenic interactions at EeV energies produce a subdominant flux that remains below current IceCube sensitivity. The $\gamma$-ray output, arising from neutral pion decay within clusters and subsequent electromagnetic cascades, matches the non-blazar component of the \textit{Fermi} extragalactic $\gamma$-ray background. Crucially, approximately 30 HSPs within 100~Mpc suffice to explain all three messengers simultaneously, while the modest source density and magnetic deflections in the large-scale structure naturally account for the observed near-isotropy of the UHECR arrival distribution.

Key constraints on HSPs as UHECR sources:

\begin{itemize}
    \item The GZK Horizon: Photo-pion production limits sources above $6 \times 10^{19}$ eV to within $\sim$$100$ Mpc for protons \cite{Greisen1966, Zatsepin1966}. The 3HSP catalog contains only $\sim$$30$ HSPs with \mbox{$z < 0.02$} \mbox{($D_L < 90$ Mpc),} including Mrk~421 \mbox{($d \approx 140$ Mpc),} Mrk~501 \mbox{($d \approx 150$ Mpc),} 1ES~1959+650 ($d \approx 210$ Mpc), and PKS~2155-304 ($d \approx 540$ Mpc). This small population creates statistical challenges: if each contributes comparably and magnetic deflections isotropize arrivals, anisotropy signals are diluted. Moreover, detector complementarity is critical—Auger covers Southern sources (PKS~2155-304) while TA covers Northern sources (Mrk~421, Mrk~501).

    \item Secondary Neutral Messengers: UHECR propagation produces cosmogenic neutrinos and $\gamma$-rays. The Fermi isotropic $\gamma$-ray background limits the proton fraction if sources evolve strongly \cite{Berezinsky2016}. This suggests either (1) hard injection spectra ($s \lesssim 2.0$--$2.3$) with negative/flat evolution matching HBL observations \cite{Ajello2014, Chang2019}, (2) heavier composition producing fewer secondaries, or (3) both. HSP luminosity evolution consistency with multi-messenger backgrounds provides circumstantial support \cite{Heinze2019, Fang2018}.

    \item Magnetic Field Constraints: The lack of clustering near Mrk~421/Mrk~501 implies either a strong EGMF or heavy composition. For pure protons and weak fields \mbox{($B \sim 0.01$--$0.1$ nG,} \mbox{$\lambda \sim 1$ Mpc),} deflections remain below $\sim$$3^\circ$ above $5 \times 10^{19}$ eV (Equation~(\ref{eq:magnetic_deflection})), which should produce detectable signatures. Observed isotropy reconciles through (1) stronger EGMF ($B \gtrsim 1$ nG), deflecting protons by $10^\circ$--$20^\circ$ \cite{Dolag2004, Sigl2004}; (2) heavier composition with rigidity $R = E/Z$ reduced by factors of $\sim$$6$--$26$, producing $>$$20^\circ$ deflections even in modest fields, consistent with Auger composition \mbox{measurements \cite{Aab2014};} or (3) multiple sources creating a quasi-isotropic background without a dominant nearby accelerator.
\end{itemize}

\subsection{The ``Extreme'' HBL Connection}\label{sec5.3}

A specific subset of the blazar population, the ``extreme'' high-frequency peaked BL Lacs (EHBLs), presents a unique opportunity to resolve the UHECR origin problem. Defined by an intrinsic synchrotron peak frequency $\nu_{\text{syn}}^{\text{peak}} \geq 10^{17}$ Hz and a $\gamma$-ray spectrum that remains hard ($s_{\gamma} \lesssim 2$) up to multi-TeV energies, these sources, exemplified by the archetype 1ES 0229+200, challenge standard emission models and suggest a distinct acceleration regime \cite{Costamante2001a, Foffano2019}.

\subsubsection{Hadronic Solutions for Hard Spectra}

The persistence of hard TeV spectra in EHBLs is difficult to reconcile with standard one-zone leptonic models, which typically predict a softening of the spectrum due to Klein--Nishina effects at the highest energies. This has motivated the development of proton-dominated scenarios where the high-energy emission is not inverse Compton scattering but rather direct proton-synchrotron radiation or synchrotron emission from secondary pairs produced in photo-hadronic interactions.

In proton-synchrotron frameworks, the jet environment is assumed to be extremely clean, lacking significant external photon fields. To reproduce the observed SED of 1ES 0229+200, protons must be accelerated to ultra-high energies \mbox{($E_{\rm p} \sim 10^{19}$--$10^{20}$ eV)} with a very hard injection index ($s_{\rm p} \sim 1.5$) and a high minimum Lorentz factor \mbox{($\gamma_{\text{p,min}} \sim 100$) \cite{Cerruti2015}.} Unlike the lepto-hadronic hybrid models applied to TXS 0506+056, these proton-synchrotron models require magnetic fields of $B \sim 10$--$100$ G, significantly higher than those inferred for standard HBLs but consistent with the parameter space shown in Figure~\ref{fig:hillas_plot} that comfortably exceeds the UHECR acceleration threshold.

The extreme magnetic fields and compact emission regions required by these models have important implications for UHECR production. The maximum energy achievable via diffusive shock acceleration scales as $E_{\text{max}} \sim Z e B R \beta_{\text{shock}} \Gamma$, so magnetic fields of \mbox{$B \sim 10$--$100$ G} in regions of size $R \sim 10^{14}$--$10^{15}$ cm, with bulk Lorentz factors $\Gamma \sim 20$--$50$, can accelerate protons to $\sim 10^{20}$ eV and iron nuclei to even higher total energies. However, the high magnetic field strength also increases synchrotron losses, requiring extremely efficient acceleration with timescales approaching the gyration time, $t_{\text{acc}}/t_{\text{gyr}} \lesssim 1$. Such efficient acceleration may be achievable through magnetic reconnection in the highly magnetized jet, where particles gain energy through the electric fields in reconnection layers rather than through first-order Fermi acceleration at shocks \cite{Sironi2015, Werner2018}.

\subsubsection{The IGMF and UHECR Propagation}

The connection between EHBLs and UHECRs is further strengthened by the study of the IGMF. The hard TeV photons from EHBLs interact with the EBL to produce electron--positron pairs ($e^\pm$). These pairs are deflected by the IGMF before inverse Compton scattering CMB photons back into the GeV band. The non-detection of this secondary GeV ``halo'' or ``echo'' around sources like 1ES 0229+200 in \textit{Fermi}-LAT data implies either a strong IGMF ($B_{\text{IGMF}} \gtrsim 10^{-15}$ G) that deflects pairs out of the line of sight or that the primary VHE emission is dominated by hadrons (UHECRs) rather than photons, which would not produce the expected electromagnetic cascade \cite{Tavecchio2010b, Dolag2011, Neronov2010}.

If EHBLs are indeed UHECR accelerators operating in the proton-synchrotron regime, the secondary emission produced during cosmic ray propagation (cosmogenic $\gamma$-rays and neutrinos) becomes a critical diagnostic. The absence of a strong GeV halo around 1ES 0229+200 could be interpreted as evidence that the TeV emission is hadronic rather than leptonic, supporting the UHECR connection. However, alternative explanations involving high IGMF strengths can also account for the observations without invoking \mbox{hadronic emission.}

Recent population studies suggest that EHBLs could contribute significantly to the UHECR flux above $10^{19}$ eV without violating the isotropic diffuse $\gamma$-ray background limits, provided their evolution is negative or flat (similar to the broader HBL population) and that the cosmic ray injection has a hard spectrum with $s \sim 1.5$--$2.0$ \cite{Biteau2020, Fang2018}. The required hard injection spectrum is consistent with both the proton-synchrotron modeling of EHBL SEDs and with the constraints from fitting UHECR composition evolution, suggesting a potential self-consistent picture linking extreme blazars, hard-spectrum particle acceleration, and the highest-energy cosmic rays.

\subsection{Composition Constraints and Source Implications}

The chemical composition of UHECRs provides crucial constraints on the source class and acceleration mechanisms. Recent analyzes of the depth-of-shower maximum ($X_\mathrm{max}$) from the Pierre Auger Observatory indicate a transition from a predominantly light composition (proton or helium) at $E \sim 10^{18}$~eV to an increasingly heavy composition (CNO or even Fe-like) above $10^{19.5}$~eV \cite{Aab2014}. The one-population model fits to the Auger spectrum and composition data consistently confirm this heavy nucleus dominance at the highest energies, with cosmogenic photon and neutrino flux predictions providing additional constraints on the source composition and evolution \cite{AlvesBatista2019b}. This trend, if interpreted at face value, makes it difficult to reconcile with simple HBL acceleration models, which predict rigidity-dependent maximum energies $E_{\max} \propto Z$. Under identical source conditions, heavier nuclei reach proportionally higher total energies than protons; therefore, protons are cut off first, leaving heavier nuclei to dominate the flux at the highest energies.

The $X_{\text{max}}$ observable provides a statistical measure of the mass of the primary cosmic ray through the atmospheric depth at which extensive air showers reach maximum development \cite{Matthews2005}. For a pure proton composition, air showers penetrate deeper into the atmosphere (larger $X_{\text{max}}$) before reaching maximum development compared to heavier nuclei like iron, which interact higher in the atmosphere (smaller $X_{\text{max}}$) \cite{Kampert2012}. The Auger data show that the mean $\langle X_{\text{max}} \rangle$ decreases with energy above $\sim$$10^{18.5}$ eV, indicating a shift toward a heavier composition \cite{PierreAuger2014b, PierreAuger2017_composition}. This is quantified by the elongation rate \cite{Gaisser1991}:
\begin{equation}
    D_{10} = \frac{d \langle X_{\text{max}} \rangle}{d \log_{10}(E/\text{eV})},
\end{equation}
which decreases from $\sim$$60$ g cm$^{-2}$ per decade (consistent with a constant proton composition) at $10^{18}$ eV to $\sim$$26$ g cm$^{-2}$ per decade (consistent with intermediate mass nuclei or a mixed composition transitioning toward heavier elements) at $10^{19.6}$ eV \cite{PierreAuger2014b, Unger2015}.

\textls[-35]{In this framework, the maximum energy scales with particle rigidity \mbox{$\mathcal{R} = E/Z$ \cite{Hillas1984, Lagage1983} as follows:}}
\begin{equation}
    E_{\max} = \frac{ZeBR\beta_{\rm shock}}{\Gamma} 
    \sim Z \times 10^{18} 
    \left(\frac{B}{1~\rm G}\right)
    \left(\frac{R}{10^{16}~\rm cm}\right)
    \left(\frac{15}{\Gamma}\right)
    ~\rm eV\,.
    \label{eq:hillas_composition}
\end{equation}
For magnetic fields and emission region sizes typically inferred for standard HBL jets ($B \sim 0.1$--$1$~G, $R \sim 10^{15}$--$10^{16}$~cm, $\Gamma \sim 10$--$20$), the corrected Hillas criterion gives
\begin{equation}
    E_{\max} = \frac{ZeBR\beta_{\rm shock}}{\Gamma} \sim 
    Z \times 10^{18}~{\rm eV}\,,
\end{equation}
so that iron nuclei ($Z = 26$) reach $E_{\max} \sim 2.6 \times 10^{19}$~eV, a factor of 26 higher than protons under identical conditions. This is a direct consequence of the rigidity dependence of the Hillas criterion: heavier nuclei with a higher charge $Z$ are accelerated to proportionally higher total energies than protons in the same magnetic environment, making heavy composition acceleration easier, not harder, than proton acceleration to the same total energy \cite{Hillas1984, Kotera2011}. The challenge for heavy nuclei is therefore not their acceleration, but rather their survival against photodisintegration on ambient photon fields during propagation from the source to Earth, as discussed in Section~\ref{sec:intro}. With standard HSP parameters, iron nuclei comfortably reach $\sim$$10^{19}$~eV, and for extreme HSPs with $B \sim 10$--$100$~G and $R \sim 10^{14}$--$10^{15}$~cm, iron can reach $\sim$$10^{20}$~eV or beyond. The Auger composition measurements indicating a transition toward heavier nuclei above $\sim$$10^{18.5}$~eV are therefore broadly consistent with HSP acceleration, provided the photodisintegration losses during propagation are accounted \mbox{for \cite{Rodrigues2021, Globus2015}.} This suggests the following interpretations of the compositional evolution:

(1) Rigidity-dependent maximum energy: Since $E_{\max} \propto Z$ \cite{Hillas1984,Kotera2011}, HSP jets naturally produce a rigidity-ordered composition at the highest energies: protons reach their acceleration limit first at $E_{\max} \sim 10^{18}$~eV, while heavier nuclei, with their proportionally higher charge, extend the spectrum to total energies $Z$ times larger. The flux above the proton cutoff is therefore dominated by progressively heavier species, consistent with the transition toward heavier composition observed by Auger above $\sim$$10^{18.5}$~eV~\cite{Rodrigues2021,Unger2015}.
(2) Alternative acceleration mechanisms: If acceleration occurs via magnetic reconnection rather than shock acceleration, the energy gain mechanism fundamentally differs. In reconnection-powered acceleration, particles gain energy through the electric field induced in the reconnection layer \cite{Zenitani2001, Werner2016}. In the comoving frame of the jet, the maximum particle energy is set by the condition that the particle's Larmor radius does not exceed the size of the reconnection layer $R'$, as follows:
\begin{equation}
    E'_{\max} = ZeB'R'\,,
    \label{eq:reconnection_emax}
\end{equation}
where $B'$ and $R'$ are the comoving magnetic field strength and size of the reconnection region, respectively. The observed maximum energy is then $E_{\max,\rm obs} = \Gamma E'_{\max} = \Gamma ZeB'R'$. Substituting typical extreme HSP comoving-frame parameters ($B' \sim 10$~G, $R' \sim 10^{14}$~cm, $\Gamma \sim 15$) yields the following:
\begin{equation}
    E_{\max,\rm obs} = \Gamma ZeB'R' \sim 15 \times Z 
    \times (4.8\times10^{-10}~{\rm esu}) \times 10~{\rm G} 
    \times 10^{14}~{\rm cm} \sim Z \times 10^{20}~{\rm eV}\,,
    \label{eq:reconnection_numerical}
\end{equation}
yielding $E_{\max,\rm obs} \sim 10^{20}$~eV for protons and $\sim$$26 \times 10^{20}$~eV for iron nuclei, consistent with the highest-energy cosmic rays observed \cite{Sironi2014, Werner2018}. This can also be expressed in terms of the magnetization parameter $\upsigma = B'^2/(4\pi\rho c^2)$ as $E'_{\max} \sim Z\upsigma^{1/2} m_p c^2 (n_p/n_e)^{1/2} (R'/r'_{L,p})$, but the direct form in Equation~(\ref{eq:reconnection_emax}) with explicit parameter values is more transparent and avoids the ambiguity in the value of $\upsigma$ required to reach UHECR energies. We note that $\upsigma \sim 10$, as quoted in the original text, is insufficient to reach $10^{20}$~eV via the $\upsigma$-parameterized form; the extreme HSP parameter values $B' \sim 10$~G and $R' \sim 10^{14}$~cm used here are those inferred from proton-synchrotron modeling of EHSPs such as 1ES~0229+200 \cite{Cerruti2015, Aguilar2022}, and are consistent with the extreme end of the blazar box shown in Figure~\ref{fig:hillas_plot}. Reconnection naturally produces hard particle spectra and can operate in highly magnetized environments, making it an attractive alternative to shock acceleration for explaining both the hard injection spectra required by UHECR fits and the heavy composition at the highest \mbox{energies \cite{Sironi2014, Werner2018}.}

(3) Mixed-source populations: The composition evolution could reflect a transition between different source populations at different energies. HSPs may dominate at intermediate energies ($10^{18}$--$10^{19}$~eV) with a mixed composition, while a separate population, such as more powerful radio galaxies, contributes at the highest energies, with the overall composition shaped by the rigidity-dependent maximum energies of each source \mbox{class \cite{Unger2015, Aloisio2015}.}

A complementary framework for interpreting the UHECR composition data invokes two distinct source populations rather than a single population with an energy-dependent composition. In these two-population models \cite{Muzio2019, Das2021, Ehlert2024}, the observed spectrum and composition are decomposed into a dominant heavy-nucleus-injecting population (e.g., starburst galaxies, radio galaxies, or galaxy clusters) that accounts for the bulk of the flux from the ankle to the suppression region and a subdominant light-element-injecting population (protons or helium) that extends to the highest energies and is responsible for the hardest part of the spectrum above $\sim$$10^{19.5}$~eV. This second population must satisfy stringent constraints: it must inject light nuclei with a hard spectrum ($s \lesssim 2$), have a negative or flat cosmological evolution to avoid overproducing cosmogenic neutrinos and $\gamma$-rays, and be locally abundant enough within the GZK horizon to contribute to the observed \mbox{flux \cite{Muzio2019, Ehlert2024}.}

HSP BL~Lacs are a compelling candidate for this second light-element-injecting population for several reasons. First, their photon-poor jet environments suppress photodisintegration losses, allowing protons and helium nuclei to escape essentially \mbox{intact \cite{Murase2012, Rodrigues2021}.} Second, their negative cosmological evolution \cite{Ajello2014} naturally limits the cosmogenic neutrino and $\gamma$-ray background contribution, consistent with the constraints from \textit{Fermi}-LAT and \mbox{IceCube \cite{Muzio2019, Fang2018}.} Third, the hard injection spectra ($s \sim 1.5$--$2.0$) inferred from proton-synchrotron modeling of extreme HSPs \cite{Cerruti2015, Aguilar2022} are consistent with the spectral requirements of the second population in two-population fits \cite{Das2021, Ehlert2024}. Fourth, the \mbox{Rodrigues et al. \cite{Rodrigues2021}} multi-messenger fit already places low-luminosity BL~Lacs at the center of UHECR phenomenology above the ankle, with a subdominant proton component that maps naturally onto the light-element second population of these models.

However, important tensions remain. The two-population models typically require the light-element population to dominate above $\sim$$10^{19.5}$~eV, implying proton acceleration to $\sim$$10^{20}$~eV. As discussed above, standard HSP parameters place proton acceleration marginally below this threshold, with only extreme HSPs ($B \sim 10$--$100$~G, $R \sim 10^{14}$--$10^{15}$~cm) or reconnection-powered acceleration comfortably reaching $\sim$$10^{20}$~eV \cite{Rodrigues2021, Sironi2014}. Furthermore, the required source emissivity of the light-element population in two-population fits ($\mathcal{L}_\mathrm{CR} \sim 10^{44}$--$10^{45}$ erg~Mpc$^{-3}$~yr$^{-1}$; \citet{Muzio2019, Das2021}) is broadly consistent with the HSP luminosity function \cite{Ajello2014, Rodrigues2021}, though the precise normalization depends sensitively on the assumed duty cycle and baryonic loading $\xi_\mathrm{CR}$. A definitive test of whether HSPs provide the second population in two-population UHECR models requires the improved composition sensitivity of AugerPrime and TA $\times$ 4 at the highest energies, combined with neutrino flux measurements from IceCube-Gen2 that can constrain the cosmogenic contribution from the light-element population \cite{IceCubeGen2_2021, Ehlert2024}.

(4) Hadronic interaction uncertainties: It must be emphasized that the interpretation of $X_{\text{max}}$ measurements depends critically on hadronic interaction models extrapolated from accelerator data at much lower energies \cite{Ostapchenko2011, Pierog2015}. Different models (EPOS-LHC, QGSJet-II, Sibyll) predict different $X_{\text{max}}$ values for the same composition, with systematic uncertainties of $\sim$$20$--$30$ g cm$^{-2}$ \cite{Kampert2012, Ulrich2011}. The apparent heavy composition at the highest energies might be partially an artifact of the interaction models, and the true composition could be somewhat lighter than currently inferred \cite{Kampert2012, AlvesBatista2019}.

Alternatively, the evolution of composition could reflect processes occurring during propagation rather than at the sources. Photodisintegration of heavier nuclei on the CMB and EBL during propagation produces a cascade of lighter daughter nuclei, which tends to make the observed composition lighter than the source composition at a given energy. The observed heavy composition at $10^{20}$ eV would then require an even heavier source composition, making the source identification problem more challenging. However, detailed propagation studies that account for magnetic field effects, source distributions, and composition-dependent GZK horizons can distinguish between source and propagation scenarios \cite{AlvesBatista2019}.

The interplay between source properties, acceleration mechanisms, and composition measurements remains one of the most active areas of UHECR research. Future experiments with enhanced composition sensitivity, including the AugerPrime upgrade with improved $X_{\text{max}}$ resolution and direct measurements of the muon content of air showers, will be crucial for resolving these degeneracies and determining whether HSPs can be reconciled with the observed heavy composition at the highest energies \cite{AugerPrime2016}.


\section{Outlook: Distinguishing Emission Mechanisms with \linebreak  Next-Generation Observatories}
\label{sec:future_outlook}

The observational landscape highlights a critical tension: while TXS~0506+056 provided proof-of-principle for blazar neutrino emission, the lack of broader HBL correlations and the absence of UHECR associations suggest that mechanisms are transient, rare, or obscured. Current neutrino statistics cannot resolve the degeneracy between leptonic (SSC) and hadronic (proton-synchrotron, photohadronic) models that fit electromagnetic SEDs \mbox{equally well.}

\begin{figure}[H]
  
    \includegraphics[width=0.9\textwidth]{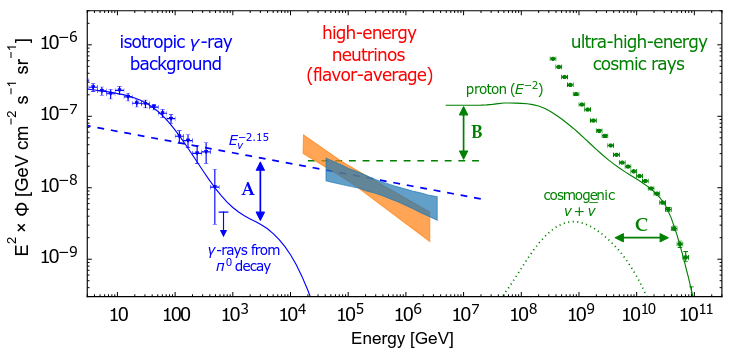}
    \caption{Multi-messenger energy generation rates: neutrino flux (orange/blue bands: IceCube with uncertainties), isotropic $\gamma$-ray background (blue points: \textit{Fermi}-LAT), and UHECR spectrum (green: Auger/TA, $E^{-2}$ scaled). Alignment across ten energy orders suggests common origin with injection rates $\sim$$10^{44}$--$10^{45}$ erg Mpc$^{-3}$ yr$^{-1}$ per channel. Dashed lines: A (cascade $\gamma$-rays $p\gamma \to \pi^0$), B (flux normalization uncertainties), and C (cosmogenic neutrinos from GZK, dotted pink). HSP dominance requires (1) hadronic neutrino production at $\sim$$1$--$10\%$ efficiency, (2) $\gamma$-ray contribution with $L_p/L_e \lesssim 10^5$, and (3) UHECR acceleration to $>$$10^{19}$ eV. Consistency with HSP parameters \linebreak  ($n \sim 10^{-6}$--$10^{-7}$ Gpc$^{-3}$, negative/flat evolution, $L_{\rm jet} \sim 10^{45}$--$10^{47}$ erg s$^{-1}$) makes them compelling multi-messenger candidates testable with IceCube-Gen2. Extracted \mbox{from \citet{IceCubeGen2_2021}.}}
    \label{fig:multimessenger_convergence}
\end{figure}

Figure~\ref{fig:multimessenger_convergence} illustrates the remarkable convergence of energy generation rates across high-energy neutrinos, the extragalactic $\gamma$-ray background, and UHECRs, spanning ten orders of magnitude. The comparable intensities suggest a common origin or closely related source populations \cite{IceCubeGen2_2021}.

However, the precise connection remains uncertain. Are HSPs responsible for all three messengers, or do different populations dominate at different energies? Does the neutrino flux arise from steady emission or rare flares? Next-generation observatories, namely IceCube-Gen2 and KM3NeT, will break this degeneracy by probing hadronic model parameters and resolving individual sources contributing to diffuse backgrounds.

\subsection{The Sensitivity Leap: IceCube-Gen2 and KM3NeT}
\label{sec:6.1}

IceCube-Gen2 will increase the instrumented ice volume by $\sim$$8\times$ relative to IceCube, improving point source sensitivity by a factor of five for detecting currently sub-threshold steady neutrino fluxes \cite{IceCubeGen2_2021}. 

\begin{figure}[H]

    \includegraphics[width=.98\textwidth]{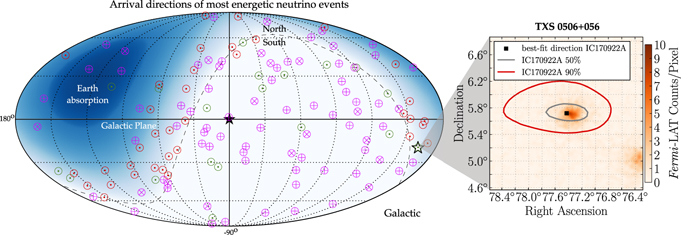}
    \caption{All-sky distribution of IceCube high-energy neutrino events in equatorial coordinates. \textbf{Left:} HESE cascades (pink circles) and tracks (pink crosses) overlaid on \textit{Fermi}-LAT AGN density. Earth absorption reduces Northern-sky ($\delta > 0^\circ$) events. Despite bright HSPs (Mrk~421, Mrk~501, PKS~2155-304), the lack of clustering beyond TXS~0506+056 indicates emission is rare/transient or below IceCube sensitivity ($E^2\Phi_\nu \sim 5 \times 10^{-13}$ GeV cm$^{-2}$ s$^{-1}$, 10 years). \textbf{Right:} TXS~0506+056 region with IC170922A ($\sim$$290$ TeV) localization (50\%/90\% contours), the only $>$$3\upsigma$ neutrino-blazar association in $\sim$$10$ years. IceCube-Gen2's $\sim$$5\times$ improved sensitivity ($E^2\Phi_\nu \sim 10^{-13}$ GeV cm$^{-2}$ s$^{-1}$) will detect HSPs contributing $\gtrsim 10\%$ to diffuse flux. Extracted from \citet{IceCubeGen2_2021}.}
    \label{fig:neutrino_skymap_current}
\end{figure}

Figure~\ref{fig:neutrino_skymap_current} shows that IceCube's high-energy neutrino events exhibit a largely isotropic pattern, with only the TXS~0506+056 association standing out. Despite prominent HSPs (Mrk~421, Mrk~501, PKS~2155$-$304), no clustering is observed, indicating either (1) predominantly transient emission during rare flares, (2) luminosities below current thresholds, or (3) limited angular resolution ($\sim$$10^\circ$--$15^\circ$ for cascades) obscuring associations. Earth absorption reduces Northern-sky events where neutrinos traverse the planet, becoming significant above $\sim$$100$ TeV.

IceCube-Gen2 will deploy $\sim$$120$ strings over $\sim$$8$ km$^3$ (vs. $\sim$$1$ km$^3$ for IceCube) with enhanced optical modules (improved photon detection, timing, and directional sensitivity via multi-PMT designs). The enlarged volume increases the effective area for TeV--PeV neutrinos by nearly 10$\times$, with denser core instrumentation improving cascade energy and angular resolution.

For Northern-sky HSPs ($\delta > 0^\circ$: Mrk~421, Mrk~501, 1ES~1959+650), Gen2 achieves $E^2 \Phi_\nu \sim 10^{-13}$ GeV cm$^{-2}$ s$^{-1}$ sensitivity for $5\upsigma$ detection after 10 years---$\sim$$5\times$ better than IceCube's $\sim 5$$\times 10^{-13}$ GeV cm$^{-2}$ s$^{-1}$ \cite{IceCubeGen2_2021}. This enables the detection of steady emissions from HSPs, contributing $\gtrsim$$10\%$ to the diffuse flux, encompassing plausible hadronic scenarios. At $\delta \sim +40^\circ$ (Mrk~421/Mrk~501), Gen2 will accumulate $\sim$$5$--$20$ events over 10 years from sources at TXS~0506+056's inferred steady-state level \mbox{($E^2\Phi_\nu \sim 3 \times 10^{-13}$ GeV cm$^{-2}$ s$^{-1}$).}

KM3NeT/ARCA in the Mediterranean offers superior Southern-sky sensitivity with denser string spacing ($\sim$$90$ m vs. $\sim$$125$ m), achieving $\sim$$0.1^\circ$ angular resolution for TeV tracks and $E^2\Phi_\upnu \sim 1$--$2 \times 10^{-13}$ GeV cm$^{-2}$ s$^{-1}$ for $\delta < -30^\circ$ sources over 10 years \cite{KM3NeT2016}. This is crucial for reducing Galactic Center confusion and identifying associations with UHECR hotspots from Auger.

Complementary coverage (Gen2: Northern, $\delta > 0^\circ$; KM3NeT: Southern, $\delta < 0^\circ$) enables all-sky HSP monitoring, including PKS~2155-304 ($\delta = -30.2^\circ$) and 1ES~0229+200 ($\delta = +20.1^\circ$). This is essential for testing UHECR--neutrino correlations with Auger's Southern-sky coverage and enables neutrino oscillation tests and flavor ratio measurements.

\subsection{Testing the Hard Spectrum Hypothesis}

One of the most definitive discriminators between emission models is the neutrino spectral shape. Current measurements show the diffuse flux has $\gamma_{\rm diffuse} = 2.37 \pm 0.09$ \cite{IceCube2021_diffuse}, but individual source spectra remain unconstrained. Next-generation detectors will enable robust spectral characterization.

\begin{figure}[H]

    \includegraphics[width=0.95\textwidth]{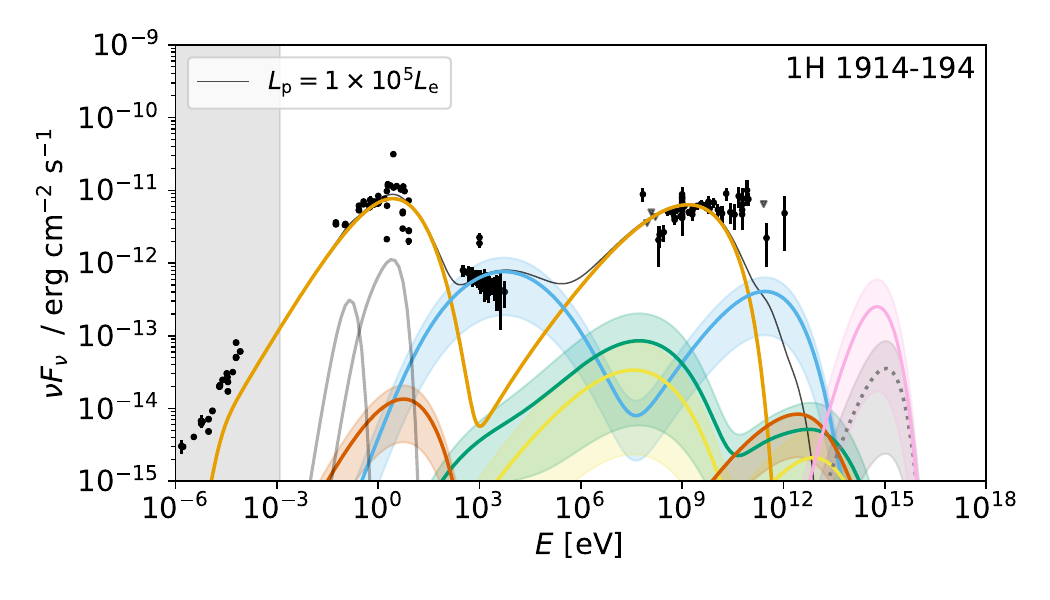}
    \caption{Multi-wavelength SED and neutrino prediction for the HSP BL~Lac 1H~1914-194 ($z = 0.137$). Black points: multi-wavelength flux observations; gray triangles: 95\% CL upper limits. Orange curve: synchrotron and inverse Compton emission from primary electrons, reproducing the double-peaked HSP structure. Gray curves: thermal emission components (accretion disk and dust torus). Blue band: Bethe-Heitler pair production ($p\gamma \rightarrow p e^+ e^-$); yellow band: photopion cascade from charged pions ($p\gamma \rightarrow \pi^+ \rightarrow \mu^+ \rightarrow e^+$); gray dotted curve: gamma rays from neutral-pion decay ($\pi^0 \rightarrow \gamma\gamma$); green band: secondary pair production from $\gamma\gamma \rightarrow e^+e^-$ interactions. The gray solid line shows a higher baryonic loading scenario with $L_{\rm p} = 10^5\,L_{\rm e}$ for reference. Pink curve: best-fit all-flavor neutrino flux from $p\gamma$ interactions, with the pink band indicating the $1\sigma$ uncertainty, corresponding to a best-fit baryonic loading of $L_{\rm p} = 10^5\,L_{\rm e}$ and a predicted muon neutrino flux of $F_{\nu_\mu} = 10^{-12.7^{+0.4}_{-1.2}}$~erg~cm$^{-2}$~s$^{-1}$. Extracted from \citet{Rodriguez2024}.}
    \label{fig:leptohadronic_sed_example}
\end{figure}

Figure~\ref{fig:leptohadronic_sed_example} shows a lepto-hadronic model for HSP 1H~1914-194 ($z = 0.137$) demonstrating multi-messenger emission under extreme baryonic loading. The electromagnetic SED (black points, radio to TeV) shows the double-peaked HSP structure (synchrotron at $\sim$$1$ eV, inverse Compton at $\sim$$10$ TeV). Gray/light blue curves show leptonic components (electron synchrotron/SSC); colored curves (teal/yellow/green/orange) show hadronic components, including proton-synchrotron, $\pi^0$ decay cascades, and secondary pairs.

\begin{itemize}
    \item Leptonic Predictions: Standard SSC models produce no neutrinos directly. Hybrid scenarios with sub-dominant hadronic components predict softer spectra \mbox{($\gamma_\upnu \sim 2.2$--$2.5$)} following diffusive shock acceleration, with $L_\upnu \lesssim 0.1 L_\upgamma$ \cite{Petropoulou2015a}.

    \item Hadronic Signatures: Proton-dominated models require hard injection spectra \linebreak  ($s_p \approx 2.0$--$2.3$) with baryonic loading $L_{\rm p}/L_{\rm e} \sim 10^3$--$10^5$ \cite{Rodriguez2024}. Extreme proton-synchrotron scenarios (e.g., 1ES~0229+200) require $s_p \approx 1.5$ and $\gamma_{\text{p,min}} \sim 100$ \cite{Cerruti2015, Petropoulou2020b}, yielding the neutrino spectral indices \cite{Mannheim1992, Mucke2003}
    \begin{equation}
    \gamma_\nu \approx s_{\rm p} \approx 1.5 \text{--} 2.3\,,
    \label{eq:neutrino_spectral_index}
    \end{equation}
    spanning hard (reconnection, turbulent acceleration) to moderate (shock acceleration) spectra.
    
\end{itemize}

IceCube-Gen2 will measure $\gamma_\upnu$ with precision $\Delta\gamma_\upnu \sim 0.3$ for sources at the threshold, sufficient for $>$$3\upsigma$ discrimination with $\sim$$20$ neutrinos \cite{IceCubeGen2_2021}. The expected event count \mbox{is \cite{Gaisser2016}}
\begin{equation}
N_\nu = T \int_{E_{\min}}^{E_{\max}} A_{\rm eff}(E)\,
\Phi_\nu(E)\,dE\,,
\label{eq:event_rate}
\end{equation}
where $A_{\rm eff}(E)$ is the energy- and declination-dependent neutrino effective area. For IceCube, $A_{\rm eff} \sim 10^4$--$10^5$~cm$^2$ at $E_\nu \sim 100$~TeV for track-like events in the Northern sky, rising to $\sim$$10^6$~cm$^2$ at $E_\nu \sim 1$~PeV \cite{Aartsen2017}. IceCube-Gen2, with its $\sim$$5\times$ improved point-source flux sensitivity, achieves $A_{\rm eff} \sim 5\times10^4$--$5\times10^5$~cm$^2$ at 100~TeV \cite{IceCubeGen2_2021}. These values are substantially below the $\sim$$1$~km$^2$ geometric footprint of the detector, reflecting losses in detection efficiency from absorption, scattering, and trigger thresholds. Scaling from the single IceCube blazar association (TXS~0506+056) detected over $\sim$10 years, Gen2's $\sim5\times$ improved flux sensitivity implies a detectable source count scaling as $N \propto 5^{3/2} \approx 11$, yielding $\sim$5--15 resolvable HSP point sources over 10 years. We emphasize that this is an HSP-specific estimate based on the observed blazar--neutrino association rate; the broader Gen2 discovery potential for all steady neutrino source populations across the full sky is separately estimated at 50--100 sources~\cite{IceCubeGen2_2021}, and the two figures should not be conflated. For nearby HSPs such as Mrk~421 and Mrk~501 ($d_L \approx 140$--$150$~Mpc) emitting at hadronic model levels ($L_p/L_e \sim 10$--$100$), Gen2 could accumulate $\sim$5--15 events over 10~years, enabling spectral index measurements with modest but meaningful \mbox{statistical power.}

\subsection{Resolving the Diffuse Background and UHECR Sources}

Stacking analyzes constrain the contribution of resolved $\gamma$-ray blazars to $\lesssim$$27\%$ of the diffuse neutrino flux \cite{Aartsen2017}, meaning that blazars may contribute anywhere from zero up to this upper limit. Future observatories will therefore either detect a blazar population signal within this allowed range or push the upper limit further down; in either case, determining whether the bulk of the diffuse flux arises from blazars, numerous faint unresolved sources, or entirely different source classes.

\begin{figure}[H]

    \includegraphics[width=0.8\textwidth]{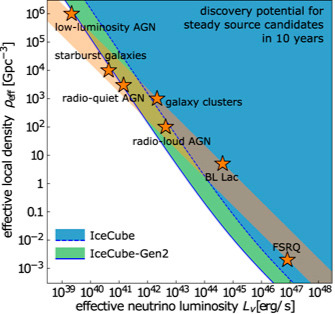}
    \caption{IceCube and Gen2 discovery potential for neutrino sources: Effective local density vs. required luminosity. Orange band: Diffuse flux compatibility \mbox{($E^2\Phi_{\rm diffuse} \sim 10^{-8}$ GeV cm$^{-2}$ s$^{-1}$ sr$^{-1}$);} edges show positive/negative evolution. Gold stars: Candidate populations (starburst galaxies, radio-quiet AGNs, clusters, BL Lacs/HSPs, FSRQs). Blue/green regions: IceCube/Gen2 10-year $>5\upsigma$ discovery zones. BL Lacs ($n \sim 10^{-6}$--$10^{-7}$~Mpc$^{-3}$, $L_{\nu} \sim 10^{43}$--$10^{44}$~erg~s$^{-1}$) sit at Gen2's boundary. Gen2 will either discover HSP point sources of order $\sim$5--15  at $z \lesssim 0.5$ (consistent with the scaling from the single TXS~0506+056 association and supportive of a blazar contribution to the diffuse flux) or constrain the HSP contribution to $\lesssim$$20\%$, redirecting searches toward other source classes. The broader Gen2 10-year discovery potential shown here ($>$5$\upsigma$ detection zones) refers to all steady neutrino source populations across the full sky~\cite{IceCubeGen2_2021}, not exclusively to HSPs. Extracted from \citet{IceCubeGen2_2021}.}
    \label{fig:source_discovery_potential}
\end{figure}

If HSPs produce UHECRs, they must also produce neutrinos via hadronic interactions ($p\gamma \to \pi^\pm, \pi^0$) or cosmogenic processes (GZK neutrinos). Figure~\ref{fig:source_discovery_potential} maps candidate populations by effective local density versus required luminosity to explain the diffuse flux ($E^2\Phi_{\text{diffuse}} \sim 10^{-8}$ GeV cm$^{-2}$ s$^{-1}$ sr$^{-1}$). The orange band shows compatibility, with edges assuming positive/negative redshift evolution.

Blue (IceCube) and green (Gen2) regions show $>$$5\upsigma$ discovery zones for steady emission over 10 years. As Figure~\ref{fig:source_discovery_potential} shows, BL Lacs fall within or near the IceCube discovery zone, meaning that IceCube already strongly disfavors BL Lacs as the dominant steady emitters of the diffuse neutrino flux if their emission is persistent \cite{Aartsen2017, Wang2014, Zhang2017}. This is a negative result for the blazar hypothesis: if HSPs were the primary steady-state sources of the diffuse flux at the luminosities required to explain it, IceCube should already have detected them. The non-detection therefore implies that either the HSP contribution is sub-dominant or neutrino emission is transient or intermittent rather than steady, with duty cycles sufficiently low that time-averaged fluxes fall below IceCube's \mbox{sensitivity \cite{Murase2016, Palladino2019}.} Gen2 will probe the remaining allowed parameter space and either detect a sub-dominant HSP population or further tighten the constraints. The 3HSP catalog contains $\sim$$2000$ sources \cite{Chang2019}, but the effective source density contributing to the diffuse flux depends critically on duty cycle and beaming. The required neutrino luminosity per HSP source can be estimated from the standard relation between the diffuse flux and the source \mbox{population \cite{Murase2016, MuraseWaxman2016}.} The neutrino luminosity density of the population is
\begin{equation}
    \mathcal{L}_\nu \equiv n_{\rm HSP} L_{\nu,\rm HSP} 
    \approx \frac{4\pi H_0}{c} 
    \frac{E^2\Phi_{\rm diffuse}}{\xi_z}\,,
    \label{eq:lnu_density}
\end{equation}
where
 $H_{0}\approx70$~km~s$^{-1}$~Mpc$^{-1}$ is the Hubble constant, $E^{2}\Phi_{\rm diffuse}\sim10^{-8}$~GeV~cm$^{-2}$~s$^{-1}$~sr$^{-1}$ is the observed diffuse neutrino flux, and $\xi_{z}$ are the cosmological evolution factors \linebreak  ($\xi_{z}\sim1$--$3$ for negative/flat evolution, $\sim$$10$--$30$ for positive evolution). The individual source luminosity is then
\begin{equation}
    L_{\nu,\rm HSP} \approx \frac{4\pi H_0}{c} 
    \frac{E^2\Phi_{\rm diffuse}}{n_{\rm HSP}\xi_z}\,,
    \label{eq:lnu_hsp}
\end{equation}
For $n_{\rm HSP}\sim10^{-6}$~Mpc$^{-3}$ and $\xi_{z}\sim1$--$3$ (appropriate for negative-to-flat cosmological evolution~\cite{Ajello2014}), this yields $L_{\nu,\rm HSP}\sim10^{43}$--$10^{44}$~erg~s$^{-1}$, consistent with hadronic models ($L_{p}/L_{e}\sim10$--$100$, $\epsilon_{\nu}\sim1\%$--$10\ $) \cite{Rodriguez2024,Murase2016}.

\subsubsection*{Outcomes}

\begin{itemize}
    \item Detection: Under optimistic assumptions about source density, duty cycle, and baryonic loading, and if HSPs follow negative evolution, (declining at $z > 1$; \mbox{\citet{Ajello2014}),} Gen2 could detect sources of order $\sim$$5$--$15$ at $z < 0.5$ spatially correlated with 3HSP at high Galactic latitudes (see Section~\ref{sec:6.1} for the corrected estimate). Such a detection would strongly support a dominant blazar contribution to the diffuse neutrino flux and provide circumstantial evidence for a link to UHECRs, though it would not constitute a definitive confirmation without independent composition and spectral measurements \cite{IceCubeGen2_2021,Murase2022}.

    \item Non-Detection: A null result would disfavor HSPs as dominant steady neutrino emitters under the model assumptions adopted here, pointing toward either sub-dominant HSP contributions, highly transient emission with low duty cycles, or alternative source classes such as starburst galaxies ($n \sim 10^{-5}$~Gpc$^{-3}$) or radio-quiet AGNs ($n \sim 10^{-4}$~Gpc$^{-3}$) \cite{Murase2022}. However, interpreting a null result requires careful accounting of the model assumptions, particularly source density, duty cycle, and baryonic loading, since these strongly affect the predicted signal level.
    \end{itemize}

Gen2 resolves sources contributing $>$$10\%$ to the diffuse flux with \mbox{$n \lesssim 10^{-6}$~Gpc$^{-3}$ \cite{MuraseWaxman2016}.} The critical resolvability condition is that the individual source flux exceeds the detector point-source sensitivity \cite{MuraseWaxman2016}, as follows:
\begin{equation}
    \Phi_{\rm source} \gtrsim \Phi_{\rm min}\,,
    \label{eq:resolvability}
\end{equation}
where $\Phi_{\rm min}$ is the minimum detectable flux of the instrument. For a population of $N_{\rm sources}$ contributing equally to the diffuse flux, each source contributes a flux of order $\Phi_{\rm source} \sim \Phi_{\rm diffuse}\Delta\Omega/N_{\rm sources}$, where $\Delta\Omega \sim 4\pi$~sr is the full sky solid angle. Resolvability therefore requires $N_{\rm sources} \lesssim \Phi_{\rm diffuse}\Delta\Omega/\Phi_{\rm min}$. For Gen2's point-source sensitivity $\Phi_{\rm min} \sim 10^{-13}$~GeV~cm$^{-2}$~s$^{-1}$ at 100~TeV and $E^2\Phi_{\rm diffuse} \sim 10^{-8}$~GeV~cm$^{-2}$~s$^{-1}$~sr$^{-1}$, populations with $n \lesssim 10^{-6}$~Gpc$^{-3}$ are resolvable over 10 years of operation \cite{IceCubeGen2_2021, MuraseWaxman2016}.

\clearpage 
\subsection{Multi-Messenger Diagnostics: X-Ray Polarization and Temporal Correlations}

\textls[-15]{\textit{IXPE}, launched in 2021, measures X-ray polarization (2--8 keV) with $\sim$$1\%$ \mbox{precision \cite{Weisskopf2022},}} distinguishing leptonic from hadronic scenarios. For synchrotron radiation from a power-law particle distribution with spectral index $s$, the linear polarization degree $\Pi$ (defined as the ratio of the polarized flux to the total flux) is given by \cite{Rybicki1986}
\begin{equation}
    \Pi = \frac{s + 1}{s + 7/3}\,,
    \label{eq:polarisation}
\end{equation}
where $\Pi = 1$ corresponds to fully polarized emission and $\Pi = 0$ to unpolarized emission. For
a power-law spectral index $s$, Equation~(\ref{eq:polarisation}) gives $\Pi\approx65\%$ for hard protons ($s_{p}\sim1.5$) and $\Pi\approx72\%$ for a softer electron distribution ($s_{e}\sim2.5$). Synchrotron polarization is a monotonically increasing function of $s$: softer spectra yield higher maximum intrinsic polarization. The practical discriminant between leptonic and proton-synchrotron scenarios therefore does not rest primarily on the absolute value of $\Pi$ alone, but also on the temporal stability of the polarization position angle (PA): proton-synchrotron emission is characterized by a slowly evolving PA (proton cooling times $t_{p}\sim R/c$$\sim$ hours to days) compared to electron synchrotron (cooling times $t_{e}$$\sim$ minutes to hours)~\cite{Zhang2013}. A stable PA observed in the hard X-ray band ($\gtrsim$$2$~keV) during a neutrino multiplet therefore provides a stronger smoking gun for hadronic emission than the absolute value of $\Pi$ alone. Conversely, a rapidly rotating PA during a neutrino-active state favors leptonic X-rays with neutrinos from a distinct, geometrically separated hadronic component~\cite{Liodakis2022}.

\subsubsection*{Correlation Scenarios}

\begin{itemize}
    \item High $\Pi$ + Neutrinos: Reconnection in ordered fields accelerates protons, produces neutrinos, and organizes magnetic fields. A stable PA indicates a quasi-steady field over days to weeks.
    \item Low $\Pi$ + Neutrinos: Turbulent regions with tangled fields; neutrinos come from an extended jet or decoupled zones.
    \item PA Swing + Neutrinos: Field reconfiguration due to jet instabilities or helical modes.
\end{itemize}

For Mrk~421 flares, (1) Gen2 detects a neutrino multiplet (3--5 events, 1--2 weeks), and alerts within hours; (2) \textit{IXPE} responds (100--200 ks, 2--3 days), and measures $\Pi/\mathrm{PA}$; (3) multi-wavelength coverage (\textit{Swift}, \textit{NuSTAR}, \textit{Fermi}, CTAO) provides a full SED; and (4) optical polarimetry (RoboPol) probes the field at lower frequencies. This ``golden event'' would establish blazars as extreme accelerators.

\subsection{Temporal Evolution and Real-Time Multi-Messenger Campaigns}

Gen2 enhances real-time monitoring, lowering the alert threshold to \mbox{$E_\upnu \sim 10$--$20$ TeV,} increasing alerts from $\sim$$10$--$15$/year to $\sim$$100$--$300$/year \cite{IceCubeGen2_2021}. This enables rapid follow-up with \textit{Swift} ($\sim$$50$ s slew), \textls[-25]{\textit{Fermi}-LAT (continuous), ZTF/LSST (optical), MeerKAT/ASKAP/SKA (radio), and CTAO/VERITAS/MAGIC/H.E.S.S. ($\sim$$30$ seconds--hours 
	response).}

\subsubsection*{Key Objectives}

\begin{enumerate}
    \item Orphan Flare Characterization: If $>$$30\%$ of neutrinos have no concurrent electromagnetic activity, it favors spatial/temporal decoupling (spine--sheath); if $<$$10\%$, it favors co-spatial production \cite{IceCube2018b, Keivani2018}.

    \item Time-Domain Correlations: Three timing signatures distinguish scenarios:
    \begin{itemize}
        \item \textit{Simultaneous} ($\Delta t \sim 0$): Co-spatial production. Variability timescale \linebreak  $\Delta t_{\rm var} \sim R/(c \delta_{\rm D})$ constrains size.
        \item \textit{Delayed neutrinos} ($\Delta t$$\sim$ hours--days): Secondary production downstream in slower material.
        \item \textit{Neutrino precursors} ($\Delta t$$\sim$ hours--days, neutrinos leading): Hadronic acceleration before leptonic emission, favoring reconnection \cite{Petropoulou2020a}.
    \end{itemize}
    
    \item Long-Term Monitoring: Under the assumption that HSPs contribute to the diffuse flux at the level inferred from TXS~0506+056, a stacking analysis combining sub-threshold signals from $\sim$$20$--$30$ HSPs drawn from the 3HSP catalog could accumulate $\sim$$10$--$20$ neutrino events in total over 10 years of Gen2 operation, potentially sufficient for a $\sim$$3\upsigma$ population-level signal \cite{Aartsen2017}. These estimates are model-dependent and sensitive to assumptions about source density, duty cycle, and baryonic loading. Correlation of neutrino arrival times with $\gamma$-ray activity states measured by \textit{Fermi}-LAT and Swift provides an additional diagnostic. As illustrative forecasting benchmarks for a specific analysis setup, correlation coefficients of $\rho > 0.5$ would be consistent with co-spatial neutrino and $\gamma$-ray production; $\rho < -0.3$ would suggest temporal decoupling consistent with orphan neutrino flares; and $|\rho| < 0.2$ would be consistent with independence or low duty cycles. However, these thresholds are not generic confirmation criteria, and their interpretation depends on the specific source sample, time baseline, and analysis methodology adopted \cite{Petropoulou2020a}.
\end{enumerate}

\subsection{The Cherenkov Telescope Array Observatory and Multi-Messenger Synergies}

The Cherenkov Telescope Array Observatory (CTAO) will provide 20 GeV--300 TeV coverage with $\sim$$0.05^\circ$ angular resolution at 1 TeV and $\sim$$10\%$ energy resolution \cite{CTA2019}. CTA-South in Chile will deploy 14 Medium-Sized Telescopes (MSTs) and 37 Small-Sized Telescopes (SSTs), while CTA-North in La Palma will feature 4 Large-Sized Telescopes (LSTs) and 9 MSTs.

CTAO's enhanced sensitivity will enable the identification of neutrino counterparts through deep observations of 50--100 h, reaching $\sim$$0.3\%$ of the Crab Nebula flux at 1~TeV ($F > 1$ TeV$\sim$$3 \times 10^{-14}$ photons cm$^{-2}$ s$^{-1}$), representing a 5--10-factor improvement over current instruments. This will uncover ``hidden'' accelerators responsible for orphan neutrino flares. Conversely, non-detections of TeV emission from neutrino-detected HSPs would imply either extreme opacity in the source ($u_{\rm ph} \gtrsim 10^3$ erg cm$^{-3}$, compactness $\ell \gtrsim 10^4$) that absorbs TeV photons while allowing neutrinos to escape, or spatial decoupling, where neutrinos originate from regions distinct from the TeV-emitting zones.

The low-energy threshold of 20 GeV provided by the LSTs will overlap with \textit{Fermi}-LAT's high-energy range (extending to 300 GeV), creating seamless spectral coverage across the critical 20--300 GeV transition region. This capability will discriminate between pure SSC models producing smooth inverse Compton components with photon indices $\Gamma_{\rm IC} \sim 1.5$--$2.5$ and hadronic models where cascade emission from $\pi^0$ decay creates spectral hardening ($\Gamma < 1.5$) or bumps in this energy range. Furthermore, CTAO's ability to detect sub-hour variability with timescales $\Delta t_{\rm var} \sim 1$--$10$ min will directly constrain the emitting region size to $R \lesssim 3 \times 10^{14} (\delta_{\rm D} / 10) (\Delta t_{\rm var} / 1\,{\rm h})$ cm. Such compact regions imply photon densities $u_{\rm ph} \sim 10^{3}$--$10^{5}$ erg cm$^{-3}$, enabling efficient $p\gamma$ interactions (optical depth $\tau_{{\rm p}\upgamma} \gtrsim 0.1$--$1$) for neutrino production while simultaneously creating strong pair-production opacity ($\tau_{\upgamma\upgamma} \gg 1$) for TeV photons, requiring high bulk Lorentz factors $\Gamma \gtrsim 20$--$50$ to reduce comoving-frame energies.

Temporal correlation studies with sub-hour resolution will distinguish between emission scenarios. Zero-lag correlation with a correlation coefficient $\rho > 0.7$ and a time lag of $|\Delta t| < 1$ days between TeV $\gamma$-rays and neutrinos would indicate co-production via $p\gamma$ interactions in the same compact acceleration zone. Delayed neutrinos lagging $\gamma$-rays by hours to days would suggest production in slower-moving material downstream or in extended jet regions. Anti-correlation or no correlation ($|\rho| < 0.2$) would imply spatial decoupling between different jet components or entirely independent physical mechanisms.

For bright nearby HSPs like Mrk~421 undergoing major flares, the combination of CTAO, IceCube-Gen2, and \textit{IXPE} observations would provide definitive multi-messenger evidence for hadronic processes. A neutrino multiplet detection would confirm hadronic acceleration to at least $E_{\rm p} \gtrsim\,\rm{few}\,10^2$ TeV, and in many $p\gamma$ blazar scenarios, to PeV or higher energies, depending on the target photon field. High X-ray polarization ($\Pi > 70\%$) would indicate either proton-synchrotron emission or ordered magnetic field structures consistent with magnetic reconnection. A hard TeV spectrum with a photon index $\Gamma_\upgamma < 2$ would be consistent with hadronic emission mechanisms (proton-synchrotron or $\pi^0$ decay). Correlated variability on similar timescales ($t_{\rm var}$$\sim$ hours to days) would confirm co-spatial production in compact regions with characteristic sizes $R \sim 10^{15}$ cm. Such a ``golden event'' would represent a breakthrough in multi-messenger astronomy, establishing HSP BL Lacs as the long-sought source of ultra-high-energy cosmic rays and confirming their role as sites of extreme particle acceleration.

The Large High Altitude Air Shower Observatory (LHAASO), located at a 4410~m altitude in Sichuan, China, provides a complementary and powerful probe of TeV--PeV $\gamma$-ray emission from HSP blazars \cite{Cao2019, Cao2023}. Unlike IACTs such as CTAO, which operate with limited duty cycles ($\sim$$10\%$) and small fields of view ($\sim$$5^\circ$--$10^\circ$), LHAASO operates as a wide-field ($\sim$$2$~sr instantaneous field of view) water Cherenkov detector array with a duty cycle exceeding $95\%$, providing continuous all-sky monitoring at energies from $\sim$$100$~GeV to beyond 1~PeV \cite{Cao2019}. This makes LHAASO particularly well suited for detecting long-duration or slowly evolving $\gamma$-ray emissions from HSP blazars that might be missed by pointed IACT observations, as well as for catching rare but extreme flaring events through its unbiased sky coverage.

LHAASO has already demonstrated sensitivity to extragalactic sources, with its WCDA (Water Cherenkov Detector Array) and KM2A (Kilometer-Squared Array) components providing complementary coverage across the TeV--PeV band \cite{Cao2023}. For nearby HSPs such as Mrk~421 ($z = 0.031$) and Mrk~501 ($z = 0.034$), LHAASO's sensitivity above $\sim$$10$~TeV is particularly relevant: emission in this energy range probes the highest-energy end of the IC component in leptonic models and the onset of proton-synchrotron or $\pi^0$ decay emission in hadronic models, providing a discriminant between the two scenarios that is complementary to CTAO's sub-TeV sensitivity \cite{Cao2019}. Furthermore, LHAASO's broad energy coverage enables direct measurement of the EBL-attenuated spectra of HSPs at multi-TeV energies, constraining both the intrinsic spectral shape and the EBL \mbox{density \cite{Franceschini2017}.}

In the context of multi-messenger astrophysics, LHAASO plays a crucial role as a real-time monitor for the $\gamma$-ray counterparts of IceCube neutrino alerts. Its wide field of view means that a significant fraction of IceCube alert directions are within LHAASO's instantaneous field of view at any given time, enabling rapid identification of $\gamma$-ray counterparts without the need for target-of-opportunity repointing \cite{Cao2023}. For HSP blazars specifically, simultaneous detection of a neutrino multiplet by IceCube-Gen2 and enhanced TeV emission by LHAASO from the same direction would provide strong evidence for hadronic activity, with the ratio of TeV $\gamma$-ray to neutrino flux constraining the $\pi^0/\pi^\pm$ production ratio and hence the dominant interaction channel ($p\gamma$ vs. $pp$) \cite{Murase2022, Rodriguez2024}. LHAASO's ongoing monitoring of the Northern sky, which includes the majority of the well-studied HSP population (Mrk~421, Mrk~501, 1ES~1959+650, RGB~J0710+591), thus makes it an essential component of the next-generation multi-messenger network alongside IceCube-Gen2, KM3NeT, \mbox{and CTAO.}

\subsection{Summary: The Path Forward}

The coming decade will address whether HSP BL Lac objects serve as primary sources of high-energy neutrinos and ultra-high-energy cosmic rays through the convergence of multiple independent diagnostics \cite{IceCubeGen2_2021, Murase2022}.

Neutrino spectral characteristics provide direct constraints on acceleration physics: magnetic reconnection and turbulent processes produce hard spectra ($\gamma_\upnu \sim 1.5$--$2.0$), while diffusive shock acceleration yields softer spectra ($\gamma_\upnu \gtrsim 2.2$) \cite{Sironi2014, Petropoulou2020a}. IceCube-Gen2's spectral resolution ($\Delta\gamma_\upnu \sim 0.3$) will enable $>3\upsigma$ discrimination with $\sim$$20$ neutrinos per source \cite{IceCubeGen2_2021}. Scaling from the single TXS~0506+056 association detected by IceCube over $\sim$$10$ years via $N \propto 5^{3/2}$, Gen2 is expected to detect HSP point sources of order $\sim$5--15 concentrated at high Galactic latitudes ($|b| > 30^\circ$) and low redshifts ($z < 0.5$), matching the 3HSP distribution, which would strongly support but not definitively confirm HSP dominance of the diffuse neutrino flux
, since the predicted source count depends sensitively on model-dependent assumptions \cite{Chang2019, IceCubeGen2_2021}. Detection at this level would confirm HSP dominance of the diffuse neutrino flux, while fewer than five resolved sources would redirect attention to starburst galaxies ($n \sim 10^{-5}$~Gpc$^{-3}$) or radio-quiet AGNs ($n \sim 10^{-4}$~Gpc$^{-3}$) \cite{Palladino2019, Murase2016}.

Temporal relationships between neutrino and electromagnetic emissions probe spatial structure \cite{Keivani2018, IceCube2018b}. Orphan neutrino flares exceeding 30\% would favor structured spine--sheath jets with decoupled regions; fractions below 10\% indicate co-spatial \mbox{production \cite{Petropoulou2020a, Garrappa2019}.} High X-ray polarization ($\Pi > 70\%$) during neutrino emission would indicate proton-synchrotron or reconnection-ordered fields; low polarization ($\Pi < 40\%$) suggests leptonic dominance with sub-dominant hadronic components \cite{Weisskopf2022, Zhang2013}. Approximately 5--10 joint \textit{IXPE}-Gen2 observations will test these scenarios.

Spatial correlations between UHECR arrival directions (accounting for $\sim$$10^\circ$--$20^\circ$ magnetic deflections) and neutrino sources would directly confirm HSPs as cosmic-ray accelerators \cite{PierreAuger2017, Kotera2011}. Null results would suggest different source populations or stronger extragalactic fields ($B \gtrsim 1$ nG) and heavier composition ($\langle A \rangle \gtrsim 20$) than currently \mbox{estimated \cite{AlvesBatista2019}.} Temporal correlations between TeV $\gamma$-rays and neutrinos, measurable through CTAO-Gen2 observations, will provide diagnostics of the emission geometry. As model-dependent illustrative benchmarks, zero-lag correlation ($\rho > 0.7$) would be consistent with co-spatial production, delays of hours to days would suggest extended-region production, and anti-correlation or no correlation ($|\rho| < 0.2$) would be consistent with spatial decoupling \cite{CTA2019, Cerruti2019}. These thresholds are illustrative, and their diagnostic power depends on the specific source, time baseline, and observational cadence.

The synthesis of these six observational constraints—neutrino spectral indices, source counts, orphan flare prevalence, X-ray polarization--neutrino correlations, UHECR--neutrino cross-correlations, and temporal patterns—will either establish extreme HSPs as dominant sources of both ultra-high-energy cosmic rays and high-energy neutrinos or reveal an entirely different class of particle accelerators hidden in electromagnetic surveys \cite{Murase2022, ANCHORDOQUI2022}. Multi-messenger observations from next-generation facilities will resolve fundamental questions about the origin of the most energetic particles in the Universe that have persisted since their discovery more than a century ago \cite{Kotera2011}.


\section{Conclusions}
\label{sec:conclusions}

High-synchrotron-peaked BL Lac objects occupy a unique position in multi-messenger astrophysics. Their extreme particle acceleration capabilities, clean radiation environments (minimal external photon fields), and relative proximity ($z < 0.5$ for most sources) make them prime candidates for solving two enduring mysteries: the origin of ultra-high-energy cosmic rays exceeding $10^{19}$ eV and the sources of high-energy astrophysical neutrinos above 100 TeV \cite{Murase2016, ANCHORDOQUI2022}.

The 2017 multi-messenger detection of TXS~0506+056 provided the first suggestive evidence that blazars may produce detectable neutrino fluxes through hadronic interactions in relativistic jets, though the $\sim$$3\upsigma$ significance of the association fell short of the $5\upsigma$ threshold conventionally required for a discovery claim \cite{IceCube2018}. However, subsequent observations revealed complexity: the 2014--2015 orphan neutrino flare (13 events without concurrent electromagnetic activity) and stringent stacking constraints ($<$$27\%$ of diffuse flux from $\gamma$-ray blazars) indicated that simple one-zone lepto-hadronic models are insufficient \cite{IceCube2018b, Aartsen2017}. The failure to detect significant HSP--neutrino or HSP--UHECR correlations, despite theoretical appeal from Hillas considerations, suggests that efficient acceleration occurs either in radiatively hidden zones not traced by TeV emission (extended jets, sheaths), during rare transient phases with low duty cycles ($\lesssim$$10\%$), or in structured environments where leptonic and hadronic processes are spatially decoupled \cite{Petropoulou2020a, Rodrigues2019}.

Four key observational challenges confront HSPs as dominant sources. First, the extreme baryonic loading ($L_{\rm p}/L_{\rm e}$$\sim$$10^3$--$10^5$) required for detectable neutrino fluxes approaches or exceeds Eddington limits, raising questions about jet power budgets \cite{Cerruti2019, Murase2016}. Second, Pierre Auger data show evolution toward heavier composition ($\langle \ln A \rangle$ increasing from \mbox{$\sim$$1$ at} \mbox{$10^{18.5}$ eV} to $\sim$$2.5$ at $10^{19.5}$ eV), whereas shock acceleration in HSP jets predicts proton-dominated emission, though magnetic reconnection may reconcile this \mbox{tension \cite{Aab2014, Unger2015}.} Third, remarkably isotropic UHECR arrival directions (dipole amplitude $<$$7\%$ above \mbox{$8 \times 10^{18}$ eV)} are difficult to reconcile with rare, beamed sources unless magnetic deflections are large ($\gtrsim$$20^\circ$) or source densities exceed electromagnetic survey \mbox{estimates \cite{PierreAuger2017, Sigl2004}.} Fourth, the measured diffuse neutrino spectrum ($\gamma_{\upnu,\rm diffuse} = 2.37 \pm 0.09$) is softer than many hadronic blazar models predict, particularly those invoking hard proton spectra ($s_{\rm p} \sim 1.5$--$1.8$) for extreme HSPs \cite{IceCube2021_diffuse, Petropoulou2020a}.

Despite these challenges, HSPs remain compelling because alternative source classes face equally severe difficulties \cite{Murase2022, Palladino2019}. Starburst galaxies and radio-quiet AGNs require implausibly high neutrino production efficiencies ($\epsilon_\upnu \gtrsim 10\%$) given their modest luminosities. Gamma-ray bursts and tidal disruption events occur at rates of $\sim$$1$--$4$~Gpc$^{-3}$~yr$^{-1}$ for long GRBs \cite{Wanderman2010, Guetta2007}, $\sim$$0.1$--$1$~Gpc$^{-3}$~yr$^{-1}$ for short GRBs \cite{Nakar2006}, and $\sim$$10^{-4}$--$10^{-3}$~Gpc$^{-3}$~yr$^{-1}$ for TDEs \cite{Stone2016, vanVelzen2018}. While GRB rates are in principle sufficient to contribute to the diffuse neutrino flux, their short emission timescales (seconds to minutes) imply that the time-averaged neutrino luminosity is greatly reduced relative to the peak, unless neutrino emission timescales extend to months or years rather than seconds or days. TDE rates, though higher than originally quoted, remain low enough that explaining the diffuse flux requires either very high neutrino production efficiencies or extended emission \mbox{timescales \cite{Murase2020, Lunardini2017}.} Radio galaxy lobes easily satisfy Hillas criteria for UHECRs but lack the compact, high-field regions required for efficient PeV neutrino production \cite{Fang2018, Heinze2019}.

Next-generation facilities will resolve these ambiguities within the coming decade \cite{IceCubeGen2_2021, CTA2019}. IceCube-Gen2 and KM3NeT, with approximately five-fold improved sensitivity and operational status in the late 2020s to early 2030s, will either detect $\sim$5--15 HSP neutrino point sources correlated with the 3HSP catalog at low redshift ($z\lesssim0.5$) or further constrain HSP contributions to less than 20\% of the diffuse flux, definitively testing the blazar hypothesis \cite{IceCubeGen2_2021, KM3NeT2016}. We note that the broader Gen2 discovery potential for all steady neutrino source populations across the full sky is estimated at \mbox{50--100 sources~\cite{IceCubeGen2_2021};}. The more conservative figure adopted here refers specifically to HSPs, derived by scaling from the TXS~0506+056 association via $N\propto(5)^{3/2}\approx11$. Neutrino spectral measurements with precision $\Delta\gamma_\upnu \sim 0.3$ will distinguish between shock acceleration, yielding spectral indices around 2.2 to 2.5, and reconnection, producing harder indices of 1.5 to 1.8 \cite{Sironi2014, Petropoulou2020a}. The Cherenkov Telescope Array Observatory, achieving five- to ten-fold improved $\gamma$-ray sensitivity with sub-hour variability capabilities and operational by approximately 2025, will measure correlated TeV and neutrino light curves, test proton-synchrotron models through hard spectral signatures with photon indices below 2.0, and identify electromagnetic counterparts of orphan neutrino flares \cite{CTA2019}. IXPE and future X-ray polarimeters will measure polarization degrees and position angles during neutrino detections, with high polarization exceeding 70\% alongside stable position angles during neutrino multiplets, providing smoking-gun evidence for hadronic emission, while low polarization below 40\% would favor spatial decoupling or leptonic-dominated scenarios \cite{Weisskopf2022, Zhang2013}. AugerPrime and TA $\times$ 4, with approximately four-fold increased aperture and enhanced composition sensitivity operational around 2025, will measure UHECR composition event-by-event at the highest energies, testing whether sources produce the observed transition to heavy nuclei or if HSPs contribute only to the proton component below $10^{19}$ eV \cite{Kawata2017, PierreAuger2017}.

By synthesizing six key observables---neutrino spectral index, source count distribution, orphan flare prevalence, X-ray polarization--neutrino correlation, UHECR--neutrino cross-correlation, and TeV--neutrino temporal patterns---we will either confirm HSP BL Lacs, particularly extreme HSPs with hard TeV spectra, as the long-sought source of ultra-high-energy cosmic rays and significant contributors to the high-energy neutrino sky or conclusively redirect the search toward alternative accelerators such as starburst galaxies, galaxy clusters, or yet-unidentified populations \cite{Murase2022, AlvesBatista2019}. The multi-messenger observations of the 2020s and 2030s will write the definitive chapter in the century-long quest to identify the cosmic accelerators responsible for the most energetic particles in the Universe, finally answering the question posed by Victor Hess in 1912 and Enrico Fermi in 1949: where do cosmic rays originate, and how does nature achieve particle acceleration to energies millions of times beyond human technological capabilities \cite{Kotera2011, ANCHORDOQUI2022}?

The study of HSP BL Lacs thus represents not merely an investigation of a specific blazar subclass but a critical test of our understanding of extreme particle acceleration, jet physics, and the multi-messenger connections linking electromagnetic radiation, neutrinos, and cosmic rays in the most powerful, persistent sources in the Universe.

Table~\ref{tab:exec_summary} synthesizes the main findings of this review, mapping each topic to its dominant physical tension and the specific next-generation observables (X-ray polarimetry with IXPE, TeV spectroscopy with CTAO, neutrino population analyzes with IceCube-Gen2 and KM3NeT, and mass-sensitive UHECR measurements with AugerPrime and TA $\times$ 4) that will either establish HSP BL~Lacs as the source of the highest-energy cosmic particles or redirect the search toward alternative accelerator classes.

\startlandscape
\begin{table}[H]

\caption{Synthesis of this review: Principal findings, open tensions, 
and next-generation observables that will resolve them.\label{tab:exec_summary}}

\begin{tabularx}{\textwidth}{>{\raggedright\arraybackslash}m{2cm}>{\raggedright\arraybackslash}m{2.5cm}>{\raggedright\arraybackslash}m{9cm}>{\raggedright\arraybackslash}m{5.5cm}>{\raggedright\arraybackslash}m{5.5cm}}
\toprule
\textbf{\S} &
\textbf{Topic} &
\textbf{Key Points} &
\textbf{Main Tension} &
\textbf{Discriminants} \\
\midrule
Section \ref{sec:properties_selection} &
HSP properties \& Hillas &
$\nu_{\rm peak} \gtrsim 10^{17}$~Hz, $\gamma_e \gtrsim 10^6$; 
clean jets and $B \sim 0.01$--$0.1$~G favor rigidity-limited UHECR acceleration. &
$\delta,B,R$ degenerate; $\nu_{\rm peak}$ state-dependent; EHSP sparsely sampled. &
Simultaneous X-ray/TeV spectra; IXPE polarization angle (shock vs.\ reconnection). \\
\midrule
Section \ref{sec:catalogs} &
Catalogs (2WHSP/3HSP) &
$\sim$2000 HSPs via WISE selection; $\sim$40\% lack $z$; 
weighting scheme critically shapes stacking results. &
Flux-limit inhomogeneity; $z$ incompleteness; $\nu_{\rm peak}$ inconsistencies. &
Spectroscopic $z$ campaigns; uniform FOM-based weighting. \\
\midrule
Section \ref{sec:emission_models} &
Leptonic vs.\ hadronic models &
SSC fails for orphan TeV flares; hadronic models require 
$L_p/L_e \sim 10^3$--$10^5$ and risk IGRB overshoot. &
SED degeneracy; rapid variability disfavors one-zone geometry. &
Time-resolved simultaneous SEDs; optical/X-ray polarization; CTAO light-curves. \\
\midrule
Section \ref{sec:neutrino_observations} &
Neutrino population &
Stacking limits constrain $\xi_{\rm CR} \lesssim 10^2$--$10^3$; 
HSP contribution likely from rare hadronic-flare subsets, not steady emission. &
Low statistics; large trial factors; $\xi_{\rm CR}$ depends on duty cycle. &
IceCube-Gen2/KM3NeT stacking; flavor ratios; coordinated EM triggers. \\
\midrule
Section \ref{sec:txs} &
TXS~0506+056 &
2017 coincidence provided compelling evidence for blazar neutrino emission ($\sim$$3\upsigma$); not a confirmation at the $5\upsigma$ level; 
2014--2015 orphan burst unexplained by standard one-zone models. &
Single-source statistics; $\delta$ uncertainty gives orders-of-magnitude range in $L_p$. &
IceCube-Gen2/KM3NeT time-dependent analyzes; IXPE during active states. \\
\midrule
Section \ref{sec:uhecr_connection} &
UHECR connection &
Low-luminosity HSP BL~Lacs reproduce PAO spectrum above ankle 
($\xi_{\rm CR} \sim 380$; \citet{Rodrigues2021});
Auger favors mixed-to-heavy nuclei; near-isotropy constrains source density. &
Composition systematics; EGMF/GMF uncertainties; heavy-nuclei pointing $\gg$$10^\circ$. &
AugerPrime/TA $\times$ 4 mass-sensitive composition; rigidity-ordered anisotropy. \\
\midrule
Section \ref{sec:future_outlook} &
Outlook &
Joint TeV timing + X-ray polarimetry + $\nu$ statistics + UHECR composition 
needed; IceCube-Gen2, KM3NeT, CTAO, IXPE, AugerPrime within $\sim$10~yr. &
Degeneracies persist with single-messenger or time-averaged data. &
Real-time alert networks linking $\nu$, $\gamma$-ray, and UHECR observatories. \\
\bottomrule
\end{tabularx}
\end{table}
\finishlandscape




\vspace{6pt}

\funding{Stuani Pereira, L.A., gratefully acknowledges financial support from FAPESP under grant numbers 2021/01089-1, 2024/02267-9, and 2024/14769-9, as well as CNPq under grant numbers 403337/2024-0, 153839/2024-4, and 200164/2025-2. Dos Anjos, R.C., acknowledges financial support from the NAPI ``Fenômenos Extremos do Universo'' of Fundação de Apoio à Ciência, Tecnologia e Inovação do Paraná. R.C.A.'s research is supported by the CAPES/Alexander von Humboldt Program (88881.800216/2022-01), Conselho Nacional de Desenvolvimento Cient\'{i}fico e Tecnol\'{o}gico (CNPq) under grant numbers (307750/2017-5) and (401634/2018-3/AWS), as well as by the Araucária Foundation (698/2022) and (721/2022). She also thanks L'Oreal Brazil for their support, with the partnership of ABC and UNESCO in Brazil.}

\dataavailability{No new data were created or analyzed in this study.}


\conflictsofinterest{The authors declare no conflicts of interest.} 



\abbreviations{Abbreviations}{
The following abbreviations are used in this manuscript:\\
\vspace{-12pt}
\noindent 
\begin{longtable}[l]{@{}ll}
2WHSP & 2nd WISE High Synchrotron Peaked catalog\\
3HSP & 3rd High Synchrotron Peaked catalog\\
4FGL-DR3 & 4th \textit{Fermi}-LAT catalog Data Release 3\\
ADAF & Advection-Dominated Accretion Flow\\
AGN & Active Galactic Nucleus/Nuclei\\
AGILE & Astrorivelatore Gamma a Immagini LEggero\\
ARCA & Astroparticle Research with Cosmics in the Abyss\\
ASKAP & Australian Square Kilometer Array Pathfinder\\
AugerPrime & Pierre Auger Observatory Upgrade\\
BL Lac & BL Lacertae object\\
BLR & Broad Line Region\\
CL & Confidence Level\\
CMB & Cosmic Microwave Background\\
CNO & Carbon, Nitrogen, Oxygen (nuclear group)\\
CTA & Cherenkov Telescope Array\\
CTAO & Cherenkov Telescope Array Observatory\\
DESI & Dark Energy Spectroscopic Instrument\\
DSA & Diffusive Shock Acceleration\\
EBL & Extragalactic Background Light\\
EC & External Compton\\
EeV & Exa-electronvolt ($10^{18}$ eV)\\
EGB & Extragalactic Gamma-ray Background\\
EGMF & Extragalactic Magnetic Field\\
EHBL & Extreme High-Frequency-Peaked BL Lac\\
EHSP & Extreme High-Synchrotron Peaked\\
FIRST & Faint Images of the Radio Sky at Twenty Centimeters\\
FOM & Figure of Merit\\
FSRQ & Flat-Spectrum Radio Quasar\\
GCN & Gamma-ray Coordinate Network\\
Gen2 & IceCube Generation 2\\
GMF & Galactic Magnetic Field\\
Gpc & Gigaparsec\\
GRB & Gamma-Ray Burst\\
GZK & Greisen--Zatsepin--Kuzmin\\
HBL & High-frequency-Peaked BL Lac\\
HESE & High-Energy Starting Events\\
H.E.S.S. & High-Energy Stereoscopic System\\
HSP & High-Synchrotron-sPeaked\\
IACT & Imaging Atmospheric Cherenkov Telescope\\
IBL & Intermediate-Frequency Peaked BL Lac\\
IC & Inverse Compton\\
ICM & Intracluster Medium\\
IGMF & Intergalactic Magnetic Field\\
IGRB & Isotropic Gamma-Ray Background\\
IR & Infrared\\
ISP & Intermediate-Synchrotron-Peaked\\
IXPE & Imaging X-ray Polarimetry Explorer\\
KM3NeT & Cubic Kilometer Neutrino Telescope\\
KM2A & Kilometer-Squared Array\\
LAT & Large-Area Telescope\\
LBL & Low-Frequency Peaked BL Lac\\
LDDE & Luminosity-Dependent Density Evolution\\
LHAASO & Large High Altitude Air Shower Observatory\\
LL GRB & Low-Luminosity Gamma-Ray Burst\\
LSP & Low-Synchrotron Peaked\\
LST & Large-Sized Telescope\\
LSST & Legacy Survey of Space and Time\\
MAGIC & Major Atmospheric Gamma Imaging Cherenkov\\
MAXI & Monitor of All-Sky X-ray Image\\
MeerKAT & Karoo Array Telescope\\
MHD & Magnetohydrodynamics\\
MJD & Modified Julian Date\\
MST & Medium-Sized Telescope\\
NLR & Narrow Line Region\\
NVSS & NRAO VLA Sky Survey\\
\textit{NuSTAR} & Nuclear Spectroscopic Telescope Array\\
OVRO & Owens Valley Radio Observatory\\
PA & Position Angle\\
PAO & Pierre Auger Observatory\\
PIC & Particle-in-Cell\\
PeV & Peta-Electronvolt ($10^{15}$ eV)\\
PMT & Photomultiplier Tube\\
P-ONE & Pacific Ocean Neutrino Experiment\\
RIAF & Radiatively Inefficient Accretion Flow\\
RMS & Root Mean Square\\
ROSAT & R\"{o}ntgensatellit\\
SDSS & Sloan Digital Sky Survey\\
SED & Spectral Energy Distribution\\
SKA & Square-Kilometer Array\\
SMBH & Supermassive Black Hole\\
SSC & Synchrotron Self-Compton\\
SST & Small-Sized Telescope\\
SUMSS & Sydney University Molonglo Sky Survey\\
SVOM & Space-Based Multi-Band Astronomical Variable Objects Monitor\\
\textit{Swift} & Neil Gehrels \textit{Swift} Observatory\\
TA & Telescope Array\\
TDE & Tidal Disruption Event\\
TeV & Tera-Electronvolt ($10^{12}$ eV)\\
UHECR & Ultra-High-Energy Cosmic Ray\\
UV & Ultraviolet\\
VERITAS & Very Energetic Radiation Imaging Telescope Array System\\
VHE & Very High Energy\\
VLBI & Very-Long-Baseline Interferometry\\
WISE & Wide-field Infrared Survey Explorer\\
WCDA & Water Cherenkov Detector Array\\
XRT & X-Ray Telescope\\
ZTF & Zwicky Transient Facility
\end{longtable}
}

\isPreprints{}{
\begin{adjustwidth}{-\extralength}{0cm}
} 

\reftitle{References}

\let\oldthebibliography=\thebibliography
\let\endoldthebibliography=\endthebibliography
\renewenvironment{thebibliography}[1]{%
  \oldthebibliography{#1}%
  \setlength{\itemsep}{0pt}%
  \setlength{\parskip}{0pt}%
  \raggedright%
  \small%
}{\endoldthebibliography}

\isPreprints{}{
\end{adjustwidth}
} 
\end{document}